\documentclass[structabstract]{aa}  
\usepackage{graphicx}
\usepackage{txfonts}
\usepackage{natbib}
\usepackage{multirow}
\usepackage{longtable}
\usepackage{pdflscape}
\usepackage{fixltx2e}
\usepackage{diagbox}
\usepackage[colorinlistoftodos,color=blue!20]{todonotes}

\def\4he{$^4$He}

\def\kms{\mathrm{km\,s}^{-1}}
\def\e#1{\times 10^{#1}}

\def\msol{\mathrm{M}_\odot}
\def\lsol{L_\odot}

\def\up#1{$^{#1}$}
\def\down#1{$_{#1}$}

\def\h2{$\mathrm{H}_2$}
\def\so2{$\mathrm{SO}_2$}

\def\spy{\;\msol~\mathrm{ yr}^{-1}}

\begin{document}

        \title{Sulphur-bearing molecules in AGB stars}
   \subtitle{II: Abundances and distributions of CS and SiS}

   \author{T. Danilovich
          \inst{1}\fnmsep\thanks{Postdoctoral Fellow of the Fund for Scientific Research (FWO), Flanders, Belgium}
          \and
          S. Ramstedt\inst{2} \and D. Gobrecht\inst{1}
\and           L. Decin\inst{1} \and E. De Beck\inst{3} \and H. Olofsson\inst{3} 
          }

   \institute{Department of Physics and Astronomy, Institute of Astronomy, KU Leuven, Celestijnenlaan 200D,  3001 Leuven, Belgium
   \and
   Department of Physics and Astronomy, Uppsala University, Box 516, 75120, Uppsala, Sweden
   \and 
   Department of Space, Earth and Environment, Chalmers University of Technology, Onsala Space Observatory, 43992, Onsala, Sweden  \\
             \email{taissa.danilovich@kuleuven.be}
             }

   \date{Received  / Accepted }

% \abstract{}{}{}{}{} 
% 5 {} token are mandatory
 
  \abstract
   {Sulphur has long been known to form different molecules depending on the chemical composition of its environment. More recently, the sulphur-bearing molecules SO and \h2S have been shown to behave differently in oxygen-rich AGB circumstellar envelopes of different densities.}
   {By surveying a diverse sample of AGB stars for CS and SiS emission, we aim to determine in which environments these sulphur-bearing molecules most readily occur. We include sources with a range of mass-loss rates and carbon-rich, oxygen-rich, and mixed S-type chemistries. Where these molecules are detected, we aim to determine their CS and SiS abundances.}
   {We surveyed 20 AGB stars of different chemical types using the APEX telescope, and combined this with an IRAM 30~m and APEX survey of CS and SiS emission towards over 30 S-type stars. For those stars with detections, we performed radiative transfer modelling to determine abundances and abundance distributions.}
   {We detect CS towards all the surveyed carbon stars, some S-type stars, and the highest mass-loss rate oxygen-rich stars ($\dot{M}\geq 5\e{-6}\spy$). SiS is detected towards the highest mass-loss rate sources of all chemical types ($\dot{M}\geq 8\e{-7}\spy$). We find CS peak fractional abundances ranging from $\sim 4\e{-7}$ to $\sim2\e{-5}$ for the carbon stars, from $\sim 3\e{-8}$ to $\sim1\e{-7}$ for the oxygen-rich stars and from $\sim 1\e{-7}$ to $\sim8\e{-6}$ for the S-type stars. We find SiS peak fractional abundances ranging from $\sim 9\e{-6}$ to $\sim 2\e{-5}$ for the carbon stars, from $\sim 5\e{-7}$ to $\sim 2\e{-6}$ for the oxygen-rich stars, and from $\sim 2\e{-7}$ to $\sim 2\e{-6}$ for the S-type stars. }%We derived Si\up{32}S/Si\up{34}S = 11.4 for AI~Vol, the only star for which we had a reliable isotopologue detection.}
  {Overall, we find that wind density plays an important role in determining the chemical composition of AGB CSEs. It is seen that for oxygen-rich AGB stars both CS and SiS are detected only in the highest density circumstellar envelopes and their abundances are generally lower than for carbon-rich AGB stars by around an order of magnitude. For carbon-rich and S-type stars SiS was also only detected in the highest density circumstellar envelopes, while CS was detected consistently in all surveyed carbon stars and sporadically among the S-type stars.}


%\titlerunning{Sulphur molecules in the circumstellar envelopes of AGB stars}
   \maketitle
%
%________________________________________________________________

\section{Introduction}

After leaving the main sequence and passing through the red giant branch, low- to intermediate-mass stars become asymptotic giant branch (AGB) stars. These stars are characterised by intense mass loss, ejecting matter in a stellar wind which forms a circumstellar envelope (CSE) around the star \citep{Hofner2018}. These CSEs are known to be rich in different molecular species and are also a site of dust formation \citep{AGB}. The matter ejected in this manner contributes to the chemical enrichment of the interstellar medium (ISM) and the chemical evolution of galaxies \citep{Herwig2005}. 

The chemical characteristics of the CSE depend in large part on the chemical type of the AGB star, classified based on the photospheric carbon-to-oxygen ratio (C/O). Carbon-rich stars and oxygen-rich (M-type) stars have larger proportions of carbon and oxygen, respectively. S-type stars are believed to be intermediary transition objects with C/O $\sim 1$. The CSEs of carbon stars are known to contain a variety of carbon-bearing molecules \citep[see for example][]{Olofsson1993a,Cernicharo2000,Gong2015}, while the CSEs of oxygen-rich stars are typified by the presence of a variety of oxygen-bearing molecules \cite[see for example][]{Velilla-Prieto2017}. S-type stars display a mixture of the more common oxygen- and carbon-bearing molecules \citep{Schoier2011,Danilovich2014}. However, it is now known that high or low C/O do not preclude the formation of oxygen- or carbon-bearing molecules, respectively. For example, \h2O and SiO have been detected and found to have unexpectedly high abundances towards carbon stars \citep{Schoier2006,Lombaert2016}, while HCN has been detected towards oxygen-rich stars \citep{Schoier2013}.

Sulphur is a relatively abundant element which forms molecular bonds with both oxygen and carbon, among other species, and is hence found in a variety of molecules in the CSEs of AGB stars. For example, SO and \so2 are commonly found in the CSEs of oxygen-rich AGB stars \citep{Danilovich2016}, while CS has been found to be very abundant in carbon-rich AGB stars \citep{Olofsson1993a}. However, molecular abundances in AGB CSEs have been found to not only depend on the C/O of the CSE, but also on other factors, such as the density of the stellar wind, which is related to the mass-loss rate. For example, \cite{Danilovich2016} found different radial distributions of SO for low mass-loss rate AGB stars compared with higher mass-loss rate AGB stars, indicating that SO was formed at larger radii in the latter case. \h2S, which contains neither carbon nor oxygen, is preferentially detected in higher mass-loss rate oxygen-rich stars \citep{Danilovich2017a} and, to date, has only been detected, weakly, towards one carbon-rich AGB star, CW~Leo \citep{Cernicharo1987,Omont1993,Cernicharo2000}. Similar patterns of different molecular occurrences between high and low mass-loss have also been seen for other molecules such as SiO, which has been found to have higher abundances for lower mass-loss rate AGB stars \citep[][]{Gonzalez-Delgado2003,Schoier2006}.

Previously, SiS was studied in a sample of oxygen-rich and carbon-rich AGB stars by \cite{Schoier2007}, in which they generally find abundances of SiS in carbon-rich stars about an order of magnitude higher than in oxygen-rich stars. This strongly suggests that SiS is preferentially formed in carbon-rich CSEs. Although their initial models were based on a Gaussian abundance distribution profile, they find they needed to include a `core' component with a higher abundance and a small radius to properly fit their SiS observations. \cite{Decin2010} found a similar result with a high-abundance inner component and a lower-abundance outer component when modelling SiS for the oxygen-rich star IK~Tau. 

\cite{Danilovich2015a} observed the SiS ($6\to5$) line concurrently with the CO ($1\to0$) line towards 29 AGB stars of various chemical types and mass-loss rates. 
They detected SiS towards 12 of the sample stars with the general trend being that SiS was only detected towards the higher mass-loss rate stars. Part of our goal in this work is to confirm whether such a mass-loss rate or density dependent trend exists for SiS.
\cite{Lindqvist1988} searched for both CS and SiS (among other molecules) in a sample of 31 oxygen-rich stars. They detected CS ($2\to1$) in only four sources and SiS ($5\to4$) in only one source (TX~Cam). \cite{Bujarrabal1994} searched for CS ($3\to2$) and ($5\to4$), and SiS ($5\to4$) in a diverse sample of evolved stars. They detected CS in six of the highest mass-loss rate oxygen-rich stars, all three S-type stars and the majority of their carbon-rich stars. They detect SiS in three nearby high mass-loss rate oxygen-rich stars, five carbon stars, and none of the S-type stars.

\cite{Olofsson1993a} surveyed a large sample of carbon-rich stars for several molecules and derived photospheric SiS abundances, and both circumstellar and photospheric CS abundances for many of them. In general, they found that the circumstellar abundances tended to be higher by factors of five to ten than the photospheric abundances. In some cases they found circumstellar CS abundances high enough to account for or exceed the total amount of sulphur expected to be present based on the solar sulphur abundance \citep{Asplund2009}. They suggest this may be due to under-predicting mass-loss rates, which affect the derived abundances, or the effect of a simplified radiative transfer analysis.

To properly constrain the occurrences of the most common sulphur-bearing molecules in AGB CSEs, we performed a survey of 20 AGB stars, which covered all three chemical types and a range of mass-loss rates {(low mass-loss rates: $\sim 10^{-8}$ to a few $10^{-7}\spy$, intermediate mass-loss rates: $\sim10^{-6}\spy$ and higher mass-loss rates: up to a few $10^{-5}\spy$). Our sample does not include the highest mass-loss rate OH/IR stars, which will be studied separately}. We focussed our search on rotational transitions of CS, SiS, SO, \so2 and \h2S and carried out the survey using the Atacama Pathfinder Experiment \citep[APEX\footnote{This publication is based on data acquired with the Atacama Pathfinder Experiment (APEX). APEX is a collaboration between the Max-Planck-Institut f\"ur Radioastronomie, the European Southern Observatory, and the Onsala Space Observatory.},][]{Gusten2006}. The first results from this survey, for \h2S, are presented in \cite{Danilovich2017a}. This survey was supplemented with a smaller survey of 9 M-type AGB stars with the Onsala 20~m telescope (OSO), focussing on \so2, SO and SiS at low frequencies. Additionally, a survey of CS and SiS towards 33 S-type stars has also been included in this study, with observations gathered from the IRAM 30~m telescope and APEX.
In this paper we focus on the detections of SiS and CS and perform radiative transfer analyses to determine the abundances and abundance distributions of these two molecules.

%__________________________________________________________________

\section{Sample and observations}\label{obs}

\subsection{APEX sulphur survey}

We surveyed several rotational emission lines of sulphur-bearing species in a chemically diverse sample of 20 AGB stars, including seven M-type stars, five S-type stars and eight carbon stars and covering mass-loss rates from $\sim9\e{-8}\spy$ to $\sim2\e{-5}\spy$. The first results from this survey have already been presented in \cite{Danilovich2017a} for \h2S, which also further describes the observing programme carried out {in March--April and August--December of 2016,} using the Swedish-ESO PI receiver \citep[SEPIA Band 5,][]{Billade2012,Belitsky2018} for APEX \citep{Gusten2006} and the Swedish Heterodyne Facility Instrument \citep[SHeFI,][]{Belitsky2006,Vassilev2008}. In this study we have focussed on the SiS and CS observations obtained during this survey. SO and \so2 results will be presented in future papers in this series.

The full sample of stars for which SiS and CS were surveyed are listed in Table \ref{fullsample} {along with the stars from the other observations discussed below}. The lines that were included in the survey are listed in Table \ref{trans}, as are other available lines from other telescopes that we used to constrain our models. Table \ref{SiSobs} includes all the detected SiS lines and their integrated main beam intensities and Table \ref{SiSnondet} lists the rms noise at $1~\kms$ for each observed SiS line, whether or not it was detected. The integrated main beam intensities for the CS lines are listed in Table \ref{CSobs}, as are the rms noise values for all observed lines.

\begin{table*}[tp]
\caption{Basic parameters of surveyed stars.}\label{fullsample}
\centering
\scalebox{0.94}{
\begin{tabular}{llcccccccc}
\hline\hline
Star    &       RA      &       Dec     &       $\upsilon_\mathrm{LSR}$ &       Distance        &               $\dot{M}$       & $T_\mathrm{eff}$ & $\upsilon_\infty$    & Ref.\\
                &               &       &       [$\kms$] & [pc] & [$\spy$] &[K]& [$\kms$]\\
\hline
\multicolumn{2}{l}{\quad\it Carbon stars}\\
\object{R Lep}  &       04:59:36.35     &       $-$14:48:22.5   &       11      &       432     &       $       8.7\e{-7}       $       & 2200 & 18 &1\\
\object{V1259 Ori}      &       06:03:59.84     &       $+$07:25:54.4   &       42      &       1600\phantom{0} &       $       8.8\e{-6}       $       & 2200 & 16  &1\\
\object{AI Vol} &       07:45:02.80     &       $-$71:19:43.2   &       $-$39   &       710     &       $       4.9\e{-6}       $       & 2100 &12 &1\\
\object{X TrA}  &       15:14:19.18     &       $-$70:04:46.1   &       $-$2    &       360     &       $       1.9\e{-7}       $       & 2200 & 6.5 &1\\
\object{II Lup} &       15:23:04.91     &       $-$51:25:59.0   &       $-$15.5 &       500     &       $       1.7\e{-5}       $       & 2400 & 21.5  &1\\
\object{V821 Her}       &       18:41:54.39     &       +17:41:08.5     &       $-$0.5  &       600     &       $       3.0\e{-6}       $       & 2200 & 13.5 &1\\
\object{U Hya} & 10:37:33.27 & $-$13:23:04.4 & $-32$ & 208 & $8.9\e{-8}$ & 2400 &6.5&1\\
\object{RV Aqr} &       21:05:51.68     &       $-$00:12:40.3   &       1       &       670     &       $       2.3\e{-6}       $       & 2200 & 15 &1\\
\multicolumn{2}{l}{\quad\it M-type stars}\\
\object{R Hor}  &       02:53:52.77     &       $-$49:53:22.7   &       37      &       310     &       $       5.9\e{-7}       $& 2200    & 4 &1\\
\object{IK Tau} &       03:53:28.87     &       +11:24:21.7     &       34      &       265     &       $       5.0\e{-6}       $       & 2100 & 17.5 & 2\\
\object{TX Cam} & 05:00:50.39 & +56:10:52.6 &11.4 &380 & $4.0\e{-6}$ & 2400 & 17.5 &2\\
\object{NV Aur} & 05:11:19.44 & +52:52:33.2     &2 & 1200\phantom{0} &  $2.5\e{-5}$ &2000&18&1\\
\object{BX Cam} & 05:46:44.10 & +69:58:25.2     &$-$2 & 500     &$4.4\e{-6}$ &2800&19&1\\
\object{GX Mon} &       06:52:46.91     &       +08:25:19.0     &       $-$9    &       550     &       $       8.4\e{-6}       $       & 2600 & 19 &1\\
\object{W Hya}  &       13:49:02.00     &       $-$28:22:03.5   &       40.5    &       \phantom{0}78   &       $       1.0\e{-7}       $       & 2500 & 7.5 & 3\\
\object{RR Aql} &       19:57:36.06     &       $-$01:53:11.3   &       28      &       530     &       $       2.3\e{-6}       $       & 2000 & 9 &1\\
\object{V1943 Sgr}      &       20:06:55.24     &       $-$27:13:29.8   &       $-$15   &       200     &       $       9.9\e{-8}       $       & 2200& 6.5 &1\\
\object{V1300 Aql}      &       20:10:27.87     &       $-$06:16:13.6   &       $-$18   &       620     &       $       1.0\e{-5}       $       & 2000 &14  &1\\
\object{T Cep} & 21:09:31.78    &+68:29:27.2    &$-$2&190       &$9.1\e{-8}$&2400&5.5&1\\
\object{R Cas} & 23:58:24.87 &  +51:23:19.7 &   25 &176 &       $8.0\e{-7}$ &1800&10.5&2\\
\multicolumn{2}{l}{\quad\it S-type stars}\\
\object{T Cet}  &       00:21:46.27     &       $-$20:03:28.9   &       22      &       240     &       $       6.0\e{-8}       $       & 2400 & 7 & 5\\
\object{R And}  &       00:24:01.95     &       +38:34:37.4     &       $-$16   &       350     &       $5.3\e{-7}$     &       1900    &       8       &       1       \\
\object{V365 Cas}       &       01:00:53.16     &       +56:36:45.2     &       $-$2.1  &       625     &       \phantom{0.}$3\e{-8}$   &       2400    &       6.2     &       7       \\
\object{S Cas}  &       01:19:41.99     &       +72:36:40.8     &$-$30  &       570     &       $2.8\e{-6}$     &       1800    &       19      &       1       \\
\object{W And}  &       02:17:32.96     &       +44:18:17.8     &$-$35  &       450     &       $2.8\e{-7}$     &       2400    &       6       &       1       \\
\object{T Cam}  &       04:40:08.88     &       +66:08:48.7     &$-$11.7        &       540     &       $1.0\e{-7}$     &       2400    &       3.8     &       7       \\
\object{DY Gem} &       06:35:57.81     &       +14:12:46.1     &$-$16.7        &       680     &       $7.0\e{-7}$     &       2400    &       8       &       7       \\
\object{R Lyn}  &       07:01:18.01     &       +55:19:49.8     &       16      &       850     &       $3.3\e{-7}$     &       2400    &       7.5     &       6       \\
\object{R Gem}  &       07:07:21.27     &       +22:42:12.7     &$-$60  &       820     &       $4.3\e{-7}$     &       2400    &       5       &       1       \\
\object{AA Cam} &       07:14:52.07     &       +68:48:15.4     &$-$46.8        &       780     &       $5.0\e{-8}$     &       3000    &       5       &       5       \\
\object{Y Lyn}  &       07:28:11.62     &       +45:59:26.2     &$-$0.5 &       253     &       $1.7\e{-7}$     &       2400    &       8       &       1       \\
\object{TT Cen} &       13:19:35.02     &       $-$60:46:46.3   &       4       &       1180\phantom{0} &       $       4.0\e{-6}       $       & 1900 & 20 & 5\\
\object{GI Lup} &       15:06:16.31     &$-$41:28:13.8  &       6       &       690     &       $5.5\e{-7}$     &       2400    &       10      &       7       \\
\object{ST Her} &       15:50:46.63     &       +48:28:58.9     &$-$4.5 &       293     &       $1.3\e{-7}$     &       2100    &       8.5     &       5       \\
\object{ST Sco} &       16:36:36.22     &$-$31:14:02.4  &$-$4.5 &       380     &       $1.5\e{-7}$     &       2400    &       5.5     &       7       \\
\object{RT Sco} & 17:03:32.55   &$-$36:55:13.7 &$-$47 & 400 & $4.5\e{-7}$ &  2100 & 11& 5\\
\object{TV Dra} &       17:08:24.50     &       +64:19:08.7     &       22      &       390     &       \phantom{0.}$5\e{-8}$   &       2400    &       4.7     &       7       \\
\object{IRC-10401}      &       18:10:24.82     &$-$10:34:16.1  &       19      &       585     &       \phantom{0.}$2\e{-6}$   &       1800    &       17      &       \phantom{*}7*   \\
\object{ST Sgr} &       19:01:29.20     &$-$12:45:34.0  &       55.7    &       540     &       $2.0\e{-7}$     &       2400    &       6       &       6       \\
\object{S Lyr}  &       19:13:11.80     &       +26:00:27.8     &       49      &       2000\phantom{0} &       $3.5\e{-6}$     &       1800    &       13      &       5       \\
\object{W Aql}  &       19:15:23.35     &       $-$07:02:50.4   &       $-$23   &       395     &       $       3.0\e{-6}       $       & 2300 & 16.5 & 4\\
\object{EP Vul} &       19:33:17.84     &       +23:39:19.6     &       0       &       510     &       $2.3\e{-7}$     &       2800    &       6       &       5       \\
\object{R Cyg}  &       19:36:49.38     &       +50:11:59.5     &$-$17  &       690     &       $9.5\e{-7}$     &       2600    &       9       &       1       \\
\object{AFGL 2425}      &       19:39:00.74     &$-$16:51:56.5  &       57      &       610     &       $3.0\e{-7}$     &       1800    &       8.7     &       7       \\
\object{CSS2 41}        &       19:39:07.77     &       +29:02:38.6     &       21.5    &       880     &       $5.8\e{-7}$     &       1800    &       17      &       7       \\
\object{$\chi$ Cyg}     &       19:50:33.92     &       +32:54:50.6     &       9       &       150     &       $7.9\e{-7}$     &       2600    &       8.5     &       8       \\
\object{AA Cyg} &       20:04:27.61     &       +36:49:00.5     &       27.5    &       480     &       $2.9\e{-7}$     &       2400    &       4.5     &       7       \\
\object{DK Vul} &       20:06:33.96     &       +24:25:60.0     &$-$15  &       750     &       $2.0\e{-7}$     &       2900    &       4.5     &       5       \\
\object{RZ Sgr} &       20:15:28.41     &       $-$44:24:37.5   &       $-$31   &       730     &       $       3.0\e{-6}       $       & 2400& 9 & 6\\
\object{AD Cyg} &       20:31:36.51     &       +32:33:52.4     &       21      &       980     &       $2.1\e{-7}$     &       1800    &       8       &       7       \\
\object{RZ Peg} &       22:05:52.97     &       +33:30:24.8     &$-$23.4        &       970     &       $4.6\e{-7}$     &       2400    &       12.6    &       6       \\
\object{RX Lac} &       22:49:56.90     &       +41:03:04.3     &$-$15.4        &       310     &       \phantom{0.}$8\e{-8}$   &       2400    &       6.5     &       7       \\
\object{V386 Cep}       &       22:53:12.33     &       +61:17:00.4     &$-$51.8        &       470     &       $2.0\e{-7}$     &       1800    &       16      &       7       \\
\object{WY Cas} &       23:58:01.31     &       +56:29:13.5     &       7       &       600     &       $1.1\e{-6}$     &       2200    &       13.5    &       6       \\
\hline
\end{tabular}
}
\tablefoot{Upper section contains APEX sulphur survey sources, lower section details S-star survey, omitting duplicates. References give details of mass-loss rate, $\dot{M}$, stellar effective temperature, $T_\mathrm{eff}$, distances, {and dust properties}. (*) indicates that stellar properties were updated in this work. (1) \cite{Danilovich2015a}; (2) \cite{Maercker2016}; (3) \cite{Khouri2014} and \cite{Danilovich2016}; (4) \cite{Danilovich2014} and \cite{Ramstedt2017}; (5) \cite{Ramstedt2014}; (6) \cite{Schoier2013}; (7) \cite{Ramstedt2009}; (8) \cite{Schoier2011}.}
\end{table*}

\begin{table}[tp]
\caption{Observational parameters for the SiS and CS lines included in this study}\label{trans}
\begin{center}
\begin{tabular}{ccccccr}
\hline\hline
Mol. & Line & Freq. & Tel.&$\theta$ & $\eta_\mathrm{mb}$& $E_\mathrm{up}$\\
 && [GHz] & & [$\arcsec$] & [K]\\
\hline
\up{28}Si\up{32}S  & $4\to3$ & \phantom{0}72.618 & OSO & 45 &0.55 &9\\
& $5\to4$ & \phantom{0}90.772 & IRAM & 27 & 0.81 &13 \\
& $6\to5$ & 108.924 & IRAM & 21 & 0.78  &18 \\
& $8\to7$ & 145.227 & IRAM & 17 & 0.65 & 31\\
& $9\to8$ & 163.377 &APEX& 38 & 0.68 &39\\
& $10\to9$ & 181.525 & APEX & 34 &0.68& 48 \\
& $11\to10$ & 199.672 & APEX & 31 & 0.68 & 58\\
 & $12\to 11$ & 217.818 &APEX& 29 & 0.75 & 68\\
& $12\to 11$ & 217.818 &IRAM& 11 &0.63 & 68\\
 & $13\to 12$ &  235.961 &IRAM& 10 &0.59 &79 \\
 & $14\to 13$ & 254.103 &APEX& 25& 0.75 & 92\\
 & $16\to 15$ & 290.381 &APEX& 22 & 0.75 & 119\\
 & $19\to 18$ & 344.779 &APEX& 18 &  0.73 &166\\
\hline
\up{28}Si\up{34}S & $10\to 9$   & 176.555 &APEX& 35 & 0.68& 38\\
& $11\to 10$    & 194.205 &APEX& 32 & 0.68 &47\\
\hline
\up{12}C\up{32}S &$3\to2$ & 146.969 &IRAM& 17 & 0.73 & 14\\
&$4\to3$ & 195.954 &APEX& 32 & 0.68 & 24\\
&$5\to4$ & 244.936 &IRAM&  10 & 0.59& 35\\
 &$6\to5$ & 293.912 &APEX& 21 & 0.75 & 49\\
 &$7\to6$ & 342.883 &APEX&18& 0.73 &  66\\
 \hline
\up{12}C\up{33}S &$6\to5$ & 289.382 &APEX&  22& 0.75 &49\\
\hline
\end{tabular}
\end{center}
\tablefoot{$\theta$ is the HPBW and $\eta_\mathrm{mb}$ is the main beam efficiency}
\end{table}

\subsection{S-star survey}

A total of 33 S-type stars were surveyed in CS and SiS emission using two telescopes. The stars from this sample are listed in the bottom part of Table \ref{fullsample}.
The CS ($3\to2$) and ($5\to4$) line emission was observed at the IRAM 30\,m telescope simultaneously with the HCN observations analysed in \cite{Schoier2013}. SiS (19$\to$18) was observed at the APEX 12\,m telescope using the SHeFI receiver in August to October, 2012. SiS (5$\to$4), (6$\to$5), (12$\to$11), and (13$\to$12) was observed at IRAM June 22--24, 2013. 

The IRAM 30m telescope observations were performed in dual beamswitch mode using a beam throw of about 2\arcmin, while at APEX position-switching was used with a reference position at +3\arcmin. The pointing was checked regularly using strong CO and continuum sources and found to be consistent within $\approx$3\arcsec~of the respective telescope pointing model.

At the telescope, the antenna temperature has been corrected for the attenuation of the atmosphere and spectra are first delivered in $T_{\rm{A}}^{\star}$-scale. They have been converted to $T_{\rm{mb}}$ scale using $T_{\rm{mb}}$=$T_{\rm{mb}}$/$\eta_{\rm{mb}}$, where $\eta_{\rm{mb}}$ is the main-beam efficiency. The adopted beam efficiencies and 
full-width at half-max beam sizes ($\theta_{\rm{mb}}$) are given in Table~\ref{trans}. The uncertainty in the absolute intensity scale is estimated to be about $\pm20$\%. 

The IRAM CS ($3\to2$) and ($5\to4$) observations are listed in Table \ref{csiramobs}, the APEX SiS ($19\to18$) observations are listed in Table \ref{sisapexobs}, and the IRAM SiS ($5 \to 4$), ($6 \to 5$), ($12 \to 11$), and ($13 \to 12$) observations are listed in Table \ref{sisiramobs}.

\subsection{OSO 4~mm observations of M-type stars}

A sample of eight northern M-type AGB stars, covering a range of mass-loss rates from $\sim9\e{-8}$ to $\sim3\e{-5}\spy$, were observed in the period 18--22 February 2016 as part of science verification for the new 4~mm HEMT amplifier receiver \citep{Belitsky2015} on the 20 m telescope at Onsala Space Observatory\footnote{The Onsala 20 m telescope is operated by the Swedish National Facility for Radio Astronomy, Onsala Space Observatory at Chalmers University of Technology.} (OSO). The observations covered the SiS ($4\to 3$) emission line at 72.618~GHz for the eight sources listed in Table \ref{osoobs}, which also includes rms noise levels. The line was undetected in all sources except for \object{BX Cam}, for which it was tentatively detected. We have not included BX~Cam in our modelling since it is too northern to observe from APEX, and one tentatively detected line forms a dataset of insufficient quality to model well. However, we included the non-detected SiS ($4\to 3$) lines in our models of IK~Tau and GX~Mon, to place additional constraints on those models.

\subsection{Supplementary observations}

For the carbon star AI Vol we included two additional SiS lines, ($11\to10$) and ($10\to9$), which were observed with APEX/SEPIA Band 5 as part of an unbiased line survey whose results are yet to be published in full (De Beck et al, in prep).
To better constrain our models, we included some previously published observations. These include IRAM 30~m observations of the SiS ($6\to5$) line at 108.924 GHz from \cite{Danilovich2015a}. 
We also included observations of IK~Tau previously published by \citet{Decin2010} and a few lines taken from the APEX archive. The full list of archival observations used in our study is found in Table \ref{supobs}.

\section{Modelling}\label{modelling}

\subsection{Established parameters}\label{modparam}

The bulk of our APEX sulphur survey and OSO samples was chosen from the stars with mass-loss rates determined through CO modelling by \cite{Danilovich2015a}, {while most of the S star survey came from \cite{Ramstedt2009} and \cite{Ramstedt2014}}. We use the circumstellar model results from those studies as the basis for our SiS and CS modelling. For the stars not included in these studies, we used a variety of previously obtained mass-loss rates, as noted in Table \ref{fullsample}. W~Aql is included in our observing sample, however, the modelling results for this star presented here are based on line radiative transfer modelling of ALMA observations of CS and SiS by \cite{Brunner2018}.

Some of the key stellar and circumstellar quantities for our sample --- systemic velocity ($\upsilon_{\mathrm{LSR}}$), distance, mass-loss rate ($\dot{M}$), stellar effective temperature ($T_\mathrm{eff}$), and terminal expansion velocity ($\upsilon_\infty$) --- are listed in Table \ref{fullsample}. In one instance, IRC -10401, we recalculated the mass-loss rate based on newer data, which is explained in more detail in Sect. \ref{irc-10401}. {A detailed discussion on the uncertainties in mass-loss modelling can be found in \cite{Ramstedt2008}.}

{The referenced studies in Table \ref{fullsample} also include models of the dust surrounding each star. Similar dust modelling methods are used in all the studies and the specific dust properties of each source can be found in its referenced study.}

\subsection{Molecular data}\label{moldat}

\begin{figure}[t]
\center
\includegraphics[height=8 cm]{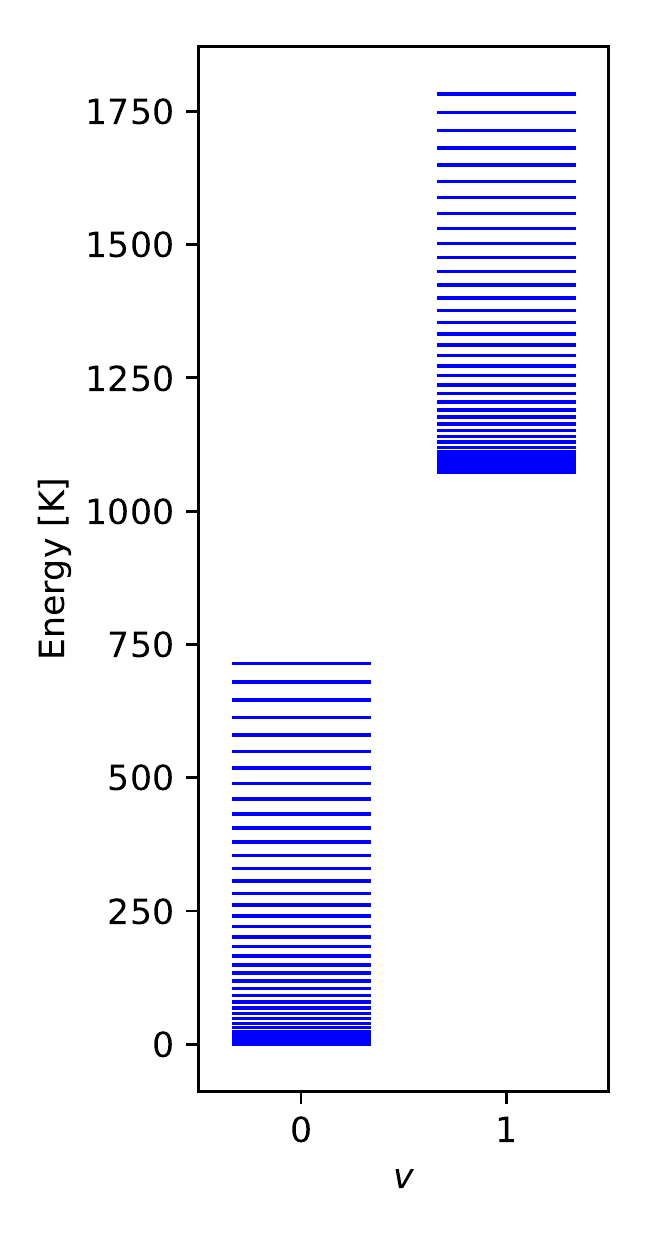}
\includegraphics[height=8 cm]{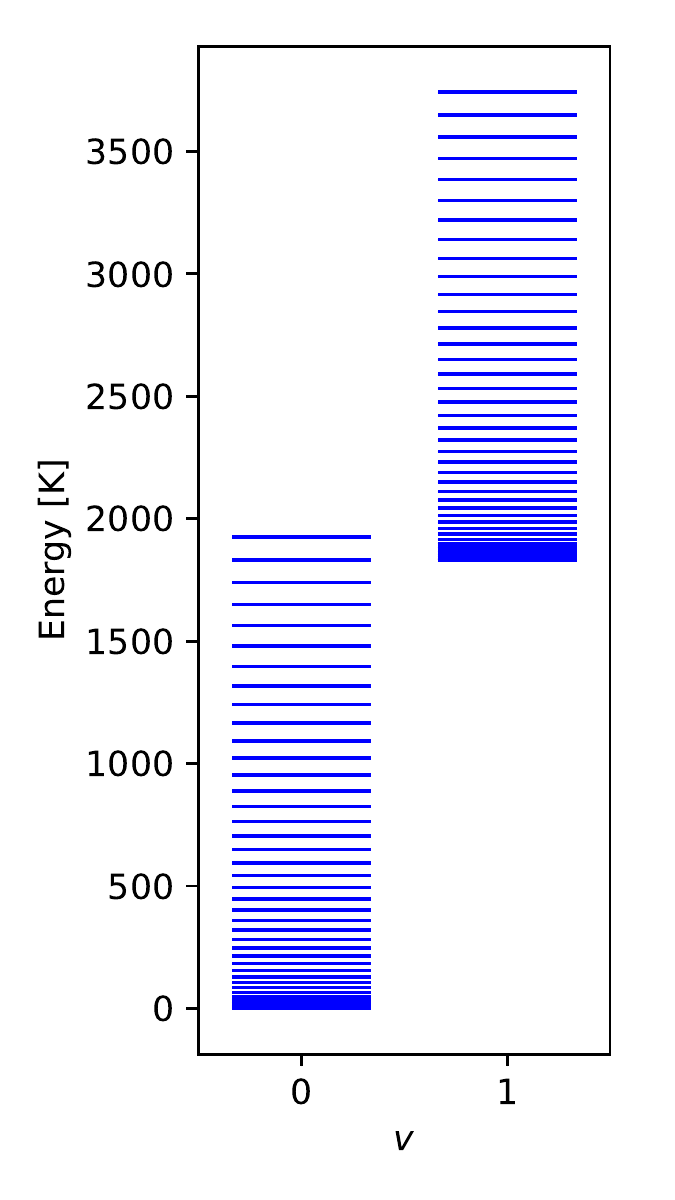}
\caption{Rotational energy levels in the ground and first vibrationally excited states for SiS (\textit{left}) and CS (\textit{right}) included in our modelling. For both molecules the most energetic rotational level included is at $J=40$.}
\label{ELD}
\end{figure}

For both SiS and CS we performed our radiative transfer analysis using molecular descriptions including rotational energy levels from $J=0$ to $J=40$ in the ground and first excited vibrational states. These energy levels are shown in Fig. \ref{ELD} and are connected by 160 radiative transitions and 820 collisional transitions. For SiS, the energy levels and radiative transition parameters were all taken from the JPL spectroscopic database\footnote{\url{https://spec.jpl.nasa.gov}} \citep{Pickett1998}, while the collisional rates for SiS-\h2 are adopted from SiO-\h2 rates, which were themselves scaled and extrapolated from the SiO-He rates of \citet{Dayou2006}. For CS, the energy levels and radiative transition parameters were all taken from the Cologne Database for Molecular Spectroscopy \citep[CDMS\footnote{\url{http://www.astro.uni-koeln.de/cdms}},][]{Muller2005,Endres2016}. The adopted collisional rates come from those of CO-\h2 computed by \citet{Yang2010} and an assumed \h2 ortho-to-para ratio of three.

\subsection{Modelling procedure}

We performed our radiative transfer modelling using a one-dimensional accelerated lambda iteration method code (ALI), which is described in detail by \cite{Maercker2008} and \cite{Schoier2011}, {and is based on the ALI scheme described by \cite{Rybicki1991}. ALI is able to deal with high optical depths} and has been used to model other S-bearing molecules such as SO, \so2, and \h2S \citep{Danilovich2016,Danilovich2017a}. 

ALI is one-dimensional so we assumed a smooth, spherically symmetric wind with a constant mass-loss rate and velocity profile based on the stellar parameters listed in Table \ref{fullsample} which are described in more detail in \citet{Danilovich2015a}. To fit our models to the observed data, we first assume a Gaussian molecular abundance distribution, 
\begin{equation}
f(r)= f_0 \exp\left(-\left(\frac{r}{R_e}\right)^2\right),
\end{equation}
where $f_0$ is the peak central abundance and $R_e$ is the $e$-folding radius at which the the abundance has dropped to $f_0/e$. As we have no {a priori} constraints on the $e$-folding radius, we leave both $f_0$ and $R_e$ as free parameters in our modelling, to be adjusted to best fit the available data. This is only possible for the sources for which we have detected at least two different transitions with sufficiently distinct emitting regions. {Gaussian abundance profiles have been shown to be adequate fits for various molecules in the past, such as SiO \citep{Gonzalez-Delgado2003}, \h2O \citep{Maercker2016}, and others \citep{Schoier2011,Danilovich2014}. In the absence of more detailed information as to the radial distributions of CS and SiS (such as spatially resolved observations), we have chosen to use Gaussian abundance distribution profiles here based on these past results and due to the ease with which they can be adjusted to find the best fit for the data.}

To determine which models best fit the data, we minimised a $\chi^2$ statistic, which we define as
\begin{equation}
\chi^2 = \sum^N_{i=1} \frac{\left(I_\mathrm{mod,i} - I_\mathrm{obs,i}\right)^2}{\sigma_i^2},
\end{equation}
where $I$ is the integrated main beam line intensity, $\sigma$ is the uncertainty in the observed line intensities and $N$ is the number of lines being modelled. In general, we assumed an uncertainty of 20\% in line intensity for our observed lines, except for those we identify as being tentatively detected, for which we assumed a 50\% uncertainty. The uncertainties calculated for our model results are for a 90\% confidence interval using this $\chi^2$ formulation. To better allow us to compare between stars for which differing numbers of observed lines might be available, we further defined a reduced-$\chi^2$ statistic: $\chi^2_\mathrm{red} = \chi^2/(N-p)$ where $p=2$ is the number of free parameters in our models (and hence for $N\leq3$ we leave $\chi_\mathrm{red}^2=\chi^2$). In the cases where only one line was detected for a particular source and molecule, we cannot calculate a $\chi^2$ value and our uncertainties are based on a 20\% shift in model integrated intensity.

\subsection{Modelling results}

We were able to successfully model the SiS and CS line emission in the CSEs of all the stars in our sample for which at least two lines per molecule were detected with only one exception. 
II~Lup proved difficult to model using a spherically symmetric model with a smoothly accelerating wind and will be discussed in more detail in Appendix \ref{iilup}.
For the stars with only one detected line or with only two lines from adjacent SiS transitions detected, it was not possible to determine an $e$-folding radius. In these cases, we obtained the $e$-folding radius from a fit to our other results. This is discussed in more detail in Sect. \ref{analysis}.

\begin{figure*}[t]
\begin{center}
\includegraphics[width=\textwidth]{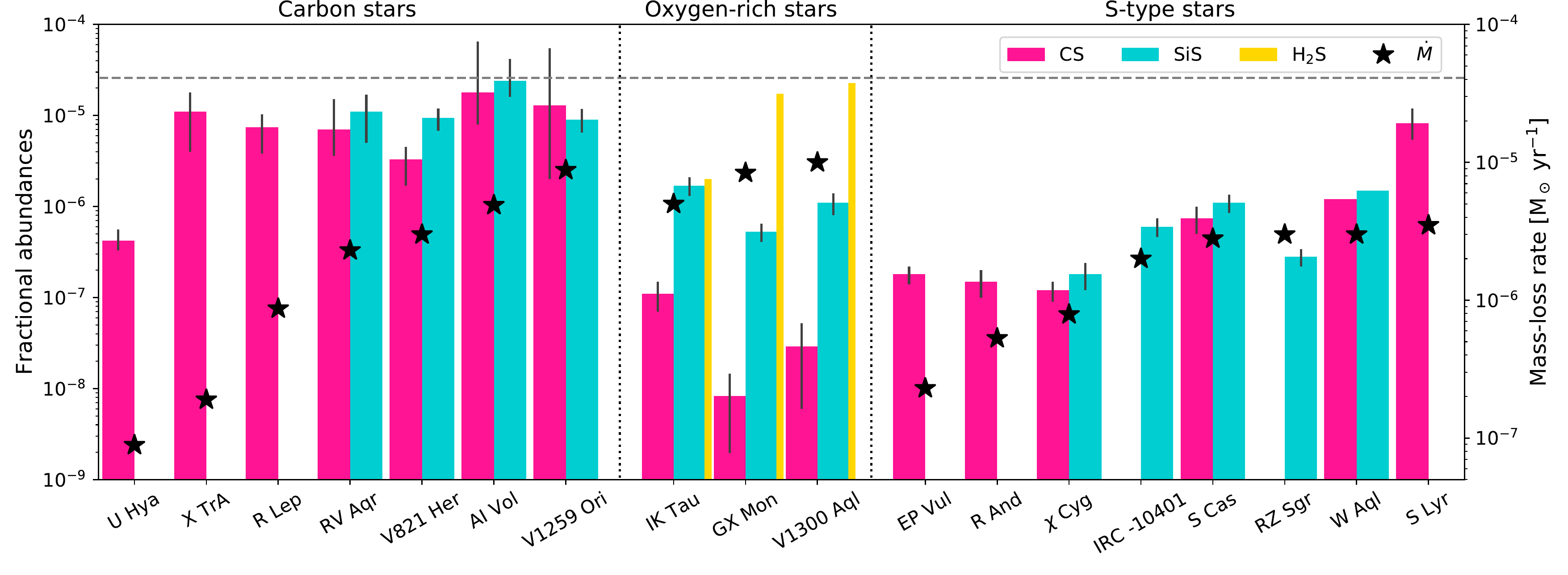}
\caption{Abundances of CS and SiS with uncertainties for the carbon-rich stars (\textit{left}), the S-type stars (\textit{right}), and including \h2S abundances from \cite{Danilovich2017a} (assuming an ortho-to-para ratio of three) for the oxygen-rich stars (\textit{middle}). Fractional abundances relative to \h2 are given by the left vertical axis and mass-loss rates are indicated by the black stars and the right vertical axis. The dashed, grey, horizontal line represents the maximum sulphur abundance expected based on the \cite{Asplund2009} solar abundance.}
\label{starab}
\end{center}
\end{figure*}

\begin{figure}[t]
\begin{center}
\includegraphics[width=0.49\textwidth]{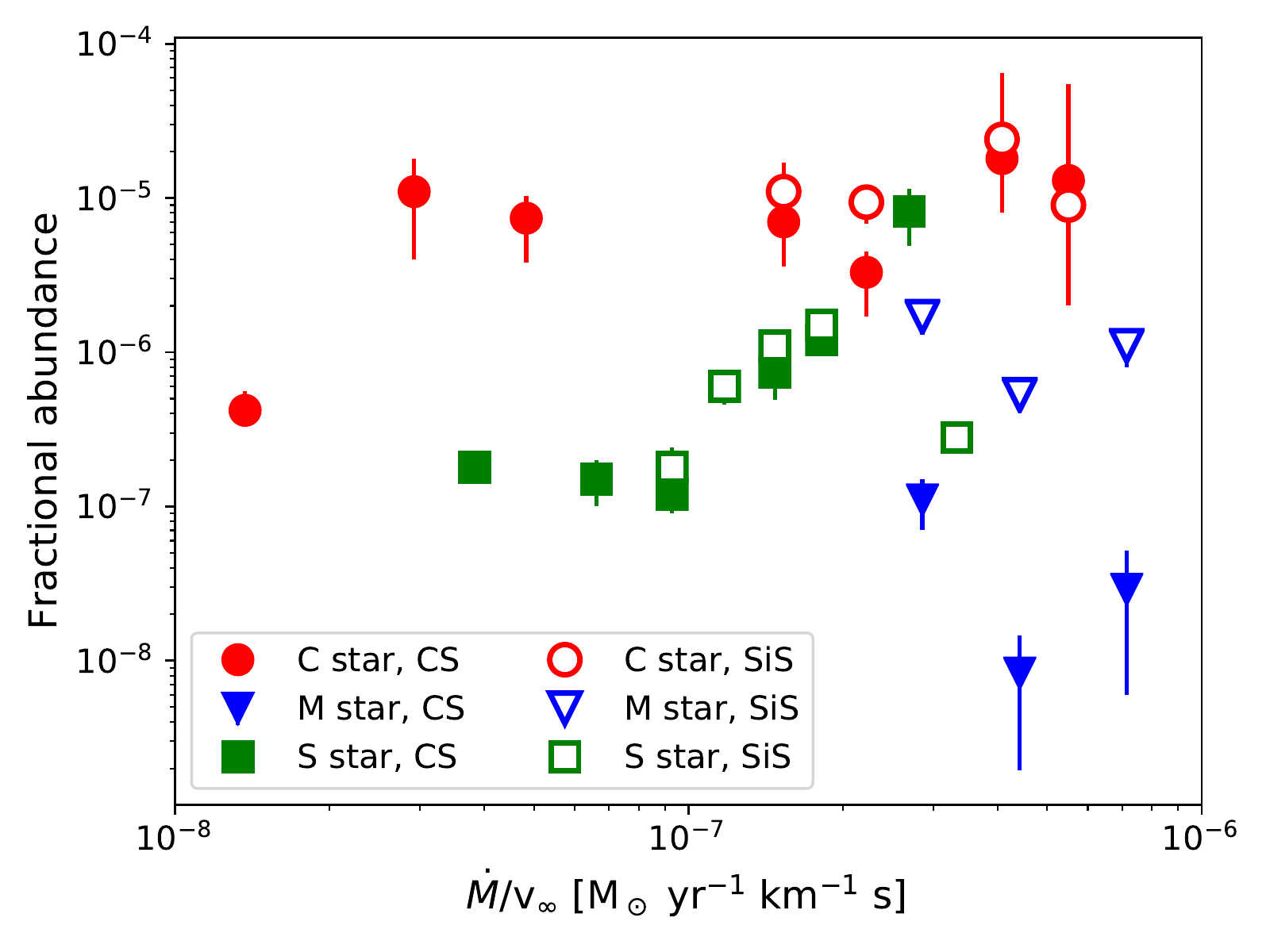}
\caption{Abundances of both SiS and CS plotted against stellar wind density, given by the mass-loss rate divided by the terminal expansion velocity.}
\label{abundancevsdensity}
\end{center}
\end{figure}

\begin{figure}[t]
\begin{center}
\includegraphics[width=0.49\textwidth]{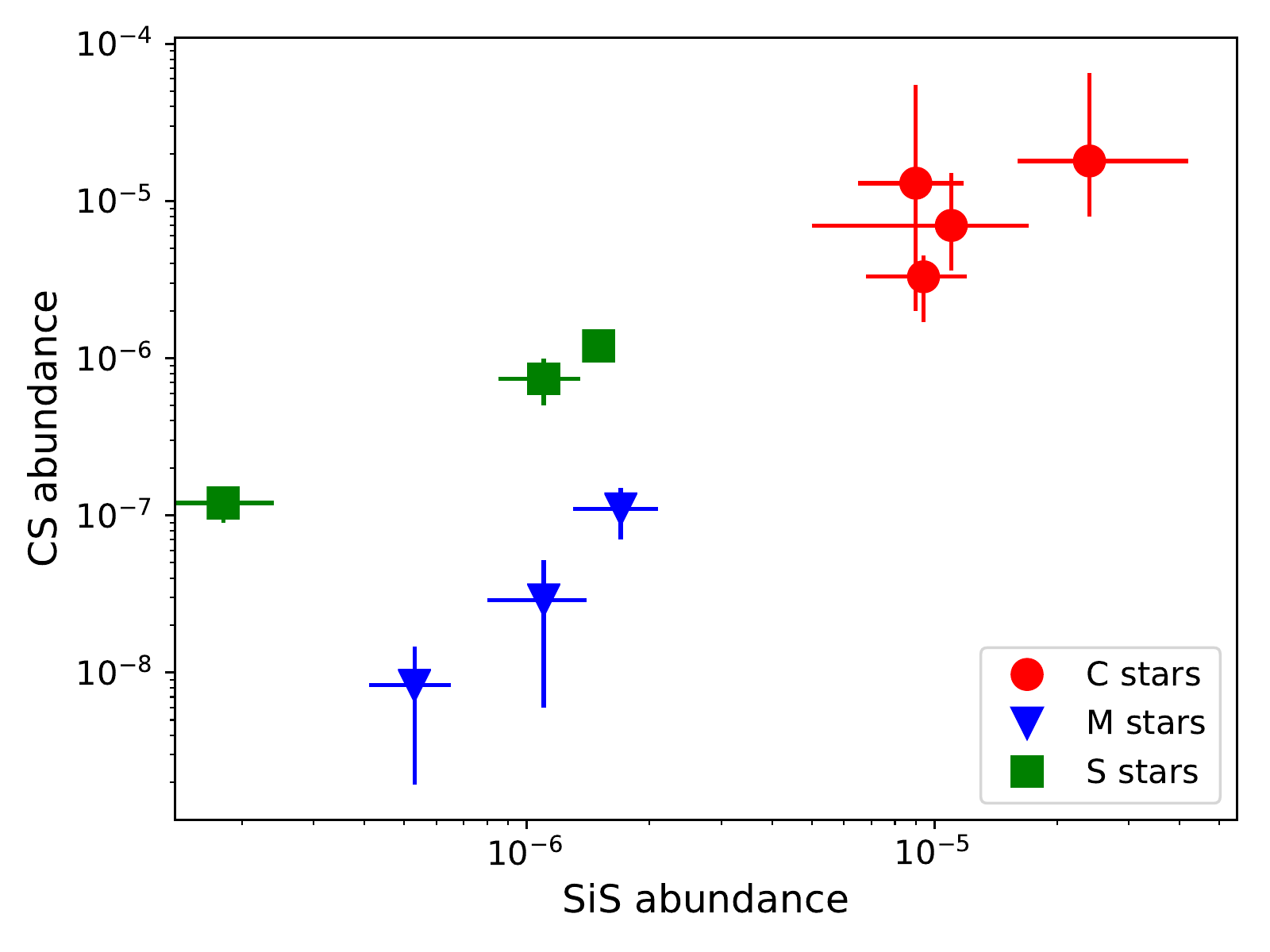}
\caption{Derived CS fractional abundances plotted against derived SiS fractional abundances.}
\label{sisvscs}
\end{center}
\end{figure}

The abundances, $f_0$, and $e$-folding radii, $R_e$, that we have derived are listed in Table \ref{results}.
A summary of the results is shown in Fig. \ref{starab}. In Fig. \ref{abundancevsdensity} we plot both SiS and CS abundances against wind density and in Fig. \ref{sisvscs} we plot CS abundance against SiS abundance for the stars towards which both molecules were detected. 
The SiS results are discussed in more detail in Sect. \ref{sisresults} and the CS results in Sect. \ref{csresults}. Isotopologue modelling is discussed in Appendix \ref{isotopologues}.

\begin{table*}[tp]
\caption{Modelling results.}\label{results}
\begin{center}
\begin{tabular}{c|ccccc|ccccc}
\hline\hline
 Star & \multicolumn{5}{c|}{SiS} & \multicolumn{5}{c}{CS}\\
 & $f_0$ & $R_e$ & $n$ & $\chi^2_\mathrm{red}$& $N_\mathrm{SiS}$& $f_0$ & $R_e$  & $n$&$\chi^2_\mathrm{red}$ &$N_\mathrm{CS}$\\
 & $[\e{-6}]$ & [$\e{15}$~cm] &  & & [cm$^{-2}$]& $[\e{-6}]$ & [$\e{15}$~cm] & & &[cm$^{-2}$]\\
\hline
\textit{Carbon stars}&&&&&&&\\
R Lep& ... &... &0 &...&...& $7.4_{-2.9}^{+3.6}$ & $9.2_{-3.7}^{+7.1}$ & 2 & 0.00&$1.2\e{17}$\\
V1259 Ori& $9.0^{+2.8}_{-2.5}$& $13\pm3$ & 5& 3.4 &$1.1\e{18}$& $1.3^{+?}_{-1.1}$ $\dagger$ & $80_{-70}^{+?}$ * & 3& 0.59&$1.7\e{18}$\\
AI Vol& $24_{-8}^{+18}$ & $6.2_{-1.1}^{+1.2}$ & 7& 3.2 & $1.2\e{18}$& $18_{-10}^{+?}$ $\dagger$& $10^{+6}_{-4}$  & 2& 0.03 & $8.9\e{17}$\\
X TrA & ... &... &0 &...&...& $11\pm7$ & $4.0_{-1.5}^{+2.5}$ & 2 & 0.01&$3.2\e{17}$\\
V821 Her & $9.4\pm2.6$& $5.7\pm1.0$ & 5 & 2.9 &$2.8\e{17}$& $3.3_{-1.2}^{+1.6}$ & $28_{-13}^{+36}$ & 3 & 0.54&$9.9\e{16}$\\
U Hya & ... &... &0 &...&... &$0.42^{+0.09}_{-0.14}$ & $4.8^{+4.7}_{-2.0}$ & 3& 7.1&$1.1\e{15}$\\
RV Aqr & $11\pm6$& $2.9^{+0.5}_{-0.6}$& 3 & 0.71 &$2.5\e{17}$& $7.0_{-3.4}^{+8.1}$& $22_{-11}^{+66}$ & 2& 0.00&$1.7\e{17}$\\
\hline
\textit{Oxygen-rich stars}&&&&&&&\\
IK Tau & $1.7\pm0.4$& $5.3\pm1.0$ & 7& 1.1&$1.1\e{17}$& $0.11\pm0.04$ & $23_{-12}^{+72}$ & 2& 0.02&$6.1\e{15}$\\
GX Mon & $0.53\pm0.12$ & $13\pm4$ & 5 & 2.0 &$3.5\e{16}$& $0.083^{+0.062}_{-0.065}$& $\gtrsim 200$ * & 2&0.36&$5.6\e{15}$\\
V1300 Aql & $1.1\pm0.3$& $12\pm3$ & 5 & 1.4 &$7.6\e{16}$& $0.029\pm0.023$&$40_{-34}^{+?}$ * & 2& 0.00&$2.5\e{15}$\\
\hline
\textit{S-type stars}&&&&&&&\\
R And &...&...&0&...&...&$0.15\pm0.5$&$18_{-8}^{+26}$&2&0.01&$1.4\e{15}$\\
S Cas &{$1.1^{+0.3}_{-0.2}$}&$3.6$ *&2&1.1& $2.4\e{16}$&$0.74_{-0.24}^{+0.26}$&$16^{+8}_{-6}$ &2&0.0&$1.7\e{16}$\\
IRC -10401 & $0.60\pm0.12$ & $3.0$ * & 1& ...& $3.7\e{15}$ &...&...&0&...&...\\
S Lyr&...&...&0&...&...&$8.2^{+3.7}_{-2.8}$& $19$ * & 1 & ...&$3.4\e{17}$\\
W Aql$^\ddagger$ & $1.5\pm0.05$ & $6.0$ &...&...& $5.9\e{16}$& $1.2\pm0.05$ & $7.0$ & ... &...& $4.7\e{16}$\\
EP Vul&...&...&0&...&...& $0.18\pm0.04$ & $8.6$ * & 1 & ...&$1.1\e{15}$\\
$\chi$ Cyg &$0.18\pm0.06$&$2.5$ *&2&1.4&$1.5\e{15}$&$0.10\pm0.03$&$15^{+34}_{-7}$&2&0.01&$5.7\e{14}$\\
RZ Sgr& $0.28\pm0.06$&$7.2$ *&1&...& $1.0\e{16}$&...&...&0&...&...\\
\hline
\end{tabular}
\end{center}
\tablefoot{$f_0$ is the peak abundance relative to \h2 and $R_e$ is the $e$-folding radius. $n$ is the number of observed lines included in our radiative transfer analysis. {$N$ is the column density obtained from our radiative transfer modelling.} (*) indicates models which cannot be radially constrained with the available data. Where only one observed line is available, $R_e$ is calculated from whichever is applicable out of Eqs. \ref{resis} and \ref{recs}. ($\dagger$) indicates models for which upper limits cannot be placed on the uncertainties due to optical depth effects that come into play at higher abundances of CS. ($^\ddagger$) W~Aql is originally modelled in Brunner et al (submitted), based in part on ALMA observations, and is only included here for completion. Hence uncertainties, $N$ and $\chi^2_\mathrm{red}$ are omitted for this source since the fit to a radial profile derived from an ALMA image is not comparable to fits based only on single-dish data.}
\end{table*}

\section{Analysis}\label{analysis}

\subsection{SiS}\label{sisresults}

\begin{figure}[t]
\begin{center}
\includegraphics[width=0.49\textwidth]{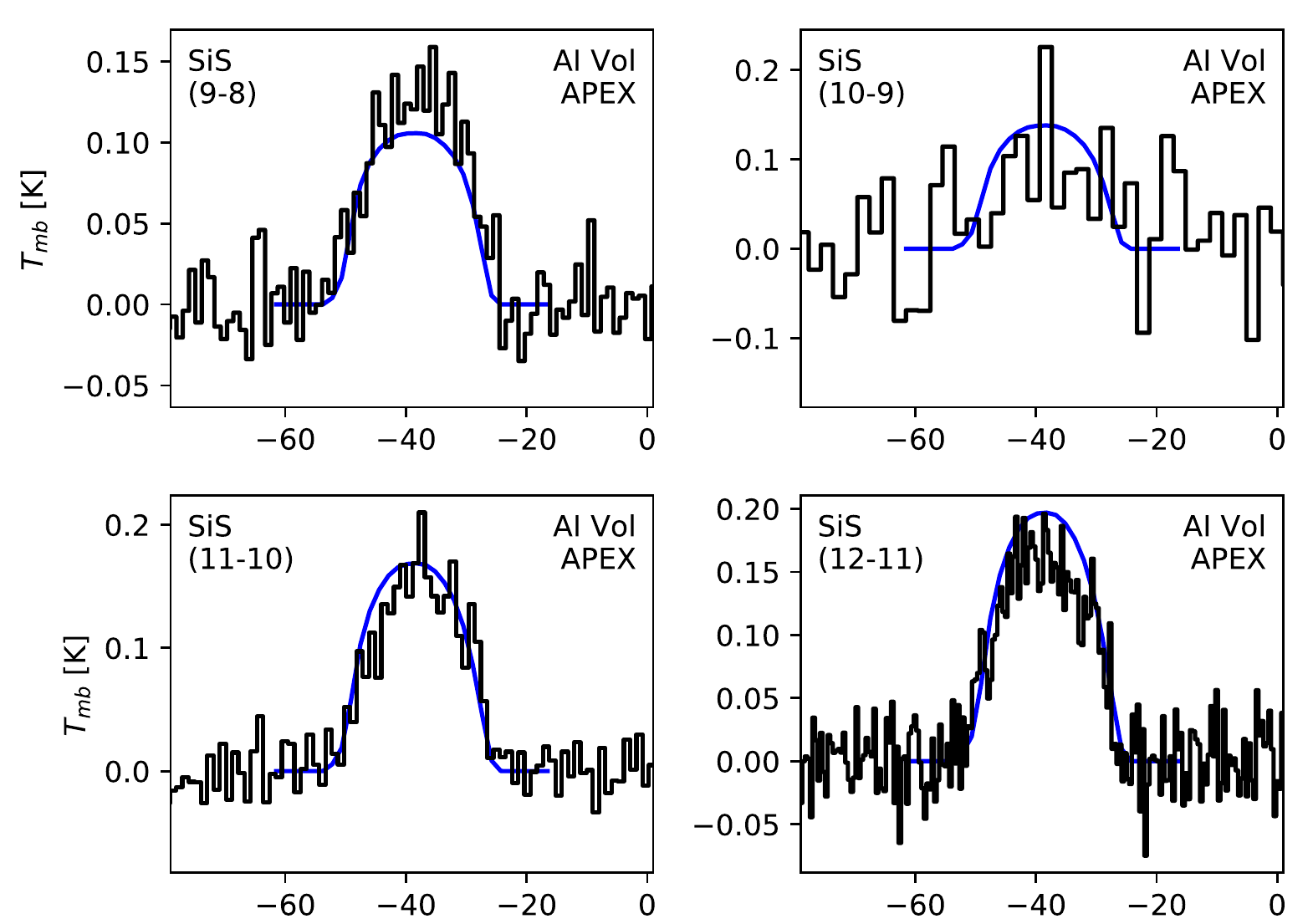}
\includegraphics[width=0.49\textwidth]{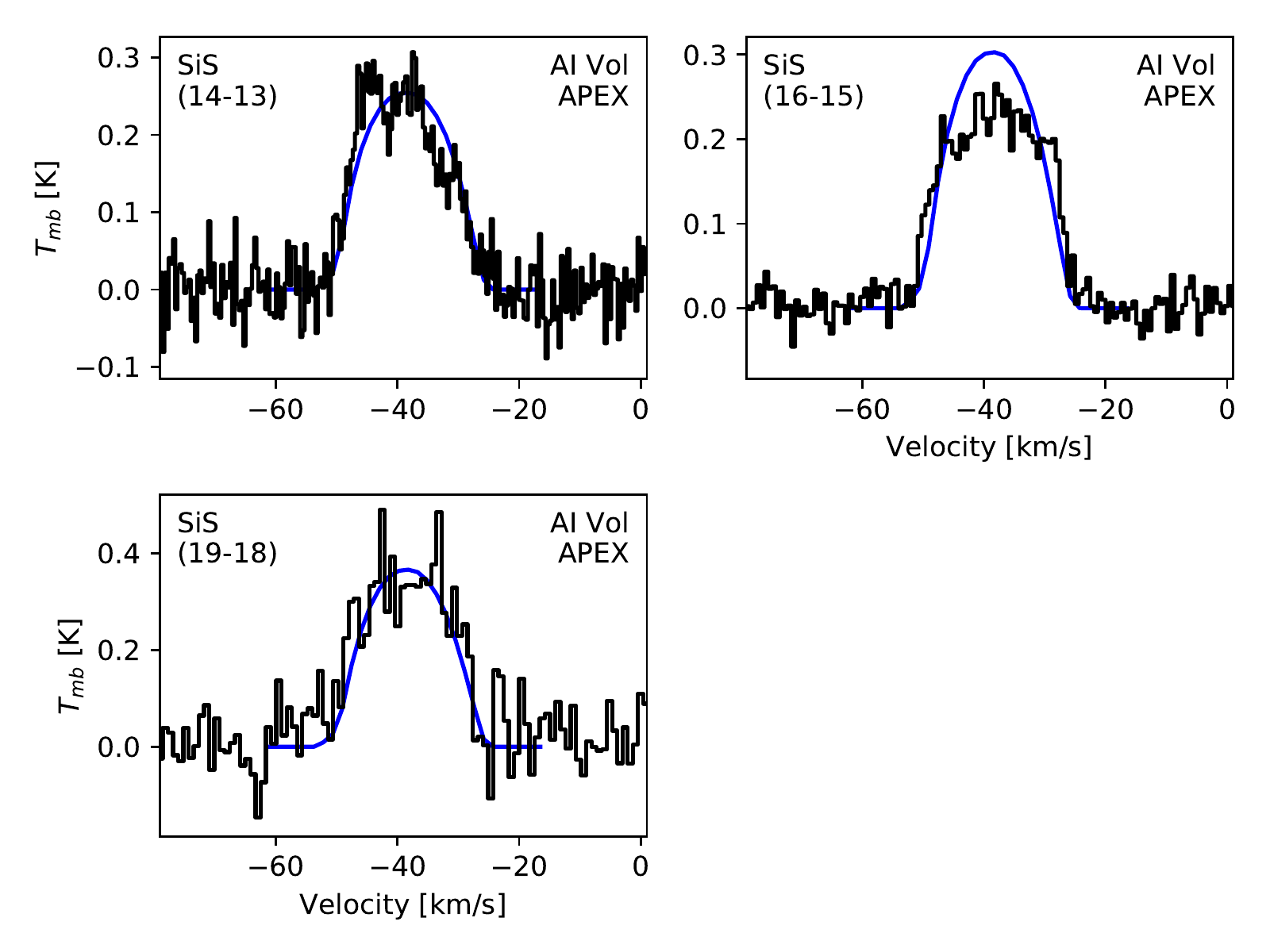}
\caption{Observations (black histograms) and model results (blue lines) for SiS towards AI Vol, a carbon star, plotted with respect to LSR velocity.}
\label{SiSAIVolplots}
\end{center}
\end{figure}

\begin{figure}[t]
\begin{center}
\includegraphics[width=0.49\textwidth]{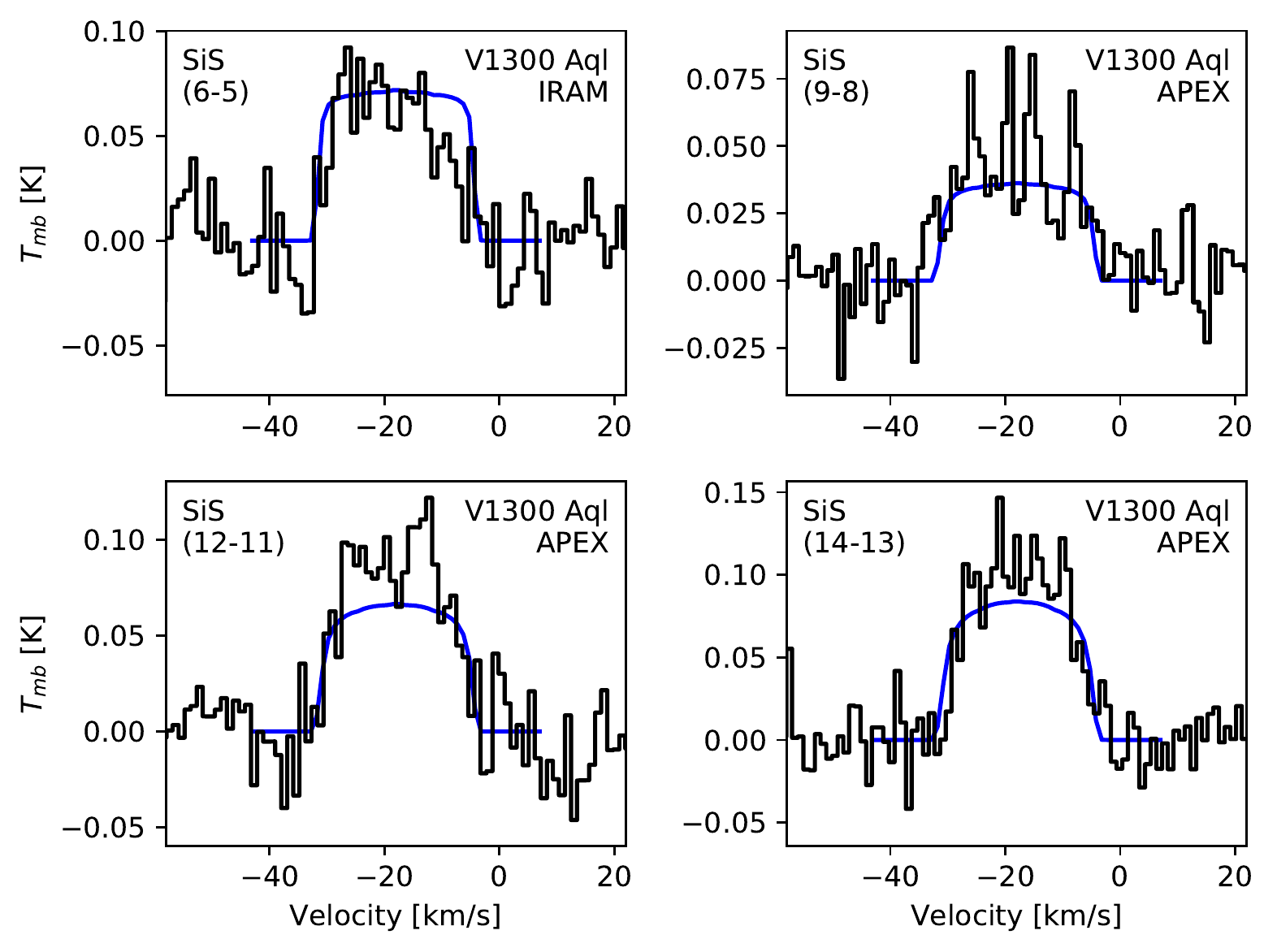}
\includegraphics[width=0.1\textwidth]{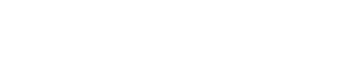}
\includegraphics[width=0.35\textwidth]{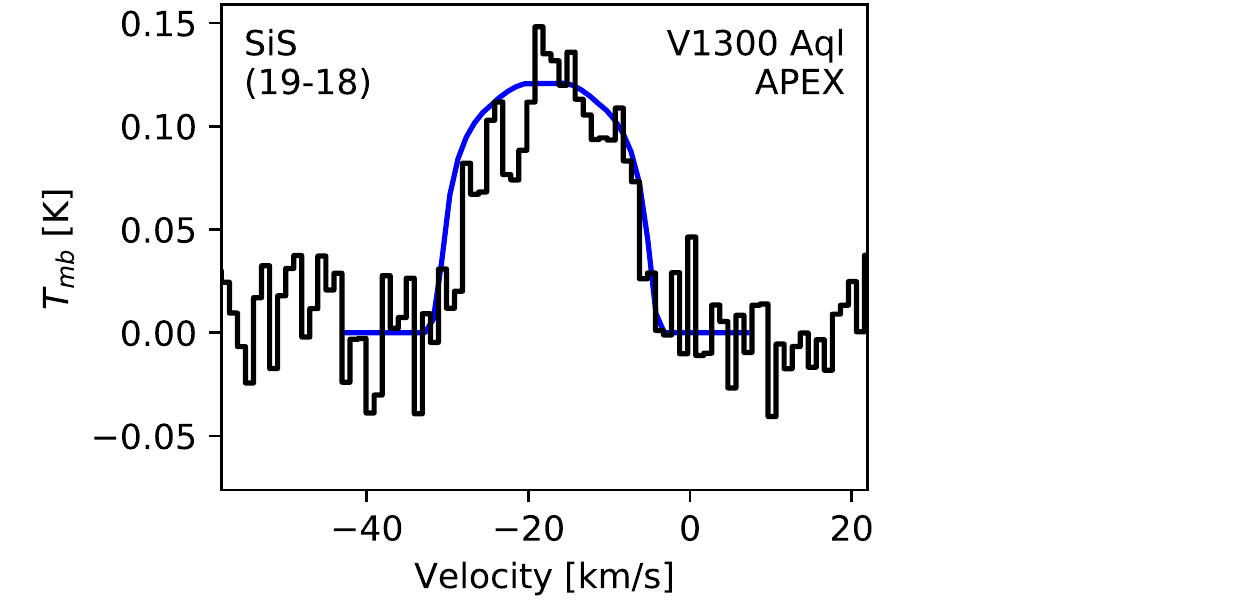}
\caption{Observations (black histograms) and model results (blue lines) for SiS towards V1300 Aql, an M-type star, plotted with respect to LSR velocity.}
\label{SiSV1300Aqlplots}
\end{center}
\end{figure}

In the APEX survey, SiS was detected towards five out of the eight surveyed carbon stars and three out of seven M-type stars. In the S star survey, SiS was detected in five sources. {The low-energy SiS ($4\to3$) line was only tentatively detected for one of the stars observed with the OSO 20m telescope, although higher-energy transitions were detected for the two stars overlapping with the APEX sulphur survey (IK~Tau and GX~Mon).} In all cases, these were among the highest mass-loss rate sources in each category. The implications of this will be discussed further in Sect. \ref{disc}.
The SiS observations and model results for a representative carbon star, AI~Vol, are plotted in Fig. \ref{SiSAIVolplots} with the same for the remaining carbon stars plotted in {Figures \ref{SiSCplots-1}, \ref{SiSCplots-2}, and \ref{SiSCplots-3}}. 
The SiS results for a representative oxygen-rich star, V1300~Aql, are plotted in Fig. \ref{SiSV1300Aqlplots} with the same for the remaining oxygen-rich stars plotted in {Figures \ref{SiSMplots-1} and \ref{SiSMplots-2}}.
In general we see higher abundances of SiS for carbon stars than for oxygen-rich or S-type stars. SiS abundances range from $\sim 9\e{-6}$ to $\sim 2\e{-5}$ for carbon stars, from $\sim3\e{-7}$ to $\sim2\e{-6}$ for oxygen rich stars, and from $\sim 2\e{-7}$ to $\sim 1\e{-6}$ for the S-type stars. 

\begin{figure*}[t]
\begin{center}
\includegraphics[width=0.49\textwidth]{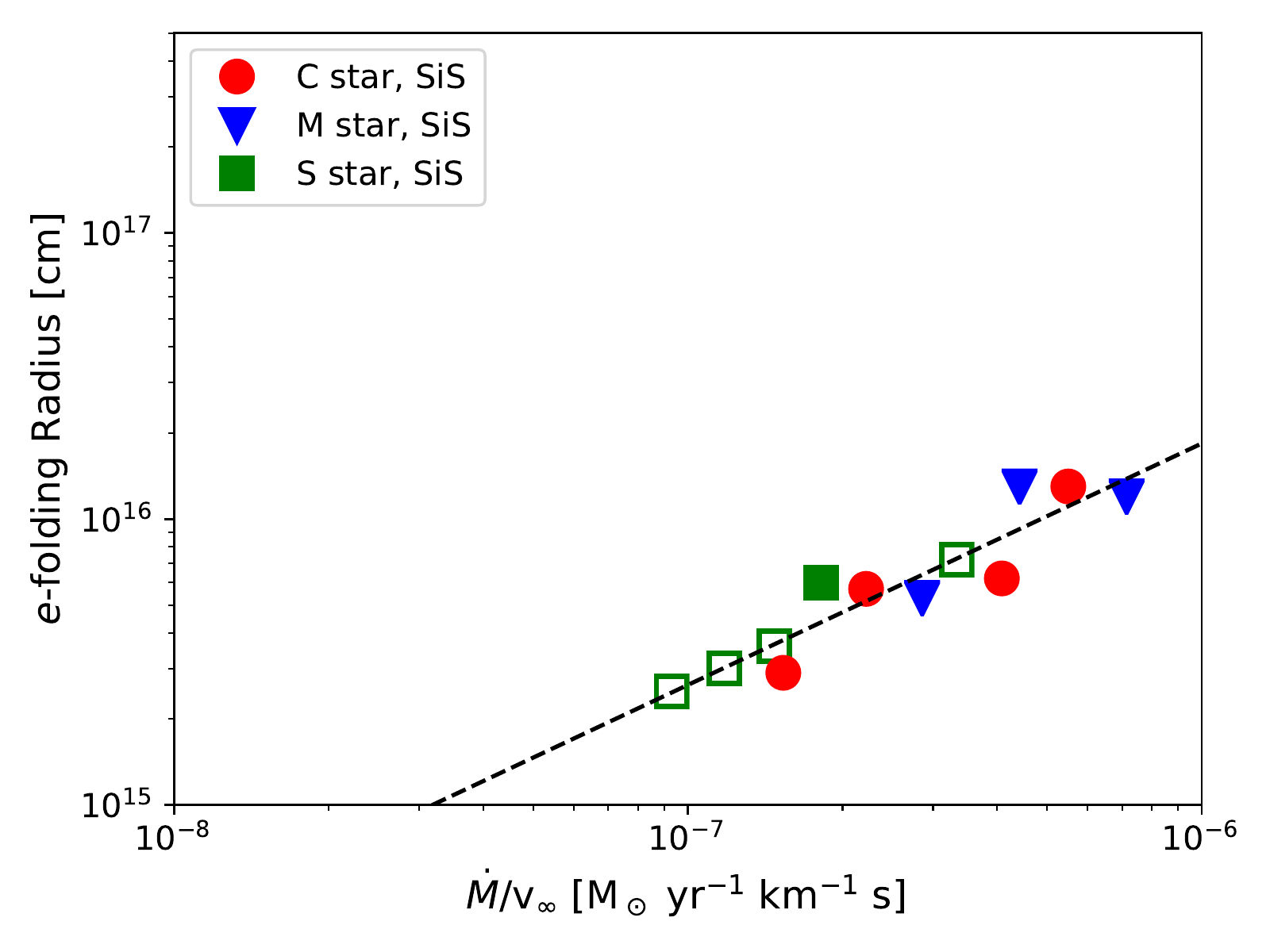}
\includegraphics[width=0.49\textwidth]{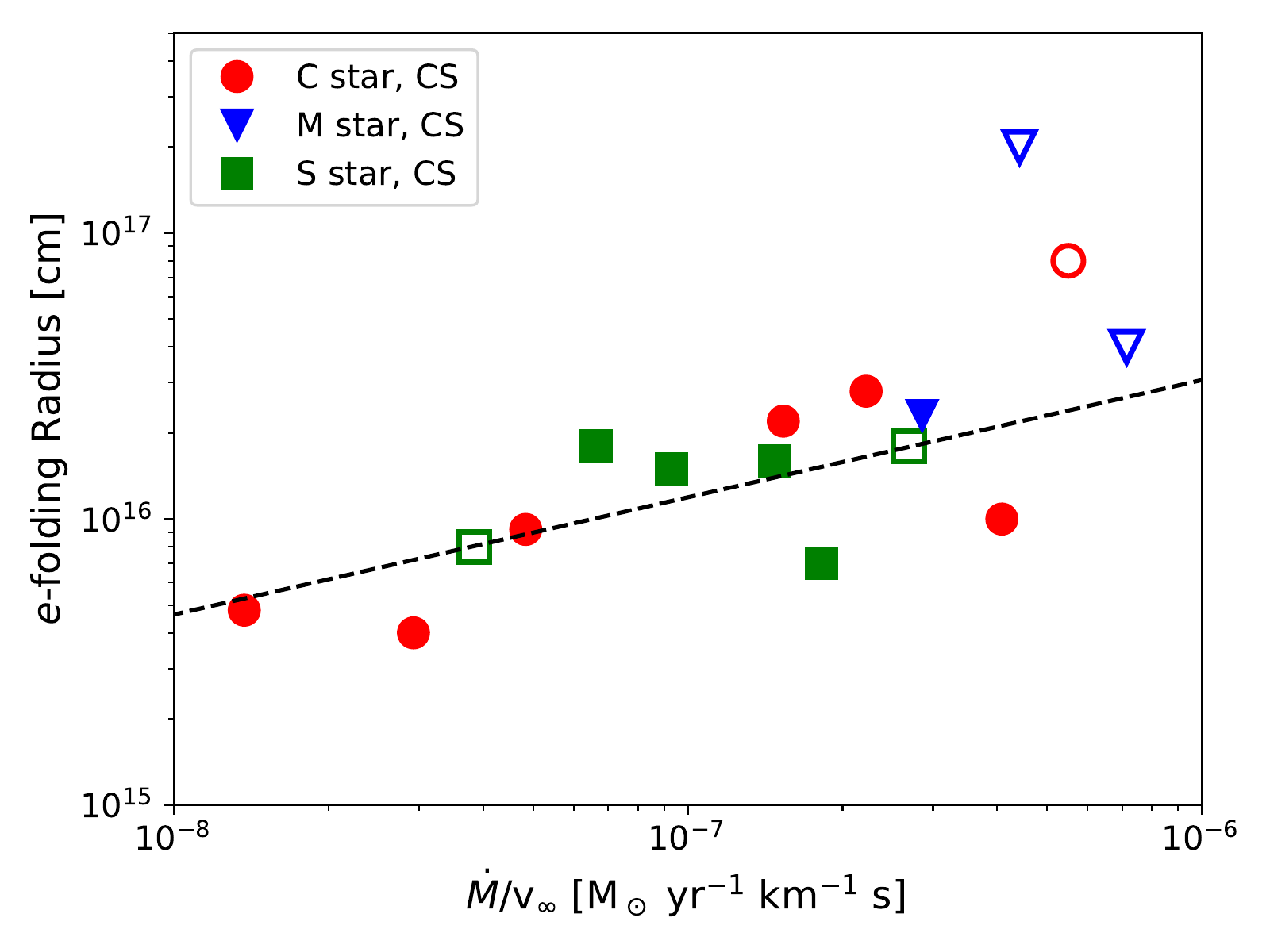}
\caption{$e$-folding radii of SiS (\textit{left}) and CS (\textit{right}) plotted against stellar wind density, given by the mass-loss rate divided by the terminal expansion velocity. The dashed lines show the best fit to the results, excluding the unfilled points, for which $R_e$ is derived from the best fit relation, or which do not have well-constrained $R_e$ for CS. See text for details.}
\label{radvsdens}
\end{center}
\end{figure*}

From the stars for which we were able to constrain the SiS $e$-folding radius, we found the following relation between $e$-folding radius and wind density {when weighting with the uncertainties listed in Table \ref{results}}
\begin{equation}\label{resis}
\log_{10}(R_{e,\mathrm{SiS}}) = (21.3\pm0.2) + (0.84\pm0.03)\log_{10}\left(\frac{\dot{M}}{\upsilon_\infty}\right),
\end{equation}
where $R_e$ is given in cm, $\dot{M}$ in $\spy$, $\upsilon_\infty$ in $\kms$, {and the errors are $1\sigma$ uncertainties}.
This fit is represented by the dashed black line in the left panel of Fig. \ref{radvsdens}. When modelling the S stars with only one SiS detection, or with only the SiS ($12\to11$) and ($13\to12$) lines detected, it was not possible to fit the $e$-folding radius with the available data. Hence, we derived $R_e$ for these sources using Eq. \ref{resis} and then fitted the peak abundance, $f_0$, to the available data.

\subsection{CS}\label{csresults}

In the APEX survey, CS is detected in all of the surveyed carbon stars and in three out of the seven surveyed M-type stars. In the S star survey it is detected towards  six sources.

\begin{figure}[t]
\begin{center}
\includegraphics[width=0.49\textwidth]{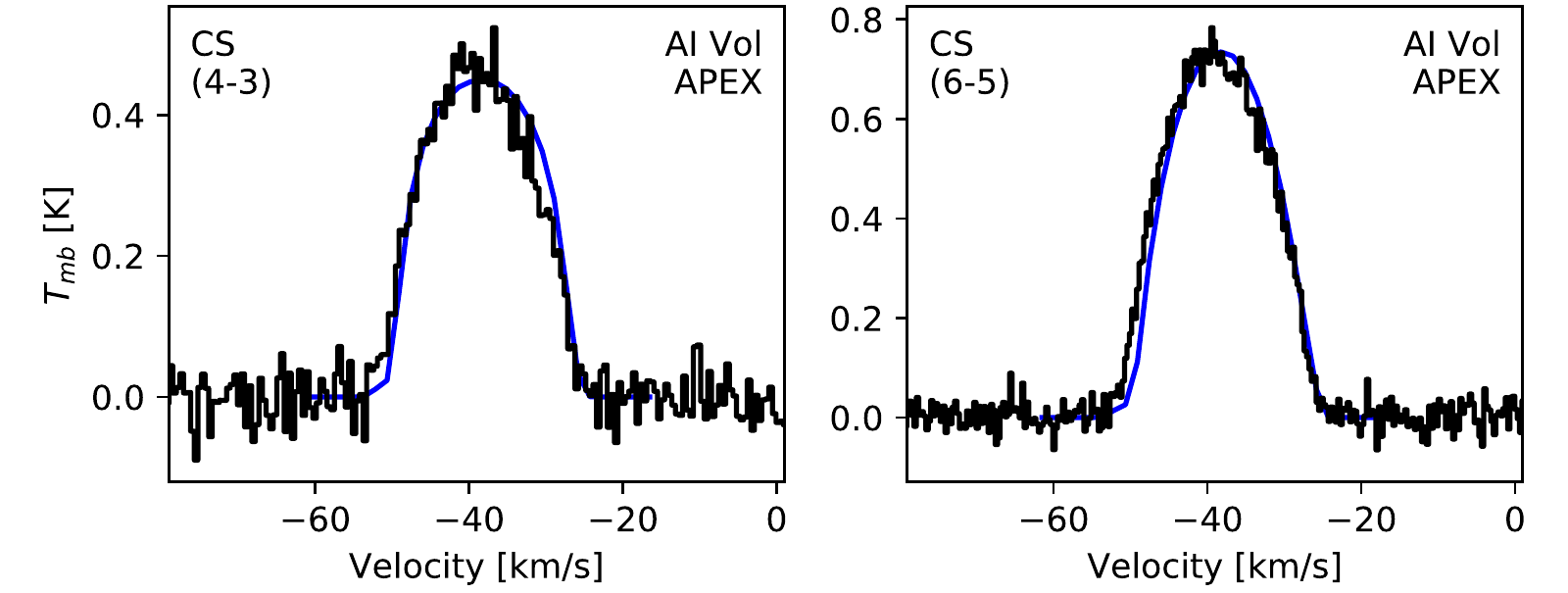}
\caption{Observations (black histograms) and model results (blue lines) for CS towards AI Vol, a carbon star, plotted with respect to LSR velocity.}
\label{CSAIVolplots}
\end{center}
\end{figure}

\begin{figure}[t]
\begin{center}
\includegraphics[width=0.49\textwidth]{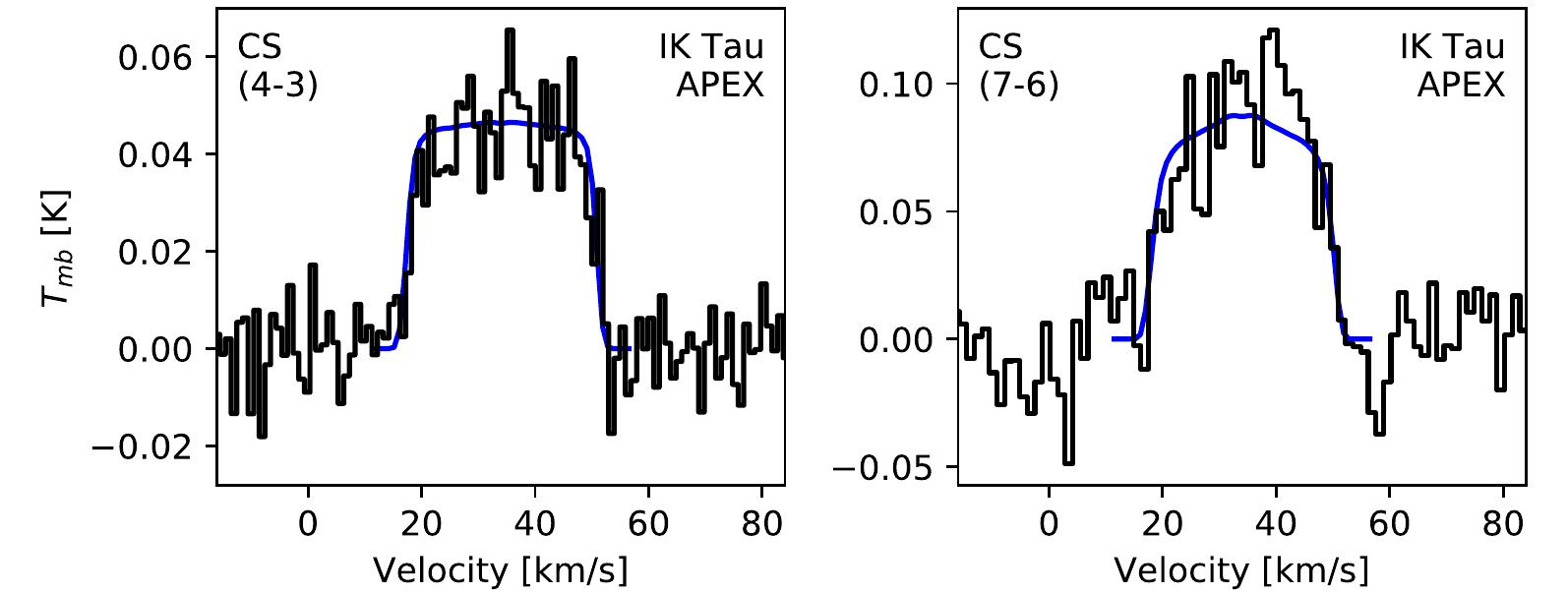}
\caption{Observations (black histograms) and model results (blue lines) for CS towards IK Tau, an M-type star, plotted with respect to LSR velocity.}
\label{CSIKTauplots}
\end{center}
\end{figure}

For each source we have only two or three observed CS lines with which to constrain our models. However, in each case there was at least a pair of lines with transitions separated by $\Delta J=2$ which corresponded to an increase in upper energy level by a factor greater than two and different emitting regions within the CSE. This proved to be sufficient to constrain the $e$-folding radius for most of our sources. Some example CS results for AI~Vol (carbon star), IK~Tau (M-type), and $\chi$~Cyg (S-type) are plotted in Figures \ref{CSAIVolplots}, \ref{CSIKTauplots}, and \ref{CSSplots}, respectively. The remaining carbon-rich CS models are plotted in {Figures \ref{CSCplots-1}, \ref{CSCplots-2}, \ref{CSCplots-3}, \ref{CSCplots-4}, \ref{CSCplots-5}, and \ref{CSCplots-6}, while the remaining oxygen-rich CS models are plotted in Figures \ref{CSMplots-1} and \ref{CSMplots-2}}. CS abundances range from $\sim 4\e{-7}$ to $\sim 2\e{-5}$ for carbon stars, from only $\sim3\e{-8}$ to $\sim1\e{-7}$ for oxygen rich stars, and from $\sim1\e{-7}$ to $\sim8\e{-6}$ for the S-type stars. 

We had some difficulty finding a conclusive $e$-folding radius for CS towards three of our sources: the carbon-rich V1259~Ori, and the oxygen-rich GX~Mon and V1300~Aql. The main difficulty with the two oxygen-rich stars was the low signal-to-noise ratio for the CS lines, leading to ambiguity in fitting the models. The case of V1259~Ori is more complicated and is discussed in more detail in Sect. \ref{v1259ori}.

From the stars for which we were able to constrain the CS $e$-folding radius, we find the following relation between $e$-folding radius and wind density {when weighting with the uncertainties listed in Table \ref{results}}
\begin{equation}\label{recs}
\log_{10}(R_{e,\mathrm{CS}}) = (18.9\pm0.2) + (0.40\pm0.03)\log_{10}\left(\frac{\dot{M}}{v_\infty}\right),
\end{equation} 
where $R_e$ is given in cm, $\dot{M}$ in $\spy$, $\upsilon_\infty$ in $\kms$, {and the errors are $1\sigma$ uncertainties}.
This fit is represented by the dashed black line in the right panel of Fig. \ref{radvsdens}.

We also ran models for V1259~Ori, GX~Mon, and V1300~Aql with the $R_e$ as obtained from Eq. \ref{recs}. While we were able to find adequate models with this added restriction, the $\chi^2$ values of the new models were consistently higher than for the models listed in Table \ref{results}. The new models also had systematically higher $f_0$ values by about 10--20\%, but this increase does not change our overall conclusions. We generally refer to the original models, with $R_e$ as a free parameter, when we discuss results.

\section{Discussion}\label{disc}

\subsection{Limitations of the modelling}

As mentioned in Sect. \ref{csresults}, we did not have access to many CS lines, making the modelling uncertain. The accuracy of all our CS models would improve with access to higher-$J$ lines or interferometric data to better constrain the extents of the CS envelopes. 
When experimenting with high CS abundances, we found the issue that very high CS abundances ($\gtrsim 2\e{-5}$) lead to the model lines becoming fainter with increased abundance rather than brighter, as would ordinarily be expected, most likely due to high optical depths leading to saturation. This explains why, for some of the highest CS abundances, we were not able to find an uncertainty to the upper limit of the abundance in our models (see the results marked with a $\dagger$ in Table \ref{results}). We did not run into similar problems with SiS {since, over a given energy range, SiS energy levels are more numerous, allowing the molecules to be spread over a larger number of states and hence reducing the optical depth of the lines. This is also why SiS lines are inherently less intense than CS lines \citep{Muller2005}}.
Two stars for which we found somewhat anomalous results, AI~Vol and II~Lup, are discussed in more detail in Sections \ref{aivol} and \ref{iilup}.

\subsection{Trends seen in our results}

The estimated SiS and CS fractional abundances are summarised in Fig. \ref{starab}, and the abundances shown as a function of CSE density in Fig. \ref{abundancevsdensity}. 
SiS is only detected for the higher mass-loss rate objects for all three chemical types. Its abundance is $\approx$\,10$^{-5}$ for the carbon stars, and hence CS and SiS combined can account for {almost} all of the sulphur in these stars. For the oxygen-rich and S-type stars the average SiS abundance is about one order of magnitude lower. Based on our detection limits, we calculated an upper limit on the fractional SiS abundance for W~Hya of $1\e{-6}$, for RR~Aql of $3\e{-7}$, for R~Lep of $6\e{-6}$ and {for S~Lyr of $1.5\e{-5}$}. W~Hya is the lowest mass-loss rate M-type star, RR~Aql and R~Lep are the highest mass-loss rate M-type and carbon stars, respectively, for which SiS was not detected, {and S~Lyr is the highest mass-loss rate S-type star with a CS detection but not an SiS detection}. These upper limits do not rule out abundances comparable to those for similar stars for which SiS was detected. {We note, in particular, that the upper limit for SiS towards S~Lyr is a similar amount (in dex) above the calculated CS abundance in Fig. \ref{abundancevsdensity} as seen for the other S-type stars as well as the carbon stars with similar densities.}

The CS abundances are $\approx$\,10$^{-5}$ for the carbon stars independent of the CSE density (only the star with the lowest mass-loss rate is a significant exception to this). For the oxygen-rich stars CS was only detected for the higher mass-loss rate objects, and here the abundances are more than two orders of magnitude lower than for the carbon stars. The upper limits we calculated for W~Hya, RR~Aql, {and RZ~Sgr} were $9\e{-8}$, $6\e{-8}$, {and $1.5\e{-7}$} respectively. {For the M-type stars this is} comparable to the abundances for the M-type stars with detected CS. {For RZ~Sgr the CS upper limit is just below the SiS abundance, giving a similar difference between SiS and (upper limit) CS abundances as found for the other S-type stars.} For the S-type stars {as a whole,} there is a trend such that the CS abundances for the lower mass-loss rate stars are almost two orders of magnitude lower than for the carbon stars, which increase to values similar to that of carbon stars at the higher mass-loss rates. {However, RZ~Sgr does not comply with this trend if the CS upper limit is taken into account. Similarly, a trend in SiS abundances could be seen with higher abundances correlated with higher densities if RZ~Sgr were to be excluded. It is unclear from the available data, especially considering the small number of detections for CS and SiS towards S stars, whether RZ~Sgr is an outlier due to inaccurate input parameters (such as mass-loss rate and/or distance) or whether it truly belies the apparent trends seen for the other S stars. Higher sensitivity observations of the S stars, providing a larger sample of detections, would help to confirm whether the trends we see here are real or a coincidental product of the sample. The potential anomaly of RZ~Sgr aside,} the abundances of CS for the S-type stars fall between those of carbon and oxygen-rich stars.

In Fig. \ref{sisvscs} we plot the modelled abundances of CS against those of SiS. The points appear to be grouped by chemical type with the carbon stars clustered in the top right, exhibiting the highest abundances of both molecular species. For the M-type and S-type stars there may be a correlation between the abundances of SiS and CS, possibly following slightly different trends. However, the small number of sources involved renders this only a tentative result. 

While it might be expected that S-type stars with greater amounts of photospheric carbon --- such as those classified as SC stars --- would be more likely to produce circumstellar CS, there is no clear relationship in our results between spectral type and the detection of CS in the S-type stars. 
For example TT~Cen and S~Lyr are both SC stars, but CS was only detected in S~Lyr, despite the expectation that (assuming a similar abundance) it should be easier to detect in TT~Cen, which is a closer source and has a similar (even slightly higher) mass-loss rate.

{As can be seen in Fig. \ref{radvsdens}, we find generally smaller $e$-folding radii for SiS than for CS. This is most likely due to the fact that the binding energy of CS is higher than that of SiS \citep[7.8~eV compared with 6.4~eV,][]{Herzberg1989,Gail2013}, making SiS more readily photodissociated by the interstellar radiation field.}

We also note the results of \cite{Gonzalez-Delgado2003}, \cite{Schoier2006}, and \cite{Ramstedt2009} who observed and modelled the abundances of SiO in oxygen-rich, carbon-rich and S-type AGB stars, respectively. Those studies find a trend of decreasing SiO abundance with increasing wind density, most clearly seen for the oxygen-rich and carbon-rich stars (since fewer high mass-loss rate S-type stars have been identified). {A similar trend was found by \cite{Massalkhi2018} for SiC$_2$ abundance decreasing with carbon star wind density.}
Although we do not see a clear trend in SiS abundance with density (see Fig. \ref{abundancevsdensity}), it is possible that the detection of SiS only in the highest mass-loss rate AGB stars is linked to the decreased abundances of SiO {and/or SiC$_2$} in the same stars. {Also, considering only the S-type stars and excluding RZ~Sgr (see discussion above), there is a possible trend of increased SiS abundance with increased density, the opposite of the trend seen for SiO by \cite{Ramstedt2009}.} For the carbon stars that have very high SiS abundances, accounting for roughly half the available S \citep[for a solar abundance of S, taken from][]{Asplund2009}, the SiS abundance can also account for a significant portion of the available Si (the solar abundance of which is approximately twice that of S). For some of the low mass-loss rate stars included in the \cite{Gonzalez-Delgado2003} and \cite{Schoier2006} studies, however, the abundances of SiO approaches the solar abundance of Si. For W~Hya, a nearby low mass-loss rate star for which we did not detect SiS, \cite{Khouri2014a} found an SiO abundance high enough to account for almost all of the Si, while \cite{Danilovich2016} found that SO and \so2 combined account for almost all of the S. {As Si, O and C are all known to play a part in dust formation, the depletion of these elements onto dust grains may play a part in the sulphur chemistry, especially if we consider that larger quantities of dust are generally associated with higher mass-loss rate AGB stars \citep{Justtanont1992}. In any case,} it seems from both earlier studies and from this work that the wind density plays an important role in determining the chemical composition of AGB CSEs.

\subsection{Comparison with other observational studies}

\cite{Schoier2007} surveyed a sample of carbon- and oxygen-rich AGB stars and detected SiS towards eleven carbon stars and eight M-type AGB stars. They do not explicitly list any non-detections, but their presented SiS lines are mostly\footnote{Indeed, the only star for which they detected SiS and which had a lower mass-loss rate ($\dot{M} = 5\e{-7}\spy$) than our lower-limit for SiS detections is the oxygen-rich R~Cas. However, a more recent study of R~Cas using higher-$J$ CO lines from \textsl{Herschel}/HIFI to determine the mass-loss rate, conducted by \cite{Maercker2016}, found $\dot{M} = 8\e{-7}\spy$, equal to the lowest mass-loss rate star with an SiS detection in our sample.} seen towards high mass-loss rate stars.  
In their radiative transfer modelling, \cite{Schoier2007} assumed the same photodissociation radius for SiS as for SiO. Therefore, they used the \cite{Gonzalez-Delgado2003} SiO empirical relation between mass-loss rate, wind velocity and photodissociation to find $e$-folding radii of SiS for their sample stars. They were unable to find good fits to the observed data using this assumption and found better fits by adding a central component with radius out to $1\e{15}$~cm and with a high SiS abundance of $2\e{-5}$. With this distribution SiS would account for most of the sulphur in the inner CSE.
In contrast, we were able to find good fits with Gaussian SiS abundance distributions by leaving the $e$-folding radius as a free parameter. This approach did not require the inclusion of a central component of higher SiS abundance. For those sources which overlap with the \cite{Schoier2007} sample we found smaller $e$-folding radii than they did and fractional abundances larger than their Gaussian components but smaller than their inner components. {It is not surprising that leaving the $e$-folding radius as a free parameter gives a better fit to the observations than using the SiO $e$-folding radius does, since the dissociation energy of SiS is 6.4~eV, compared with 8.28~eV for SiO \citep{Gail2013}. These are sufficiently different that the extents of the corresponding molecular envelopes ought not to be identical.}

\cite{Decin2010} modelled several molecules, including SiS and CS, for the oxygen-rich AGB star IK~Tau. They find an inner abundance for CS of $8\e{-8}$, relative to the \h2 abundance, and use a non-Gaussian distribution based partly on the results of chemical modelling. Their CS result is {in good agreement with ours,} within a factor of $\sim1.4$, although we find a smaller extent for the CS envelope. This is mostly likely due to the addition of the lower-$J$ line CS ($4\to3$) in our study, compared with the use of only the ($7\to6$) and ($6\to5$) in the \cite{Decin2010} study. For SiS, \cite{Decin2010} found a similar core and extended plateau abundance distribution to that used by \cite{Schoier2007}. They found a high inner abundance of SiS of $1.1\e{-5}$, relative to \h2, which drops to $8\e{-9}$ at $\sim1.5\e{15}$~cm and does not decrease again until $\sim3\e{16}$~cm. While this agrees well with  \cite{Schoier2007}, despite having been calculated using different methodology, it does not agree with our result for the same reasons discussed above. Our abundance of $1.7\e{-6}$ is intermediate to their two extremes and the extent of our SiS envelope is significantly smaller.

\cite{Olofsson1993a} surveyed a sample of about 40 carbon stars and detected the CS ($2\to1$) line in 11 of them. The only stars in both their sample and ours were U~Hya, X~TrA, and R~Lep, for which they did indeed also detect CS. The CS abundances they calculated for all sources were upper limits, and our results fall well below these for U~Hya and X~TrA. For R~Lep our calculated CS peak fractional abundance is in good agreement with the upper limit they found, although we find the $e$-folding radius to be larger by about a factor of two.
In addition to the radio data, \cite{Olofsson1993a} also collected infrared photometry for about 60 stars and, based on LTE model atmospheres, calculated photospheric molecular abundances for a few species, including CS and SiS. For the three overlapping stars they found relatively high photospheric SiS abundances, while we did not detect SiS in their CSEs (in fact, they were the only three carbon stars for which we did not detect SiS). For CS, the photospheric abundance found for R~Lep is more than two orders of magnitude smaller than our CSE abundance, for U~Hya it is an order of magnitude larger than our CSE abundance, and for X~TrA it is comparable with our CSE abundance, being only a factor of two larger. As \cite{Olofsson1993a} note, the photospheric CS abundance is very sensitive to temperature, and hence is likely to change significantly between different phases of pulsation.

In a search for both CS and SiS (among other molecules) in a sample of 31 oxygen-rich stars, \cite{Lindqvist1988} detected the CS ($2\to1$) line towards only four sources and the SiS ($5\to4$) line only towards TX~Cam. They estimated CS abundances in the order of a few $10^{-7}$, with their IK~Tau result a factor of about two higher than what we found for that source, and an SiS abundance of $\sim 1\e{-6}$, in agreement with our results. 
\cite{Bujarrabal1994} surveyed a sample of evolved stars (mostly AGB stars) using the IRAM 30~m telescope to observe several molecular species including CS and SiS. They estimated molecular abundances based on the integrated intensities of their observed lines and assuming a constant abundance within a given radius. They note that their estimates only hold for optically thin lines. Where their sample overlapped with ours, we found higher SiS abundances for IK~Tau and V1300~Aql by factors of four and two, respectively, while our CS abundances were in agreement for IK~Tau and $\chi$~Cyg, a factor of a half smaller for V1300~Aql, V1259~Ori, and W~Aql, a factor of about three smaller for R~And, and almost a factor of 6 higher for V821~Her. In the case of SiS in the oxygen-rich stars (no carbon star SiS observations overlapped with our sample and they did not detect SiS towards any S-type stars) and CS towards the carbon stars and W~Aql, the discrepancies are most likely due to optical depth effects since our models indicate optically thick emission in these cases. For CS towards the oxygen-rich stars and in $\chi$~Cyg, which our models indicate to be optically thin, our results are in better agreement, with the discrepancy in the V1300~Aql and R~And abundances most likely due to uncertainties caused by the weak emission in the case of both our observations and those of \cite{Bujarrabal1994}.
It should also be noted that \cite{Olofsson1993a}, \cite{Lindqvist1988} and \cite{Bujarrabal1994} all used similar and simple methods for estimating abundances.

\subsection{Comparison with chemical models}

Since CS and SiS have long been known to occur in AGB CSEs, they are regularly included in chemical models of these stellar winds. Indeed, SiS and CS are commonly assumed to be parent species --- molecules formed in the innermost regions, from which other species are subsequently formed --- sometimes for CSEs of all chemical types. In this section we discuss some different existing chemical modelling results and compare them with our results.

\subsubsection{LTE models of the mid- and outer-CSE}

\cite{Willacy1997} modelled the chemistry in the cooler outer regions of the CSE of an oxygen-rich AGB star. They used the characteristics of TX~Cam ($\dot{M} = 3\e{-6}\spy$, $\upsilon_\infty=18~\kms$) as the basis for their models, although they also give some model results based on IK~Tau and R~Dor. They take SiS and \h2S as parent species and model the CSE from $2\e{15}$~cm outwards. They predict a more extended SiS envelope than we find, by about an order of magnitude, but with a peak SiS abundance in reasonable agreement with ours: about twice the abundance we find for IK~Tau. {The SiS column density they find for IK~Tau, using a similar mass-loss rate and slightly higher expansion velocity than those used in our study ($\dot{M}=4.5\e{-6}\spy$, $\upsilon_\infty = 20~\kms$), is more than an order of magnitude lower than ours.} They also predict CS to be located in a shell around the star, with the peak in abundance falling at a few $10^{16}$~cm for TX~Cam. The peak appears to be roughly in agreement with the $e$-folding radii we find for the oxygen-rich stars in our sample. For IK~Tau they find a peak abundance for CS of $\sim3\e{-7}$, which is about three times what we calculate for our centrally peaked CS distribution. {Their column density for IK~Tau is about two orders of magnitude lower than what we find.} The CS lines that we observe are formed in relatively cool regions, so they are not sensitive to a possible lower CS abundance in the inner regions.

\cite{Li2016} model a similar outer region of an oxygen-rich CSE, focussing on UV photochemistry, and using IK~Tau as their fiducial model (with a similar mass-loss rate, $\dot{M}=4.5\e{-6}\spy$, and higher expansion velocity, $\upsilon_\infty = 24~\kms$, than used in the present study). Their parent species abundances are taken from observations, where available, and shock-induced non-LTE predictions otherwise. They include several S-bearing molecules in their list of parent species: SiS, CS, \h2S, SO, \so2, and HS. Their initial abundance of SiS, in particular, accounts for a significant portion of the sulphur budget and is about an order of magnitude higher than the peak abundance we found for IK~Tau. Like \cite{Willacy1997}, they find a more extended SiS envelope, which declines more slowly than our IK~Tau model, resulting in their model being about an order of magnitude more extended than ours. Considering CS, however, their results are in reasonable agreement with ours, with a similar (centrally peaked) inner abundance of CS and a similarly large extent, although at very large extents their model deviates from our Gaussian assumption.

\subsubsection{Chemistry in a clumpy medium}

\cite{Agundez2010} investigated the effects of clumpiness on AGB chemistry and, in particular, the penetration of UV photons. They included SiS and CS as parent species for the carbon-rich models and SiS, SO and \h2S in their oxygen-rich models. They concluded that UV penetration in clumpy and low mass-loss rate CSEs (up to a few $10^{-7}\spy$) can trigger photochemistry in the warm inner regions of the CSEs, allowing the formation of CS (and HCN and NH\down{3}) in oxygen-rich CSEs (as well as \h2O and NH\down{3} in carbon-rich CSEs). Their predictions of CS abundances in the range $10^{-8}-10^{-7}$ for all mass-loss rates are in agreement with the abundances we find for the higher mass-loss rate oxygen-rich stars and are not ruled out by our non-detections for the lower mass-loss rate stars.

\cite{Van-de-Sande2018} developed a chemical model incorporating a porosity formalism to treat the increased penetration of UV photons in a clumpy CSE, and considering the relative overdensity of the clumps. They run models for different clump parameters, including the density contrast and clump size, and find that larger deviations from a smooth (non-clumpy) outflow generally result in higher abundances of CS for a range of mass-loss rates ($\dot{M}=10^{-7}$, $10^{-6}$, and $10^{-5}\spy$) in the case of oxygen-rich CSEs. Although their derived radial abundance distribution profiles differ from the Gaussian profiles we use in this study, their peak CS abundances are in good agreement with the peak abundances that we found for our higher mass-loss rate oxygen-rich stars. {Their CS column densities for oxygen-rich AGB stars, which vary significantly with clumpiness, are in agreement for the most extreme clumpy models, and up to three orders of magnitude lower than our results, depending on the specific clumpiness of the outflow.}

\subsubsection{Our results from a chemical perspective}

Regarding possible chemical mechanisms to explain our results, the only neutral-neutral formation rate for the SiS molecule is the radiative association 
Si + S $\to$ SiS reported by \cite{Andreazza2007}, which has an activation barrier of only 66~K. 
The Arrhenius rate\footnote{An Arrhenius reaction rate $k(T)$ is generally defined as
\\
$k(T) = A ({T}/{300~\mathrm{K}})^B \times \exp(- E_a/T)$
\\
where A is the pre-exponential factor, $B$ the temperature dependence and $E_a$ the activation barrier in K.} has a small pre-exponential factor and very weak temperature dependence. 
As a consequence, the process occurs efficiently at high densities and 
is negligible at low densities, in agreement with our SiS detection pattern. A temperature independent SiS photodissociation rate is reported by \cite{Prasad1980}, {but based on the difference between the \cite{Prasad1980} rates for SiO and SiH photodissociation and those calculated more recently \citep[e.g. by][]{Heays2017}, it is likely to be off by at least an order of magnitude}.
However, apart from these two reaction rates, the neutral SiS chemistry is poorly characterised.
A more detailed description is required to construct an accurate chemical-kinetic network
(i.e. interactions of SiS with SiO, silicates, other S-bearing compounds, etc).

To model the SiS abundance without specific rate prescriptions,  
we performed thermodynamic equilibrium (TE) calculations \citep{White1958,Tsuji1973}, using the same method as implemented by \citet{Gobrecht2016} but extended to a larger sample space of temperatures and densities. The TE calculations are based on the minimisation of the total gas Gibbs free energy whose components are tabulated (NIST-JANAF Thermochemical Tables\footnote{\url{http://kinetics.nist.gov/janaf/}}) for a given elemental mixture.
For the conditions of an oxygen-rich CSE (C/O $= 0.75$) we find a strong density 
dependence of the SiS fractional abundance (see Table \ref{davidtable}). In particular, 
at $T=1500$~K and densities of \mbox{$n=10^{12}$ to $10^{14}$~cm$^{-3}$}, which typically apply for 1--2 $R_*$, we find good agreement with the observed abundances in higher mass-loss rate sources. A similar argument could be applied to the carbon-rich sources.

\begin{table*}[tp]
\caption{SiS thermodynamic equilibrium fractional abundances for different number densities and C/O = 0.75.}\label{davidtable}
\centering
\begin{tabular}{c|ccccc}
\hline\hline
\backslashbox{T [K]}{$n$~[cm$^{-3}$]}    &$10^{14}$&  $10^{13}$ & $10^{12}$ &  $10^{11}$&   $10^{10}$ \\ 
\hline
2000 &  $3.4\e{-8}$ & $2.1\e{-9}$ & $3.4\e{-10}$ & $2.7\e{-10}$ & $2.5\e{-10}$\\
1500 & $4.6\e{-6}$ & $1.0\e{-6}$ & $2.2\e{-7}$  & $2.3\e{-8}$ &  $1.7\e{-9}$\\
1000 & $1.2\e{-5}$ & $1.2\e{-5}$ & $1.2\e{-5}$ &  $1.2\e{-5}$ &  $1.1\e{-5}$ \\
\hline
\end{tabular}
\end{table*}

Regarding the presence of CS in oxygen-rich stars, its formation is thought to be greatly enhanced by shock chemistry \citep{Duari1999,Cherchneff2006,Gobrecht2016}. The extreme conditions in shocks free up C from CO, allowing CS to form even in carbon-deficient environments. We expect more extreme shock conditions in the higher mass-loss rate sources \citep{Mattsson2007}, which can drive non-equilibrium reactions. Hence CS is more likely to form in these sources. In carbon stars, where there is abundant C, CS can form in thermal equilibrium and hence such extreme conditions are not required to form CS. This also explains the presence of CS in even the lowest mass-loss rate carbon stars. {The roughly constant abundance of CS in most of the carbon stars is most likely due to CS having a high binding energy and forming readily in the presence of abundant C.}

{Although sulphur does not readily condense onto dust grains, as evidenced by the lack of sulphur depletion found by studies of post-AGB stars \citep{Reyniers2007,Waelkens1991}, it is possible that dust-grain interactions may play a part in determining the abundances of CS and SiS. \cite{Gobrecht2016} performed a theoretical study on a inner winds of an oxygen-rich AGB star using an extensive chemical-kinetic network including the species CS, SiS, SO, \so2, \h2S, SH, and OCS. By comparing models with and without dust condensation, we find no significant difference in the abundances of CS and SiS. }

\section{Conclusions}

In this study we observed SiS and CS towards a large number of AGB stars. CS was detected towards all observed carbon stars, some S-type stars, and the highest mass-loss rate M-type stars. SiS was only detected towards the highest mass-loss rate sources for all chemical types.

We find higher abundances of both CS and SiS in carbon stars than S-type stars or M-type stars. More specifically, we found SiS abundances ranging from $\sim 9\e{-6}$ to $\sim 2\e{-5}$ for the carbon stars, from $\sim 5\e{-7}$ to $\sim 2\e{-6}$ for the oxygen-rich stars, and from $\sim 2\e{-7}$ to $\sim 2\e{-6}$ for the S-type stars. Our CS abundances ranged from $\sim 4\e{-7}$ to $\sim2\e{-5}$ for the carbon stars, from $\sim 3\e{-8}$ to $\sim1\e{-7}$ for the oxygen-rich stars and from $\sim 1\e{-7}$ to $\sim8\e{-6}$ for the S-type stars.

A correlation between CS abundance and CSE density for S-type stars is indicated by our results, {and may also be seen for SiS if one star is excluded from our sample. However,} no similar correlation can be seen for the carbon- or oxygen-rich stars, {although this could be partly due to the small number of sources.}
Thermodynamic equilibrium calculations predict that SiS should form more readily in denser environments, in agreement with our observational results of only detecting SiS in such environments. Also, CS formation is thought to be strongly enhanced by shock chemistry, which would explain why it forms more readily in the higher mass-loss rate oxygen rich sources --- where shocks are expected to be more extreme --- than in the low mass-loss rate oxygen-rich sources. This also explains the trend for higher CS abundances with higher densities seen in the S-type stars. Carbon stars, with their plentiful C, do not require shocks to free up C and drive CS formation and can instead form CS in thermal equilibrium.

\begin{acknowledgements}

Based on observations made with APEX under programme IDs O-097.F-9318, O-098.F-9305, 077.F-9310 and 079.F-9325.

This work is based on observations carried out under project number 010-13 with the IRAM 30m telescope. IRAM is supported by INSU/CNRS (France), MPG (Germany) and IGN (Spain).

The Onsala 20 m telescope is operated by the Swedish National Facility for Radio Astronomy, Onsala Space Observatory at Chalmers University of Technology.

LD acknowledges support from the ERC consolidator grant 646758 AEROSOL and the FWO Research Project grant G024112N. EDB acknowledges financial support from the Swedish National Space Board. HO acknowledges financial support from the Swedish Research Council.

\end{acknowledgements}

\bibliographystyle{aa}
\bibliography{33317}

\begin{thebibliography}{67}
\expandafter\ifx\csname natexlab\endcsname\relax\def\natexlab#1{#1}\fi

\bibitem[{{Ag{\'u}ndez} {et~al.}(2010){Ag{\'u}ndez}, {Cernicharo}, \&
  {Gu{\'e}lin}}]{Agundez2010}
{Ag{\'u}ndez}, M., {Cernicharo}, J., \& {Gu{\'e}lin}, M. 2010, \apjl, 724, L133

\bibitem[{{Andreazza} \& {Marinho}(2007)}]{Andreazza2007}
{Andreazza}, C.~M. \& {Marinho}, E.~P. 2007, \mnras, 380, 365

\bibitem[{{Asplund} {et~al.}(2009){Asplund}, {Grevesse}, {Sauval}, \&
  {Scott}}]{Asplund2009}
{Asplund}, M., {Grevesse}, N., {Sauval}, A.~J., \& {Scott}, P. 2009, \araa, 47,
  481

\bibitem[{{Belitsky} {et~al.}(2018){Belitsky}, {Lapkin}, {Fredrixon},
  {Meledin}, {Sundin}, {Billade}, {Ferm}, {Pavolotsky}, {Rashid}, {Strandberg},
  {Desmaris}, {Ermakov}, {Krause}, {Olberg}, {Aghdam}, {Shafiee}, {Bergman},
  {Beck}, {Olofsson}, {Conway}, {Breuck}, {Immer}, {Yagoubov},
  {Montenegro-Montes}, {Torstensson}, {P{\'e}rez-Beaupuits}, {Klein}, {Boland},
  {Baryshev}, {Hesper}, {Barkhof}, {Adema}, {Bekema}, \&
  {Koops}}]{Belitsky2018}
{Belitsky}, V., {Lapkin}, I., {Fredrixon}, M., {et~al.} 2018, \aap, 612, A23

\bibitem[{{Belitsky} {et~al.}(2015){Belitsky}, {Lapkin}, {Fredrixon}, {Sundin},
  {Helldner}, {Pettersson}, {Ferm}, {Pantaleev}, {Billade}, {Bergman},
  {Olofsson}, {Lerner}, {Strandberg}, {Whale}, {Pavolotsky}, {Flygare},
  {Olofsson}, \& {Conway}}]{Belitsky2015}
{Belitsky}, V., {Lapkin}, I., {Fredrixon}, M., {et~al.} 2015, \aap, 580, A29

\bibitem[{{Belitsky} {et~al.}(2006){Belitsky}, {Lapkin}, {Monje}, {Vassilev},
  {Risacher}, {Pavolotsky}, {Meledin}, {Olberg}, {Pantaleev}, \&
  {Booth}}]{Belitsky2006}
{Belitsky}, V., {Lapkin}, I., {Monje}, R., {et~al.} 2006, in \procspie, Vol.
  6275, Society of Photo-Optical Instrumentation Engineers (SPIE) Conference
  Series, 62750G

\bibitem[{Billade {et~al.}(2012)Billade, Nystrom, Meledin, Sundin, Lapkin,
  Fredrixon, Desmaris, Rashid, Strandberg, Ferm, Pavolotsky, \&
  Belitsky}]{Billade2012}
Billade, B., Nystrom, O., Meledin, D., {et~al.} 2012, IEEE Transactions on
  Terahertz Science and Technology, 2, 208

\bibitem[{{Brunner} {et~al.}(2018){Brunner}, {Danilovich}, {Ramstedt},
  {Marti-Vidal}, {De Beck}, {Vlemmings}, {Lindqvist}, \&
  {Kerschbaum}}]{Brunner2018}
{Brunner}, M., {Danilovich}, T., {Ramstedt}, S., {et~al.} 2018, ArXiv e-prints
  [\eprint[arXiv]{1806.01622}]

\bibitem[{{Bujarrabal} {et~al.}(1994){Bujarrabal}, {Fuente}, \&
  {Omont}}]{Bujarrabal1994}
{Bujarrabal}, V., {Fuente}, A., \& {Omont}, A. 1994, \aap, 285, 247

\bibitem[{{Cernicharo} {et~al.}(2000){Cernicharo}, {Gu{\'e}lin}, \&
  {Kahane}}]{Cernicharo2000}
{Cernicharo}, J., {Gu{\'e}lin}, M., \& {Kahane}, C. 2000, \aaps, 142, 181

\bibitem[{{Cernicharo} {et~al.}(1987){Cernicharo}, {Kahane}, {Guelin}, \&
  {Hein}}]{Cernicharo1987}
{Cernicharo}, J., {Kahane}, C., {Guelin}, M., \& {Hein}, H. 1987, \aap, 181, L9

\bibitem[{{Cherchneff}(2006)}]{Cherchneff2006}
{Cherchneff}, I. 2006, \aap, 456, 1001

\bibitem[{{Danilovich} {et~al.}(2014){Danilovich}, {Bergman}, {Justtanont},
  {Lombaert}, {Maercker}, {Olofsson}, {Ramstedt}, \& {Royer}}]{Danilovich2014}
{Danilovich}, T., {Bergman}, P., {Justtanont}, K., {et~al.} 2014, \aap, 569,
  A76

\bibitem[{{Danilovich} {et~al.}(2016){Danilovich}, {De Beck}, {Black},
  {Olofsson}, \& {Justtanont}}]{Danilovich2016}
{Danilovich}, T., {De Beck}, E., {Black}, J.~H., {Olofsson}, H., \&
  {Justtanont}, K. 2016, \aap, 588, A119

\bibitem[{{Danilovich} {et~al.}(2015){Danilovich}, {Teyssier}, {Justtanont},
  {Olofsson}, {Cerrigone}, {Bujarrabal}, {Alcolea}, {Cernicharo},
  {Castro-Carrizo}, {Garc{\'{\i}}a-Lario}, \& {Marston}}]{Danilovich2015a}
{Danilovich}, T., {Teyssier}, D., {Justtanont}, K., {et~al.} 2015, \aap, 581,
  A60

\bibitem[{{Danilovich} {et~al.}(2017){Danilovich}, {Van de Sande}, {De Beck},
  {Decin}, {Olofsson}, {Ramstedt}, \& {Millar}}]{Danilovich2017a}
{Danilovich}, T., {Van de Sande}, M., {De Beck}, E., {et~al.} 2017, \aap, 606,
  A124

\bibitem[{{Dayou} \& {Balan{\c c}a}(2006)}]{Dayou2006}
{Dayou}, F. \& {Balan{\c c}a}, C. 2006, \aap, 459, 297

\bibitem[{{Decin} {et~al.}(2010){Decin}, {De Beck}, {Br{\"u}nken},
  {M{\"u}ller}, {Menten}, {Kim}, {Willacy}, {de Koter}, \&
  {Wyrowski}}]{Decin2010}
{Decin}, L., {De Beck}, E., {Br{\"u}nken}, S., {et~al.} 2010, \aap, 516, A69

\bibitem[{{Duari} {et~al.}(1999){Duari}, {Cherchneff}, \&
  {Willacy}}]{Duari1999}
{Duari}, D., {Cherchneff}, I., \& {Willacy}, K. 1999, \aap, 341, L47

\bibitem[{{Endres} {et~al.}(2016){Endres}, {Schlemmer}, {Schilke}, {Stutzki},
  \& {M{\"u}ller}}]{Endres2016}
{Endres}, C.~P., {Schlemmer}, S., {Schilke}, P., {Stutzki}, J., \&
  {M{\"u}ller}, H.~S.~P. 2016, Journal of Molecular Spectroscopy, 327, 95

\bibitem[{{Gail} \& {Sedlmayr}(2013)}]{Gail2013}
{Gail}, H.-P. \& {Sedlmayr}, E. 2013, {Physics and Chemistry of Circumstellar
  Dust Shells} (Cambridge University Press)

\bibitem[{{Gobrecht} {et~al.}(2016){Gobrecht}, {Cherchneff}, {Sarangi},
  {Plane}, \& {Bromley}}]{Gobrecht2016}
{Gobrecht}, D., {Cherchneff}, I., {Sarangi}, A., {Plane}, J.~M.~C., \&
  {Bromley}, S.~T. 2016, \aap, 585, A6

\bibitem[{{Gong} {et~al.}(2015){Gong}, {Henkel}, {Spezzano}, {Thorwirth},
  {Menten}, {Wyrowski}, {Mao}, \& {Klein}}]{Gong2015}
{Gong}, Y., {Henkel}, C., {Spezzano}, S., {et~al.} 2015, \aap, 574, A56

\bibitem[{{Gonz{\'a}lez Delgado} {et~al.}(2003){Gonz{\'a}lez Delgado},
  {Olofsson}, {Kerschbaum}, {Sch{\"o}ier}, {Lindqvist}, \&
  {Groenewegen}}]{Gonzalez-Delgado2003}
{Gonz{\'a}lez Delgado}, D., {Olofsson}, H., {Kerschbaum}, F., {et~al.} 2003,
  \aap, 411, 123

\bibitem[{{G{\"u}sten} {et~al.}(2006){G{\"u}sten}, {Nyman}, {Schilke},
  {Menten}, {Cesarsky}, \& {Booth}}]{Gusten2006}
{G{\"u}sten}, R., {Nyman}, L.~{\AA}., {Schilke}, P., {et~al.} 2006, \aap, 454,
  L13

\bibitem[{Habing \& Olofsson(2003)}]{AGB}
Habing, H.~J. \& Olofsson, H., eds. 2003, Asymptotic Giant Branch Stars, A\&A
  Library (Springer)

\bibitem[{{Heays} {et~al.}(2017){Heays}, {Bosman}, \& {van
  Dishoeck}}]{Heays2017}
{Heays}, A.~N., {Bosman}, A.~D., \& {van Dishoeck}, E.~F. 2017, \aap, 602, A105

\bibitem[{{Herwig}(2005)}]{Herwig2005}
{Herwig}, F. 2005, \araa, 43, 435

\bibitem[{Herzberg(1989)}]{Herzberg1989}
Herzberg, G. 1989, Molecular Spectra and Molecular Structure: Spectra of
  diatomic molecules, Molecular Spectra and Molecular Structure (R.E. Krieger
  Publishing Company)

\bibitem[{{H{\"o}fner} \& {Olofsson}(2018)}]{Hofner2018}
{H{\"o}fner}, S. \& {Olofsson}, H. 2018, \aapr, 26, 1

\bibitem[{{Justtanont} \& {Tielens}(1992)}]{Justtanont1992}
{Justtanont}, K. \& {Tielens}, A.~G.~G.~M. 1992, \apj, 389, 400

\bibitem[{Kazarovets \& Pastukhova(2016)}]{Kazarovets2016}
Kazarovets, E.~V. \& Pastukhova, E.~N. 2016, Peremennye Zvezdy, Prilozhenie, 16

\bibitem[{{Khouri} {et~al.}(2014{\natexlab{a}}){Khouri}, {de Koter}, {Decin},
  {Waters}, {Lombaert}, {Royer}, {Swinyard}, {Barlow}, {Alcolea}, {Blommaert},
  {Bujarrabal}, {Cernicharo}, {Groenewegen}, {Justtanont}, {Kerschbaum},
  {Maercker}, {Marston}, {Matsuura}, {Melnick}, {Menten}, {Olofsson},
  {Planesas}, {Polehampton}, {Posch}, {Schmidt}, {Szczerba}, {Vandenbussche},
  \& {Yates}}]{Khouri2014}
{Khouri}, T., {de Koter}, A., {Decin}, L., {et~al.} 2014{\natexlab{a}}, \aap,
  561, A5

\bibitem[{{Khouri} {et~al.}(2014{\natexlab{b}}){Khouri}, {de Koter}, {Decin},
  {Waters}, {Maercker}, {Lombaert}, {Alcolea}, {Blommaert}, {Bujarrabal},
  {Groenewegen}, {Justtanont}, {Kerschbaum}, {Matsuura}, {Menten}, {Olofsson},
  {Planesas}, {Royer}, {Schmidt}, {Szczerba}, {Teyssier}, \&
  {Yates}}]{Khouri2014a}
{Khouri}, T., {de Koter}, A., {Decin}, L., {et~al.} 2014{\natexlab{b}}, \aap,
  570, A67

\bibitem[{{Li} {et~al.}(2016){Li}, {Millar}, {Heays}, {Walsh}, {van Dishoeck},
  \& {Cherchneff}}]{Li2016}
{Li}, X., {Millar}, T.~J., {Heays}, A.~N., {et~al.} 2016, \aap, 588, A4

\bibitem[{{Lindqvist} {et~al.}(1988){Lindqvist}, {Nyman}, {Olofsson}, \&
  {Winnberg}}]{Lindqvist1988}
{Lindqvist}, M., {Nyman}, L.-A., {Olofsson}, H., \& {Winnberg}, A. 1988, \aap,
  205, L15

\bibitem[{{Lombaert} {et~al.}(2016){Lombaert}, {Decin}, {Royer}, {de Koter},
  {Cox}, {Gonz{\'a}lez-Alfonso}, {Neufeld}, {De Ridder}, {Ag{\'u}ndez},
  {Blommaert}, {Khouri}, {Groenewegen}, {Kerschbaum}, {Cernicharo},
  {Vandenbussche}, \& {Waelkens}}]{Lombaert2016}
{Lombaert}, R., {Decin}, L., {Royer}, P., {et~al.} 2016, \aap, 588, A124

\bibitem[{{Maercker} {et~al.}(2016){Maercker}, {Danilovich}, {Olofsson}, {De
  Beck}, {Justtanont}, {Lombaert}, \& {Royer}}]{Maercker2016}
{Maercker}, M., {Danilovich}, T., {Olofsson}, H., {et~al.} 2016, \aap, 591, A44

\bibitem[{{Maercker} {et~al.}(2008){Maercker}, {Sch{\"o}ier}, {Olofsson},
  {Bergman}, \& {Ramstedt}}]{Maercker2008}
{Maercker}, M., {Sch{\"o}ier}, F.~L., {Olofsson}, H., {Bergman}, P., \&
  {Ramstedt}, S. 2008, \aap, 479, 779

\bibitem[{{Massalkhi} {et~al.}(2018){Massalkhi}, {Ag{\'u}ndez}, {Cernicharo},
  {Velilla Prieto}, {Goicoechea}, {Quintana-Lacaci}, {Fonfr{\'{\i}}a},
  {Alcolea}, \& {Bujarrabal}}]{Massalkhi2018}
{Massalkhi}, S., {Ag{\'u}ndez}, M., {Cernicharo}, J., {et~al.} 2018, \aap, 611,
  A29

\bibitem[{{Mattsson} {et~al.}(2007){Mattsson}, {H{\"o}fner}, \&
  {Herwig}}]{Mattsson2007}
{Mattsson}, L., {H{\"o}fner}, S., \& {Herwig}, F. 2007, \aap, 470, 339

\bibitem[{{M{\"u}ller} {et~al.}(2005){M{\"u}ller}, {Schl{\"o}der}, {Stutzki},
  \& {Winnewisser}}]{Muller2005}
{M{\"u}ller}, H.~S.~P., {Schl{\"o}der}, F., {Stutzki}, J., \& {Winnewisser}, G.
  2005, Journal of Molecular Structure, 742, 215

\bibitem[{{Olofsson} {et~al.}(1993){Olofsson}, {Eriksson}, {Gustafsson}, \&
  {Carlstroem}}]{Olofsson1993a}
{Olofsson}, H., {Eriksson}, K., {Gustafsson}, B., \& {Carlstroem}, U. 1993,
  \apjs, 87, 305

\bibitem[{{Omont} {et~al.}(1993){Omont}, {Lucas}, {Morris}, \&
  {Guilloteau}}]{Omont1993}
{Omont}, A., {Lucas}, R., {Morris}, M., \& {Guilloteau}, S. 1993, \aap, 267,
  490

\bibitem[{{Pickett} {et~al.}(1998){Pickett}, {Poynter}, {Cohen}, {Delitsky},
  {Pearson}, \& {M{\"u}ller}}]{Pickett1998}
{Pickett}, H.~M., {Poynter}, R.~L., {Cohen}, E.~A., {et~al.} 1998, \jqsrt, 60,
  883

\bibitem[{{Prasad} \& {Huntress}(1980)}]{Prasad1980}
{Prasad}, S.~S. \& {Huntress}, Jr., W.~T. 1980, \apjs, 43, 1

\bibitem[{{Ramstedt} {et~al.}(2017){Ramstedt}, {Mohamed}, {Vlemmings},
  {Danilovich}, {Brunner}, {De Beck}, {Humphreys}, {Lindqvist}, {Maercker},
  {Olofsson}, {Kerschbaum}, \& {Quintana-Lacaci}}]{Ramstedt2017}
{Ramstedt}, S., {Mohamed}, S., {Vlemmings}, W.~H.~T., {et~al.} 2017, \aap, 605,
  A126

\bibitem[{{Ramstedt} \& {Olofsson}(2014)}]{Ramstedt2014}
{Ramstedt}, S. \& {Olofsson}, H. 2014, \aap, 566, A145

\bibitem[{{Ramstedt} {et~al.}(2009){Ramstedt}, {Sch{\"o}ier}, \&
  {Olofsson}}]{Ramstedt2009}
{Ramstedt}, S., {Sch{\"o}ier}, F.~L., \& {Olofsson}, H. 2009, \aap, 499, 515

\bibitem[{{Ramstedt} {et~al.}(2008){Ramstedt}, {Sch{\"o}ier}, {Olofsson}, \&
  {Lundgren}}]{Ramstedt2008}
{Ramstedt}, S., {Sch{\"o}ier}, F.~L., {Olofsson}, H., \& {Lundgren}, A.~A.
  2008, \aap, 487, 645

\bibitem[{{Reyniers} \& {van Winckel}(2007)}]{Reyniers2007}
{Reyniers}, M. \& {van Winckel}, H. 2007, \aap, 463, L1

\bibitem[{{Rybicki} \& {Hummer}(1991)}]{Rybicki1991}
{Rybicki}, G.~B. \& {Hummer}, D.~G. 1991, \aap, 245, 171

\bibitem[{{Sch{\"o}ier} {et~al.}(2007){Sch{\"o}ier}, {Bast}, {Olofsson}, \&
  {Lindqvist}}]{Schoier2007}
{Sch{\"o}ier}, F.~L., {Bast}, J., {Olofsson}, H., \& {Lindqvist}, M. 2007,
  \aap, 473, 871

\bibitem[{{Sch{\"o}ier} {et~al.}(2011){Sch{\"o}ier}, {Maercker}, {Justtanont},
  {Olofsson}, {Black}, {Decin}, {de Koter}, \& {Waters}}]{Schoier2011}
{Sch{\"o}ier}, F.~L., {Maercker}, M., {Justtanont}, K., {et~al.} 2011, \aap,
  530, A83

\bibitem[{{Sch{\"o}ier} {et~al.}(2006){Sch{\"o}ier}, {Olofsson}, \&
  {Lundgren}}]{Schoier2006}
{Sch{\"o}ier}, F.~L., {Olofsson}, H., \& {Lundgren}, A.~A. 2006, \aap, 454, 247

\bibitem[{{Sch{\"o}ier} {et~al.}(2013){Sch{\"o}ier}, {Ramstedt}, {Olofsson},
  {Lindqvist}, {Bieging}, \& {Marvel}}]{Schoier2013}
{Sch{\"o}ier}, F.~L., {Ramstedt}, S., {Olofsson}, H., {et~al.} 2013, \aap, 550,
  A78

\bibitem[{{Smith}(2014)}]{Smith2014}
{Smith}, C.~L. 2014, PhD thesis, University of Manchester, Faculty of
  Engineering and Physical Sciences, School of Physics and Astronomy

\bibitem[{{Tsuji}(1973)}]{Tsuji1973}
{Tsuji}, T. 1973, \aap, 23, 411

\bibitem[{{Van de Sande} {et~al.}(2018){Van de Sande}, {Sundqvist}, {Millar},
  {Keller}, {Homan}, {de Koter}, {Decin}, \& {De Ceuster}}]{Van-de-Sande2018}
{Van de Sande}, M., {Sundqvist}, J.~O., {Millar}, T.~J., {et~al.} 2018, \aap,
  Forthcoming [\eprint[arXiv]{1803.01796}]

\bibitem[{{Vassilev} {et~al.}(2008){Vassilev}, {Meledin}, {Lapkin}, {Belitsky},
  {Nystr{\"o}m}, {Henke}, {Pavolotsky}, {Monje}, {Risacher}, {Olberg},
  {Strandberg}, {Sundin}, {Fredrixon}, {Ferm}, {Desmaris}, {Dochev},
  {Pantaleev}, {Bergman}, \& {Olofsson}}]{Vassilev2008}
{Vassilev}, V., {Meledin}, D., {Lapkin}, I., {et~al.} 2008, \aap, 490, 1157

\bibitem[{{Velilla Prieto} {et~al.}(2017){Velilla Prieto}, {S{\'a}nchez
  Contreras}, {Cernicharo}, {Ag{\'u}ndez}, {Quintana-Lacaci}, {Bujarrabal},
  {Alcolea}, {Balan{\c c}a}, {Herpin}, {Menten}, \&
  {Wyrowski}}]{Velilla-Prieto2017}
{Velilla Prieto}, L., {S{\'a}nchez Contreras}, C., {Cernicharo}, J., {et~al.}
  2017, \aap, 597, A25

\bibitem[{{Waelkens} {et~al.}(1991){Waelkens}, {Van Winckel}, {Bogaert}, \&
  {Trams}}]{Waelkens1991}
{Waelkens}, C., {Van Winckel}, H., {Bogaert}, E., \& {Trams}, N.~R. 1991, \aap,
  251, 495

\bibitem[{White {et~al.}(1958)White, Johnson, \& Dantzig}]{White1958}
White, W.~B., Johnson, S.~M., \& Dantzig, G.~B. 1958, The Journal of Chemical
  Physics, 28, 751

\bibitem[{{Whitelock} {et~al.}(2008){Whitelock}, {Feast}, \& {van
  Leeuwen}}]{Whitelock2008}
{Whitelock}, P.~A., {Feast}, M.~W., \& {van Leeuwen}, F. 2008, \mnras, 386, 313

\bibitem[{{Willacy} \& {Millar}(1997)}]{Willacy1997}
{Willacy}, K. \& {Millar}, T.~J. 1997, \aap, 324, 237

\bibitem[{{Woods} {et~al.}(2003){Woods}, {Sch{\"o}ier}, {Nyman}, \&
  {Olofsson}}]{Woods2003}
{Woods}, P.~M., {Sch{\"o}ier}, F.~L., {Nyman}, L.-{\AA}., \& {Olofsson}, H.
  2003, \aap, 402, 617

\bibitem[{{Yang} {et~al.}(2010){Yang}, {Stancil}, {Balakrishnan}, \&
  {Forrey}}]{Yang2010}
{Yang}, B., {Stancil}, P.~C., {Balakrishnan}, N., \& {Forrey}, R.~C. 2010,
  \apj, 718, 1062

\end{thebibliography}

%\Online
\clearpage
\appendix

\section{Observed lines and model plots}

\subsection{Observed lines}

The integrated intensities of the SiS observations from the APEX sulphur survey are listed in Table \ref{SiSobs} and the corresponding rms noise levels in mK are given in Table \ref{SiSnondet}. The integrated intensities and rms noise levels in mK for CS observations from the APEX sulphur survey are given in Table \ref{CSobs}.
The rms noise levels for the stars in the S star survey are listed in Table \ref{csiramobs} for the IRAM CS observations of ($5\to4$) and ($3\to2$), Table \ref{sisiramobs} for the IRAM SiS observations of ($5\to4$), ($6\to5$), ($12\to11$), and ($13\to12$), and Table \ref{sisapexobs} for the APEX SiS ($19\to18$) observations. For the stars with detected lines, integrated intensities and peak temperatures are also included.
Observations of the SiS ($4\to3$) line at 72.618 GHz, carried out by the 4mm receiver at Onsala Space Observatory (OSO), are listed in Table \ref{osoobs}. Previously published supplementary observations are listed in Table \ref{supobs}.

\begin{table*}[tp]
\caption{Main beam integrated intensities for SiS detections from the APEX survey.}\label{SiSobs}
\centering
\begin{tabular}{cccccc|c|c}
\hline\hline
& \multicolumn{5}{c|}{SiS} & \multicolumn{1}{c|}{\up{29}SiS} & Si\up{34}S\\
Star    & $9\to8$       &       $12\to11$       &       $14\to13$       &       $16\to15$       &       $19\to18$       & $10\to9$ & $11\to10 $\\
& [K $\kms$]& [K $\kms$]& [K $\kms$]& [K $\kms$]& [K $\kms$]&  [K $\kms$]& [K $\kms$]\\
\hline
\multicolumn{2}{c}{\it Carbon stars}&&&&&\\
R Lep   &       x       &       x       &       ...     &       x       &       ...     &       ...     &       x       \\
V1259 Ori       &       1.61    &       1.82    &       ...     &       2.03    &       1.78    &       x       &       x       \\
AI Vol  &       2.54    &       2.95    &       4.61    &       5.00    &       ...     &       x       &       \phantom{$^T$}0.61$^{T}$        \\
X TrA   &       ...     &       x       &       ...     &       x       &       ...     &       ...     &       x       \\
II Lup  &       ...     &       5.22    &       ...     &       7.78    &       ...     &       ...     &       x       \\
V821 Her        &       1.66    &       2.39    &       ...     &       3.64    &       3.27    &       x       &       x       \\
U Hya   &       ...     &       ...     &       ...     &       x       &       x       &       ...     &       x       \\
RV Aqr  &       x       &       0.69    &       ...     &       1.09    &       ...     &       ...     &       x       \\
\multicolumn{2}{c}{\it M-type stars}&&&&&\\
R Hor   &       x       &       ...     &       x       &       ...     &       ...     &       ...     &       ...     \\
IK Tau  &       ...     &       3.41    &       4.33    &       ...     &       5.61    &       x       &       x       \\
GX Mon  &       \phantom{$^T$}0.22$^T$  &       1.03    &       0.85    &       ...     &       1.02    &       x       &       x       \\
W Hya   &       x       &       ...     &       x       &       ...     &       ...     &       x       &       x       \\
RR Aql  &       x       &       x       &       x       &       ...     &       x       &       x       &       x       \\
V1943 Sgr       &       x       &       ...     &       x       &       ...     &       ...     &       x       &       x       \\
V1300 Aql       &       1.18    &       1.90    &       2.23    &       ...     &       2.33    &       x       &       x       \\
\multicolumn{2}{c}{\it S-type stars}&&&&&\\
T Cet   &       x       &       x       &       ...     &       x       &       ...     &       ...     &       x       \\
TT Cen  &       x       &       ...     &       x       &       x       &       x       &       x       &       ...     \\
RT Sco  &       x       &       ...     &       ...     &       x       &       ...     &       ...     &       x       \\
W Aql   &       ...     &       0.95    &       ...     &       ...     &       ...     &       x       &       ...     \\
RZ Sgr  &       x       &       ...     &       x       &       x       &       x       &       x       &       x       \\
\hline
\end{tabular}
\tablefoot{($^T$) indicates a tentative detection; x indicates a non-detection (see Table \ref{SiSnondet} for rms); (...) indicates lines which were not observed.}
\end{table*}

\begin{table*}[tp]
\caption{Rms noise in mK for a velocity resolution of $1~\kms$ for SiS observations from the APEX survey.}\label{SiSnondet}
\centering
\begin{tabular}{cccccc|c|c}
\hline\hline
& \multicolumn{5}{c|}{SiS} & \multicolumn{1}{c|}{\up{29}SiS} & Si\up{34}S\\
Star    & $9\to8$       &       $12\to11$       &       $14\to13$       &       $16\to15$       &       $19\to18$       & $10\to9$ & $11\to10 $\\
\hline
\multicolumn{2}{c}{\it Carbon stars}&&&&&\\
R Lep   &       9.6     &       16      &       ...     &       19      &       ...     &...&   8.9     \\
V1259 Ori       &       19$^{D}$        &       14$^D$  &       ...     &       13$^D$  &       18$^D$  & 31      & 15\\
AI Vol  &       21$^{D}$        &       16$^D$  &       20$^D$  &       14$^D$  &       ...             &39& 22$^T$                                  \\
X TrA   &       ...     &       12      &       ...     &       14      &       ...     &...&   14      \\
II Lup  &       ...     &       9.9$^D$ &       ...     &       13$^D$  &       ...     &...&   20\\
V821 Her        &       21$^{D}$        &       17$^D$  &       ...     &       18$^D$  &       19$^D$&29&16\\
U Hya   &       ...     &       ...     &       ...     &       12      &       13      &...&           9.0             \\
RV Aqr  &       19      &       11$^D$  &       ...     &       19$^D$  &       ...     &...&   14      \\
\multicolumn{2}{c}{\it M-type stars}&&&&&\\
R Hor   &       12      &       ...     &       16      &       ...     &       ...&19&13\\
IK Tau  &       ...     &       15$^D$  &       18$^D$  &       ...     &       18$^D$ & 26&9.0\\
GX Mon  &       18$^{T}$        &       11$^D$  &       20$^D$  &       ...     &       21$^D$&43&21\\
W Hya   &       16      &       ...     &       18      &       ...     &       ... &40&20\\
RR Aql  &       21      &       13      &       19      &       ...     &       20&21&11\\
V1943 Sgr       &       16      &       ...     &       13      &       ...     &       ...&15&13\\
V1300 Aql       &       14$^{D}$        &       14$^D$  &       19$^D$  &       ...     &20$^D$ &20 &13\\
\multicolumn{2}{c}{\it S-type stars}&&&&&\\
T Cet   &       15      &       13      &       ...     &       11      &       ... &...&11\\
TT Cen  &       14      &       ...     &       16      &       13      &       17 &50&...\\
RT Sco  &       25      &       ...     &       ...     &       11      &       ... &... &23\\
W Aql   &       ...     &       14$^D$  &       ...     &       ...     &       ... & 15&...\\
RZ Sgr  &       13      &       ...     &       19      &       18      &       20 &23&14\\
\hline
\end{tabular}
\tablefoot{rms values given in mK at a velocity resolution of $1~\kms$. (...) indicates lines which were not observed, ($^D$) and ($^T$) indicate detected or tentatively detected lines, respectively (see Table \ref{SiSobs} for integrated intensities).}
\end{table*}

\begin{table*}[tp]
\caption{CS detections (\textit{left}) and corresponding rms noise (\textit{right}) from the APEX survey.}\label{CSobs}
\centering
\begin{tabular}{c|ccc|c||ccc|c}
\hline\hline
 & \multicolumn{4}{c||}{Detections} & \multicolumn{4}{c}{rms noise levels}\\
 \hline
 & \multicolumn{3}{c|}{CS} & C\up{33}S& \multicolumn{3}{c|}{CS} & C\up{33}S\\
Star    &       $4\to3$ &       $6\to5$ &       $7\to6$ &       $6\to5$ &       $4\to3$ &       $6\to5$ &       $7\to6$ &       $6\to5$ \\
& [K $\kms$]& [K $\kms$]& [K $\kms$]& [K $\kms$] & [mK] &[mK]&[mK]&[mK]\\
\hline
{\it Carbon stars}&&&&&&&\\
R Lep   &       2.30    &       4.14    &       ...     &       x       &       8.3$^D$ &       18$^D$  &       ...     &       18      \\
V1259 Ori       &       4.15    &       5.20    &       4.28    &       x       &       15$^D$  &       11$^D$  &       19$^D$  &       11      \\
AI Vol  &       8.42    &       12.84   &       ...     &       x       &       20$^D$  &       15$^D$  &       ...     &       15      \\
X TrA   &       0.81    &       1.58    &       ...     &       x       &       14$^D$  &       14$^D$  &       ...     &       14      \\
II Lup  &       21.03   &       29.65   &       ...     &       0.78    &       19$^D$  &       12$^D$  &       ...     &       12$^T$  \\
V821 Her        &       6.01    &       10.83   &       9.58    &       x       &       15$^D$  &       18$^D$  &       21$^D$  &       18      \\
U Hya   &       0.17    &       0.19    &       0.42    &       x       &       8.8$^T$ &       12$^T$& 13$^D$  &       12      \\
RV Aqr  &       4.61    &       6.70    &       ...     &       x       &       14$^D$  & 18$^D$  &               ...     &       18      \\
{\it M-type stars}&&&&&&&\\
R Hor   &       x       &       ...     &       ...     &       ...     &       12      &       ...     &       ...     &       ...     \\
IK Tau  &       1.46    &       ...     &       2.57    &       ...     &       8.8$^D$ &       ...     &       29$^D$         &       ...     \\
GX Mon  &       0.94    &       ...     &       \phantom{$^T$}0.71$^T$  &       ...     &       20$^D$  &       ...     &       20$^T$  &       ...     \\
W Hya   &       x       &       ...     &       ...     &       ...     &       19      &       ...     &       ...     &       ...     \\
RR Aql  &       x       &       ...     &       x       &       ...     &       11      &       ...     &       20      &       ...     \\
V1943 Sgr       &       x       &       ...     &       ...     &       ...     &       12      &       ...     &       ...     &       ...     \\
V1300 Aql       &       \phantom{$^T$}0.31$^T$  &       ...     &       \phantom{$^T$}0.48$^T$  &       ...     &       13$^T$  &       ...     &       19$^T$  &       ...     \\
{\it S-type stars}&&&&&&&\\
T Cet   &       x       &       x       &       ...     &       x       &       11      &       10      &       ...     &       10      \\
TT Cen  &       ...     &       x       &       x       &       x       &       ...     &       13      &       16      &       13      \\
RT Sco  &       x       &       x       &       ...     &       x       &       21      &       9.8     &       ...     &       9.8     \\
RZ Sgr  &       x       &       x       &       x       &       x       &       13      &       17      &       22      &       17      \\
\hline
\end{tabular}
\tablefoot{\textit{Left:} x indicates a non-detection; (...) indicates lines which were not observed. \textit{Right:} rms values given in mK at a velocity resolution of $1~\kms$. (...) indicates lines which were not observed, ($^D$) and ($^T$) indicate detected or tentatively detected lines, respectively.}
\end{table*}

\begin{table}[tp]
\caption{IRAM 30m observations of CS ($5\to4$) and ($3\to2$) towards S type stars.}\label{csiramobs}
\centering
\begin{tabular}{ccccc}
\hline\hline
Source  &       Transition      &       rms     &       $I_\mathrm{mb}$ &       $T_\mathrm{peak}$       \\
        &       (CS)    &       [mK, $T_A^*$]   &       [K~$\kms$]      &       [K, $T_A^*$]        \\
        \hline
$\chi$ Cyg      &       ($5\to4$)       &       36.836  &       {4.93}  &       0.23         \\
        &       ($3\to2$)       &       32.389  &       {2.55}  &       0.15         \\
AA Cyg  &       ($5\to4$)       &       11.639  &       ...     &       ...     \\
        &       ($3\to2$)       &       9.690   &       ...     &       ...     \\
AD Cyg  &       ($5\to4$)       &       10.588  &       ...     &       ...     \\
        &       ($3\to2$)       &       9.795   &       ...     &       ...     \\
RX Lac  &       ($5\to4$)       &       16.155  &       ...     &       ...     \\
        &       ($3\to2$)       &       15.322  &       ...     &       ...     \\
WY Cas  &       ($5\to4$)       &       14.398  &       ...     &       ...     \\
        &       ($3\to2$)       &       11.886  &       ...     &       ...     \\
R And   &       ($5\to4$)       &       13.773  &       1.4     &       ...     \\
        &       ($3\to2$)       &       11.079  &       0.74    &       ...     \\
W And   &       ($5\to4$)       &       17.882  &       ...     &       ...     \\
        &       ($3\to2$)       &       13.771  &       ...     &       ...     \\
TV Dra  &       ($5\to4$)       &       16.410  &       ...     &       ...     \\
        &       ($3\to2$)       &       12.780  &       ...     &       ...     \\
IRC-10401       &       ($5\to4$)       &       31.302  &       ...     &       ...     \\
        &       ($3\to2$)       &       20.778  &       ...     &       ...     \\
ST Sgr  &       ($5\to4$)       &       19.392  &       ...     &       ...     \\
        &       ($3\to2$)       &       13.862  &       ...     &       ...     \\
W Aql   &       ($5\to4$)       &       31.641  &       9.0     &       0.20         \\
        &       ($3\to2$)       &       23.270  &       6.9     &       0.20         \\
EP Vul  &       ($5\to4$)       &       9.545   &       0.25    &       ...     \\
        &       ($3\to2$)       &       9.205   &       ...     &       ...     \\
CSS2 41 &       ($5\to4$)       &       23.547  &       ...     &       ...     \\
        &       ($3\to2$)       &       16.491  &       ...     &       ...     \\
RZ Peg  &       ($5\to4$)       &       10.691  &       ...     &       ...     \\
        &       ($3\to2$)       &       9.572   &       ...     &       ...     \\
V365 Cas        &       ($5\to4$)       &       19.618  &       ...     &       ...     \\
        &       ($3\to2$)       &       14.244  &       ...     &       ...     \\
S Cas   &       ($5\to4$)       &       20.352  &       6.9     &       0.15         \\
        &       ($3\to2$)       &       14.621  &       2.9     &       0.07         \\
T Cam   &       ($5\to4$)       &       13.347  &       ...     &       ...     \\
        &       ($3\to2$)       &       11.082  &       ...     &       ...     \\
S Lyr   &       ($5\to4$)       &       15.05   &       1.1     &       0.045         \\
        &       ($3\to2$)       &       14.063  &       ...     &       ...     \\
ST Her  &       ($5\to4$)       &       16.807  &       ...     &       ...     \\
        &       ($3\to2$)       &       13.845  &       ...     &       ...     \\
R Cyg   &       ($5\to4$)       &       23.594  &       ...     &       ...     \\
        &       ($3\to2$)       &       19.983  &       ...     &       ...     \\
DK Vul  &       ($5\to4$)       &       19.357  &       ...     &       ...     \\
        &       ($3\to2$)       &       13.983  &       ...     &       ...     \\
V386 Cep        &       ($5\to4$)       &       18.576  &       ...     &       ...     \\
        &       ($3\to2$)       &       14.628  &       ...     &       ...     \\
AA Cam  &       ($5\to4$)       &       14.88   &       ...     &       ...     \\
        &       ($3\to2$)       &       14.28   &       ...     &       ...     \\
R Lyn   &       ($5\to4$)       &       7.522 $^a$      &       ...     &       ...     \\
        &       ($3\to2$)       &       6.538 $^b$      &       ...     &       ...     \\
Y Lyn   &       ($5\to4$)       &       8.362 $^a$      &       ...     &       ...     \\
        &       ($3\to2$)       &       6.571 $^b$      &       ...     &       ...     \\
\hline
\end{tabular}
\tablefoot{Rms is given in units of antenna temperature at 1~$\kms$ except where specified. (\up{a}) indicates rms given at 1.2~$\kms$ and (\up{b}) indicates rms given at 2~$\kms$. Peak flux, $T_\mathrm{peak}$, is also given in units of antenna temperature, while integrated intensity, $I_\mathrm{mb}$, is given in units of main beam temperature.}
\end{table}

\begin{table}[tp]
\caption{APEX observations of SiS ($19\to18$) towards S type stars.}\label{sisapexobs}
\centering
\begin{tabular}{ccccc}
\hline\hline
Source  &       Transition      &       rms     &       $I_\mathrm{mb}$ &       $T_\mathrm{peak}$       \\
        &       (SiS)   &       [mK, $T_A^*$]   &       [K~$\kms$]      &       [K, $T_A^*$]        \\
        \hline
AFGL 2425       &       ($19\to18$)     &       8.266   &       ...     &       ...     \\
DY Gem  &       ($19\to18$)     &       5.878   &       ...     &       ...     \\
EP Vul  &       ($19\to18$)     &       11.777  &       ...     &       ...     \\
GI Lup  &       ($19\to18$)     &       7.755   &       ...     &       ...     \\
IRC-10401       &       ($19\to18$)     &       6.579   &       0.11    &       0.012         \\
R Gem   &       ($19\to18$)     &       7.280   &       ...     &       ...     \\
RT Sco  &       ($19\to18$)     &       9.301   &       ...     &       ...     \\
RZ Sgr  &       ($19\to18$)     &       8.283   &       0.10    &       0.010         \\
S Lyr   &       ($19\to18$)     &       9.771   &       ...     &       ...     \\
ST Sco  &       ($19\to18$)     &       6.227   &       ...     &       ...     \\
TT Cen  &       ($19\to18$)     &       7.868   &       ...     &       ...     \\
W Aql   &       ($19\to18$)     &       23.260  &       1.8 &   0.08    \\
\hline
\end{tabular}
\tablefoot{Rms is given in units of antenna temperature at 1~$\kms$. Peak flux, $T_\mathrm{peak}$, is also given in units of antenna temperature, while integrated intensity, $I_\mathrm{mb}$, is given in units of main beam temperature.}
\end{table}

\begin{table}[tp]
\caption{IRAM observations of SiS ($5\to4$), ($6\to5$), ($12\to11$), and ($13\to12$) towards S type stars.}\label{sisiramobs}
\centering
\begin{tabular}{ccccc}
\hline\hline
Source  &       Transition      &       rms     &       $I_\mathrm{mb}$ &       $T_\mathrm{peak}$       \\
        &       (SiS)   &       [mK, $T_A^*$]   &       [K~$\kms$]      &       [K, $T_A^*$]        \\
        \hline
DY Gem  &       ($5\to4$)       &       5.973   &       ...     &       ...     \\
        &       ($6\to5$)       &       7.179   &       ...     &       ...     \\
        &       ($12\to11$)     &       13.759  &       ...     &       ...     \\
        &       ($13\to12$)     &       16.150  &       ...     &       ...     \\
R And   &       ($5\to4$)       &       5.080   &       ...     &       ...     \\
        &       ($6\to5$)       &       6.476   &       ...     &       ...     \\
        &       ($12\to11$)     &       9.333   &       ...     &       ...     \\
        &       ($13\to12$)     &       14.126  &       ...     &       ...     \\
R Cyg   &       ($5\to4$)       &       4.772   &       ...     &       ...     \\
        &       ($6\to5$)       &       6.023   &       ...     &       ...     \\
        &       ($12\to11$)     &       8.341   &       ...     &       ...     \\
        &       ($13\to12$)     &       12.810  &       ...     &       ...     \\
S Cas   &       ($5\to4$)       &       5.436   &       ...     &       ...     \\
        &       ($6\to5$)       &       6.812   &       ...     &       ...     \\
        &       ($12\to11$)     &       11.203  &       0.54    &       0.013         \\
        &       ($13\to12$)     &       16.152  &       1.1     &       0.026         \\
W Aql   &       ($5\to4$)       &       5.999   &       0.55    &       0.017         \\
        &       ($6\to5$)       &       7.953   &       0.69    &       0.021         \\
        &       ($12\to11$)     &       12.022  &       2.9     &       0.085         \\
        &       ($13\to12$)     &       18.523  &       3.8     &       0.11         \\
WY Cas  &       ($5\to4$)       &       5.304   &       ...     &       ...     \\
        &       ($6\to5$)       &       6.528   &       ...     &       ...     \\
        &       ($12\to11$)     &       8.277   &       ...     &       ...     \\
        &       ($13\to12$)     &       12.678  &       ...     &       ...     \\
$\chi$ Cyg      &       ($5\to4$)       &       5.471   &       ...     &       ...     \\
        &       ($6\to5$)       &       6.618   &       ...     &       ...     \\
        &       ($12\to11$)     &       9.220   &       0.73    &       0.036         \\
        &       ($13\to12$)     &       13.837  &       1.2     &       0.030         \\
\hline
\end{tabular}
\tablefoot{Rms is given in units of antenna temperature at 1~$\kms$. Peak flux, $T_\mathrm{peak}$, is also given in units of antenna temperature, while integrated intensity, $I_\mathrm{mb}$, is given in units of main beam temperature.}
\end{table}

\begin{table}[tp]
\caption{OSO observations of SiS ($4\to3$) at 72.618 GHz}\label{osoobs}
\centering
\begin{tabular}{ccc}
\hline\hline
Star    &       $\int T_A^* \mathrm{d}\upsilon$ & rms at $1\kms$        \\
  & [K $\kms$] & [mK]\\
\hline
IK Tau & x & 11\\
TX Cam & x & 7.2\\
NV Aur & x & 13\\
BX Cam & 0.13$^T$ & 7.0\\
GX Mon & x & 17 \\
V1111 Oph & x & 10\\
T Cep & x & 12\\
R Cas & x & 8.1\\
\hline
\end{tabular}
\tablefoot{($^T$) indicates a tentative detection, x indicates a non-detection.}
\end{table}

\begin{table*}[tp]
\caption{Supplementary observations.}\label{supobs}
\begin{center}
\begin{tabular}{cccccc}
\hline\hline
Star    &       Transition      &       Telescope       &       $I_\mathrm{mb}$ & $\theta$&       Ref.    \\
 &  & & [K $\kms$] & [\arcsec]\\
\hline
\multicolumn{2}{c}{\it Carbon stars}\\
V1259 Ori & SiS ($6\to5$) & IRAM & 1.40 & 21 & 1\\
AI Vol & SiS ($10\to9$) & APEX & 2.7 & 34 & 2\\
& SiS ($11\to10$) & APEX & 3.0 & 31 & 2\\
& SiS ($19\to18$) & APEX & 7.7 & 18 & 3\\
& Si\up{34}S ($10\to9$) & APEX & 0.76 & 35 & 2\\
II Lup &CS ($7\to6$)& APEX & 20.3 & 18 &3\\
        &SiS ($19\to18$)& APEX & 3.5 & 18 &3\\
V821 Her & SiS ($6\to5$) & IRAM & 1.33 & 21 & 1\\
RV Aqr & SiS ($6\to5$) & IRAM & 0.39 & 21& 1\\
\multicolumn{2}{c}{\it M-type stars}\\
IK Tau & SiS ($5\to4$) & OSO & 0.32 & 42 & 5\\
        & SiS ($6\to5$) & IRAM & 0.52 & 21 & 1\\
        & SiS ($8\to 7$) & IRAM & 7.5 & 17 & 4\\
GX Mon & SiS ($6\to5$) & IRAM & 1.04 & 21 & 1\\
V1300 Aql & SiS ($6\to5$) & IRAM & 1.53 & 21 & 1\\
\hline
\end{tabular}
\end{center}
\tablefoot{References: (1) \cite{Danilovich2015a}, (2) De Beck et al (in prep), (3) APEX archive, (4) \cite{Decin2010}, (5) \cite{Schoier2007}}
\end{table*}

\subsection{Plots}

\begin{figure}[t]
\includegraphics[width=0.49\textwidth]{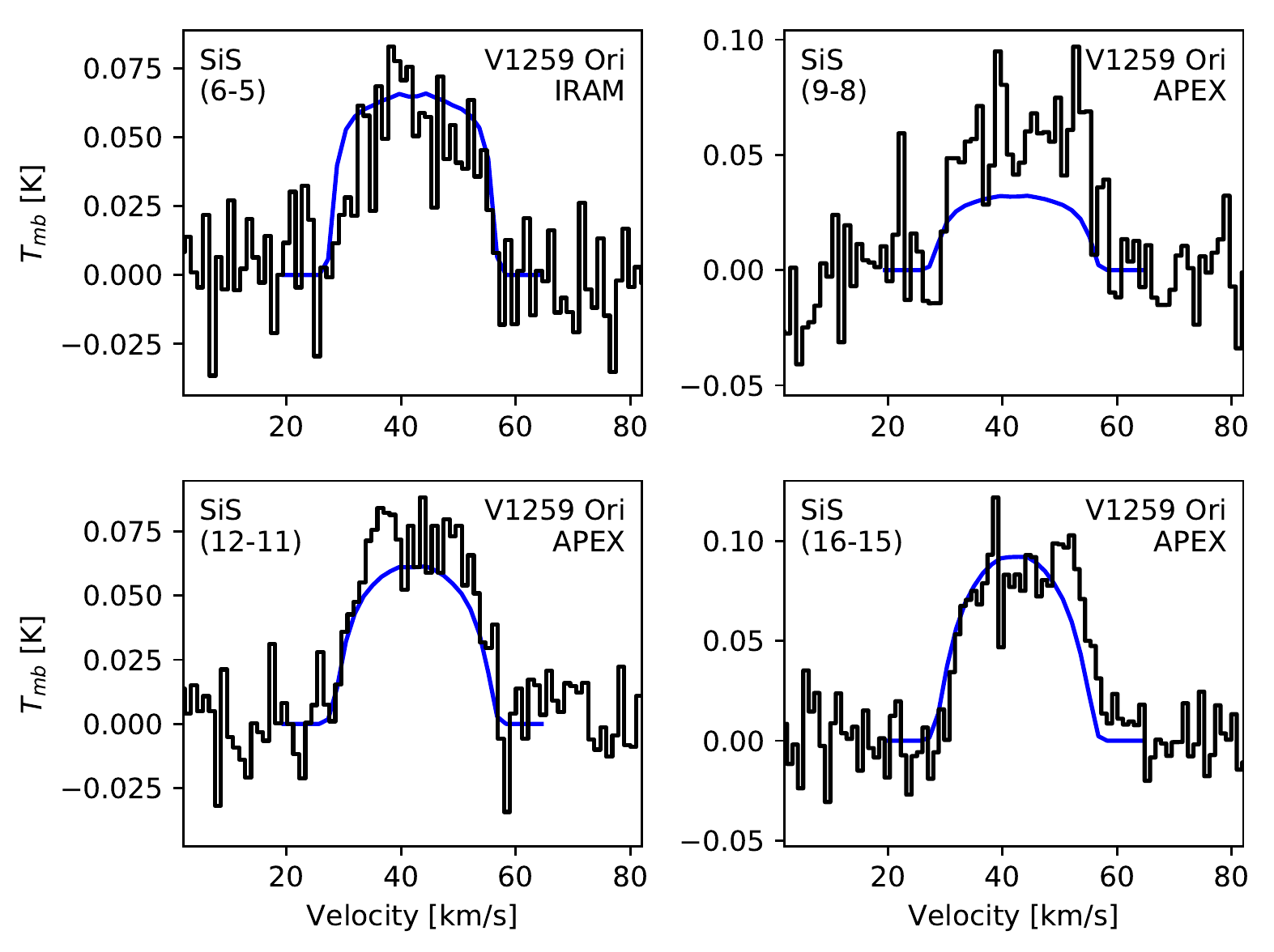}
\includegraphics[width=0.49\textwidth]{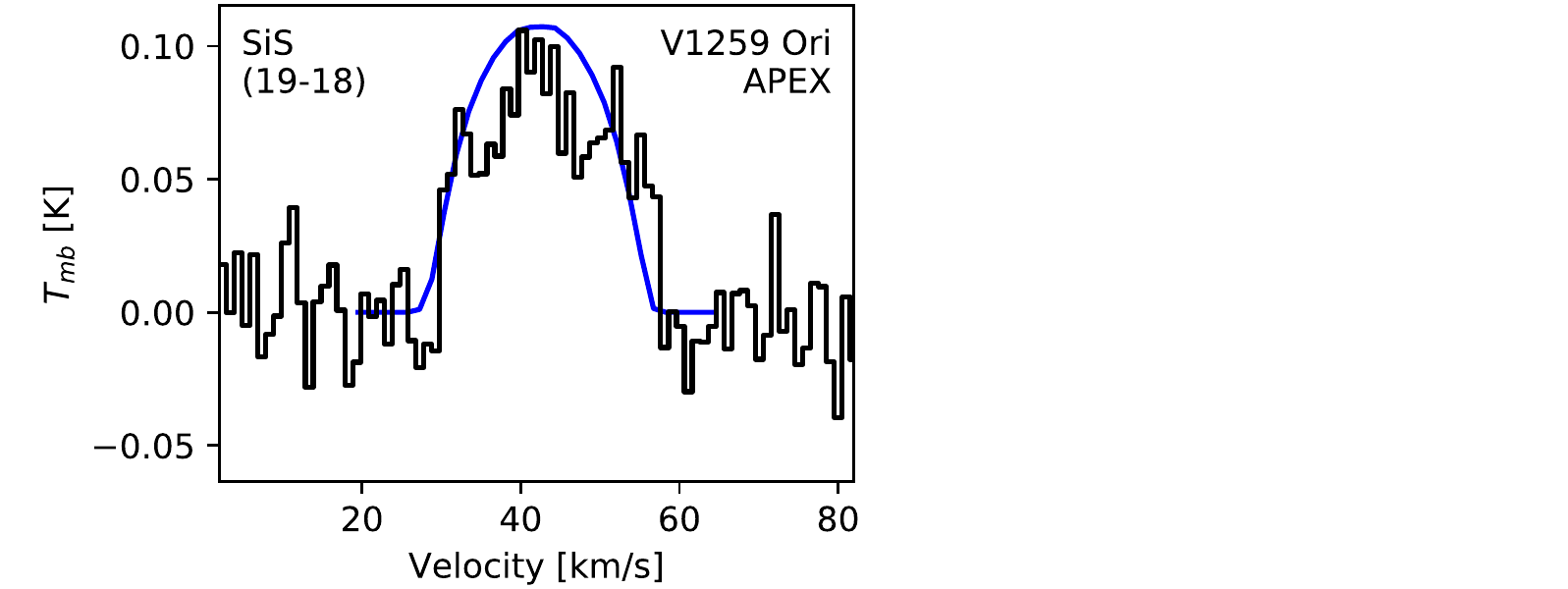}
\caption{Observations (black histograms) and model results (blue lines) for SiS towards the carbon star V1259~Ori, plotted with respect to LSR velocity.}
\label{SiSCplots-1}
\end{figure}

\begin{figure}[t]
\includegraphics[width=0.49\textwidth]{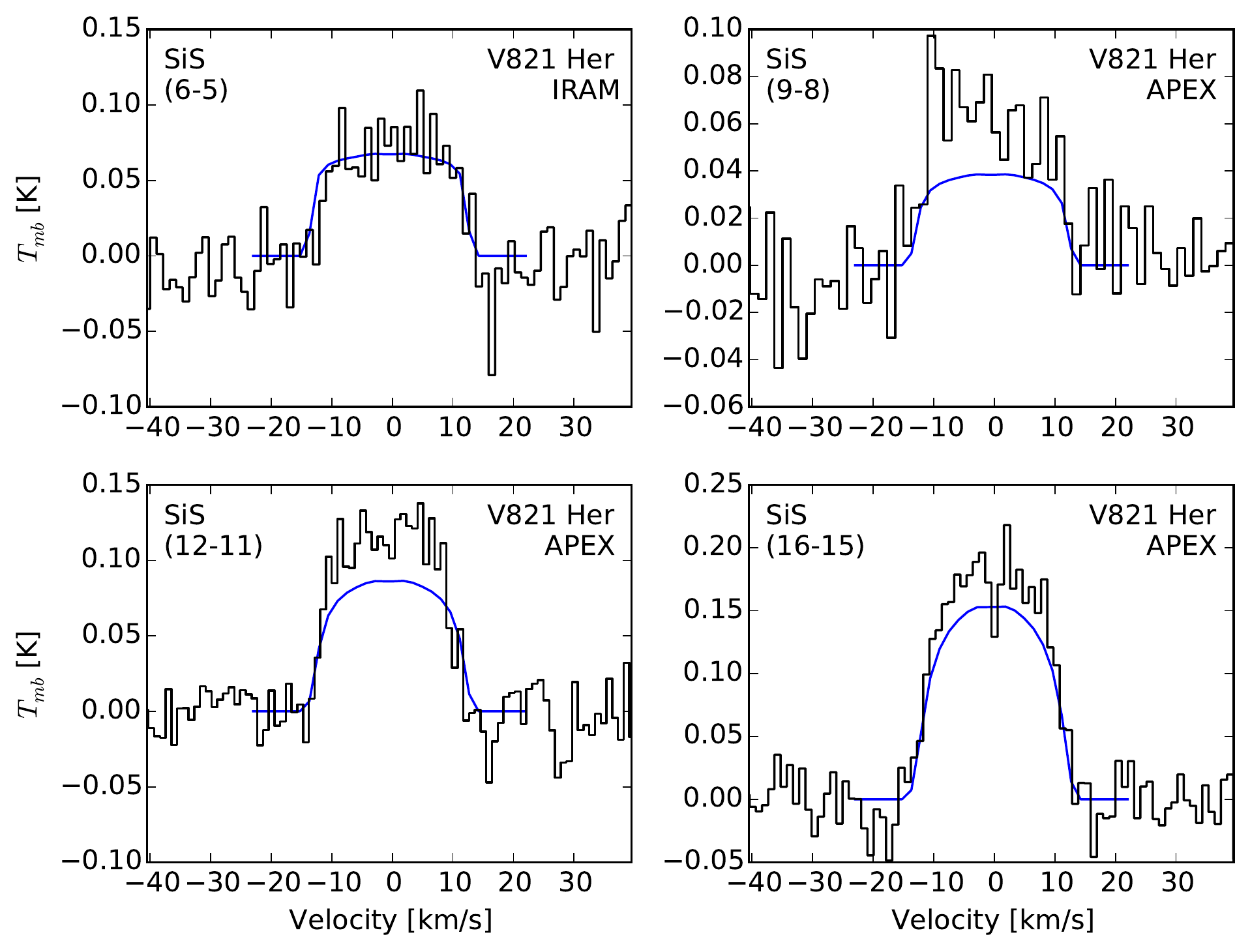}
\includegraphics[width=0.49\textwidth]{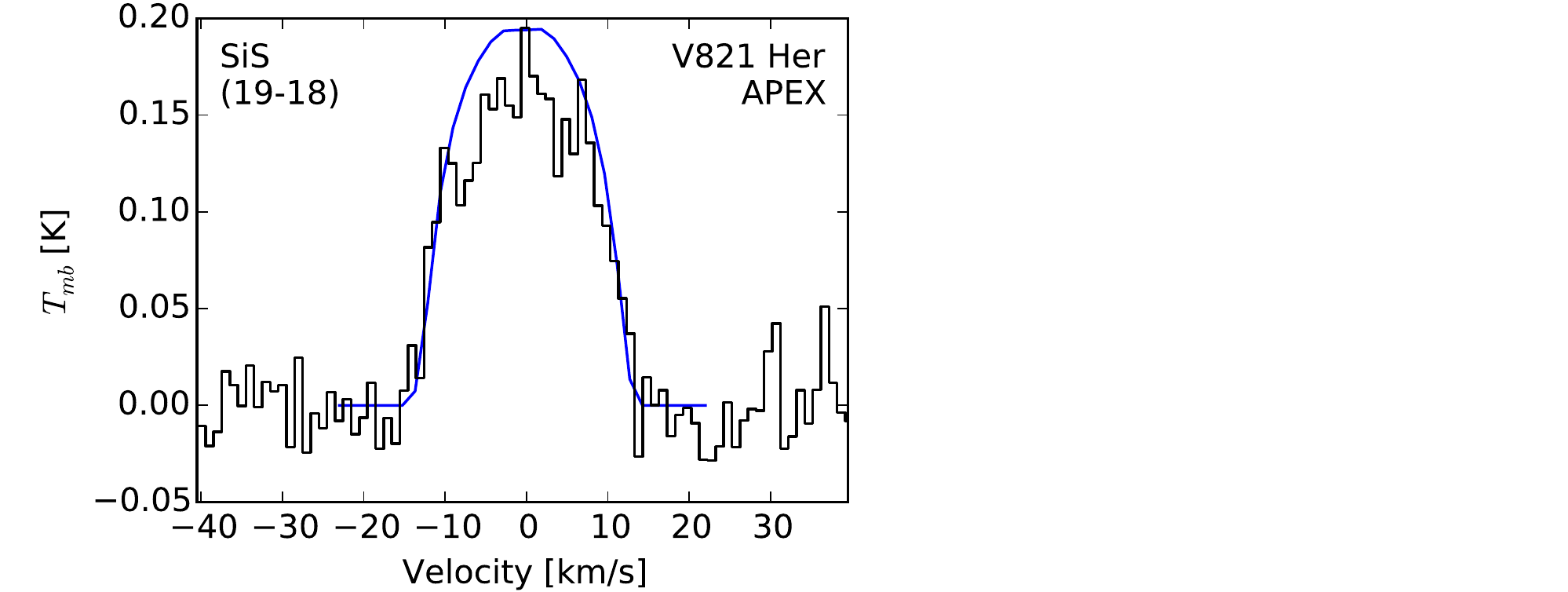}
\caption{Observations (black histograms) and model results (blue lines) for SiS towards the carbon star V821~Her, plotted with respect to LSR velocity.}
\label{SiSCplots-2}
\end{figure}

\begin{figure}[t]
\includegraphics[width=0.49\textwidth]{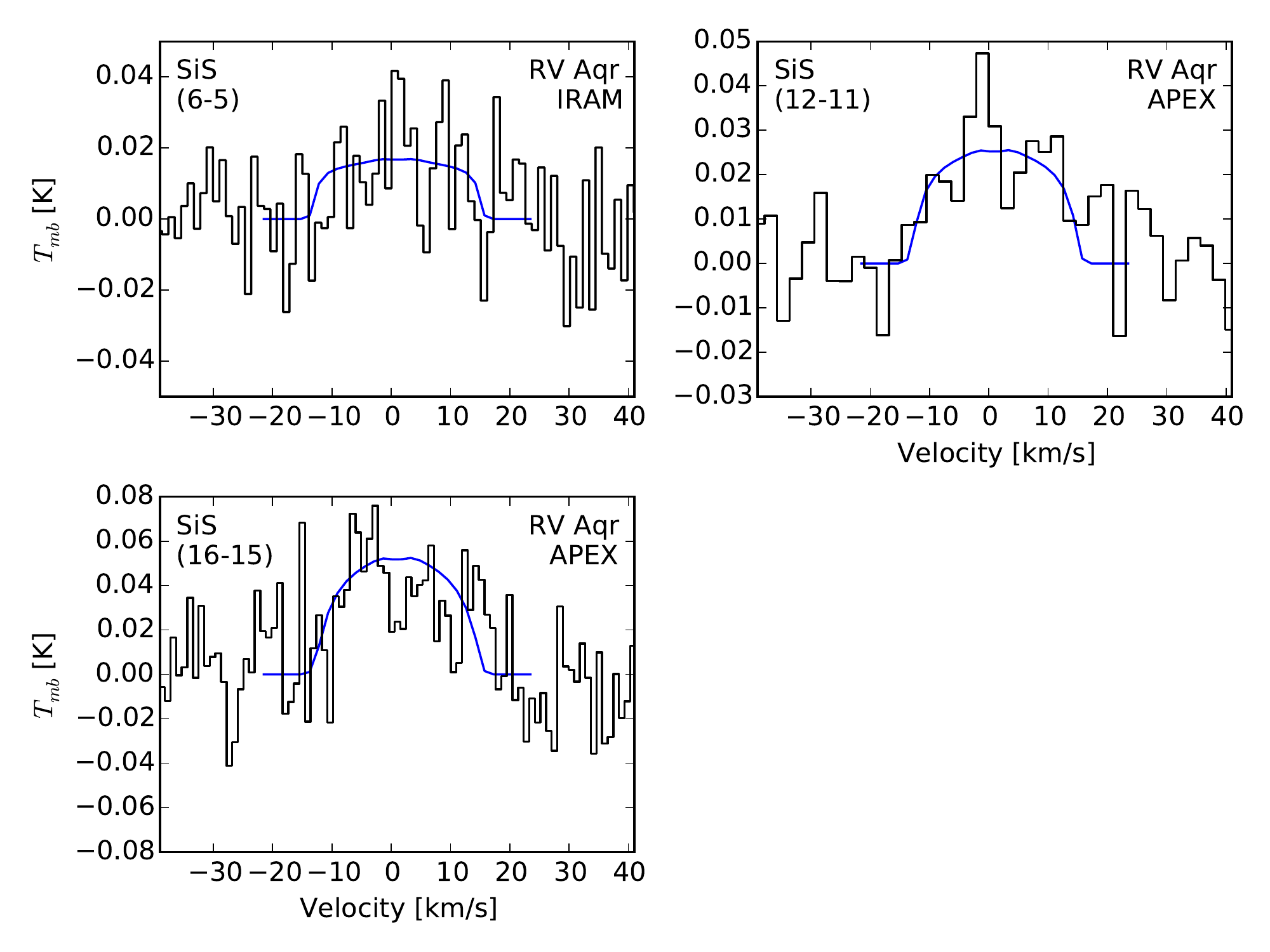}
\caption{Observations (black histograms) and model results (blue lines) for SiS towards the carbon star RV Aqr, plotted with respect to LSR velocity.}
\label{SiSCplots-3}
\end{figure}

\begin{figure}[t]
\includegraphics[width=0.49\textwidth]{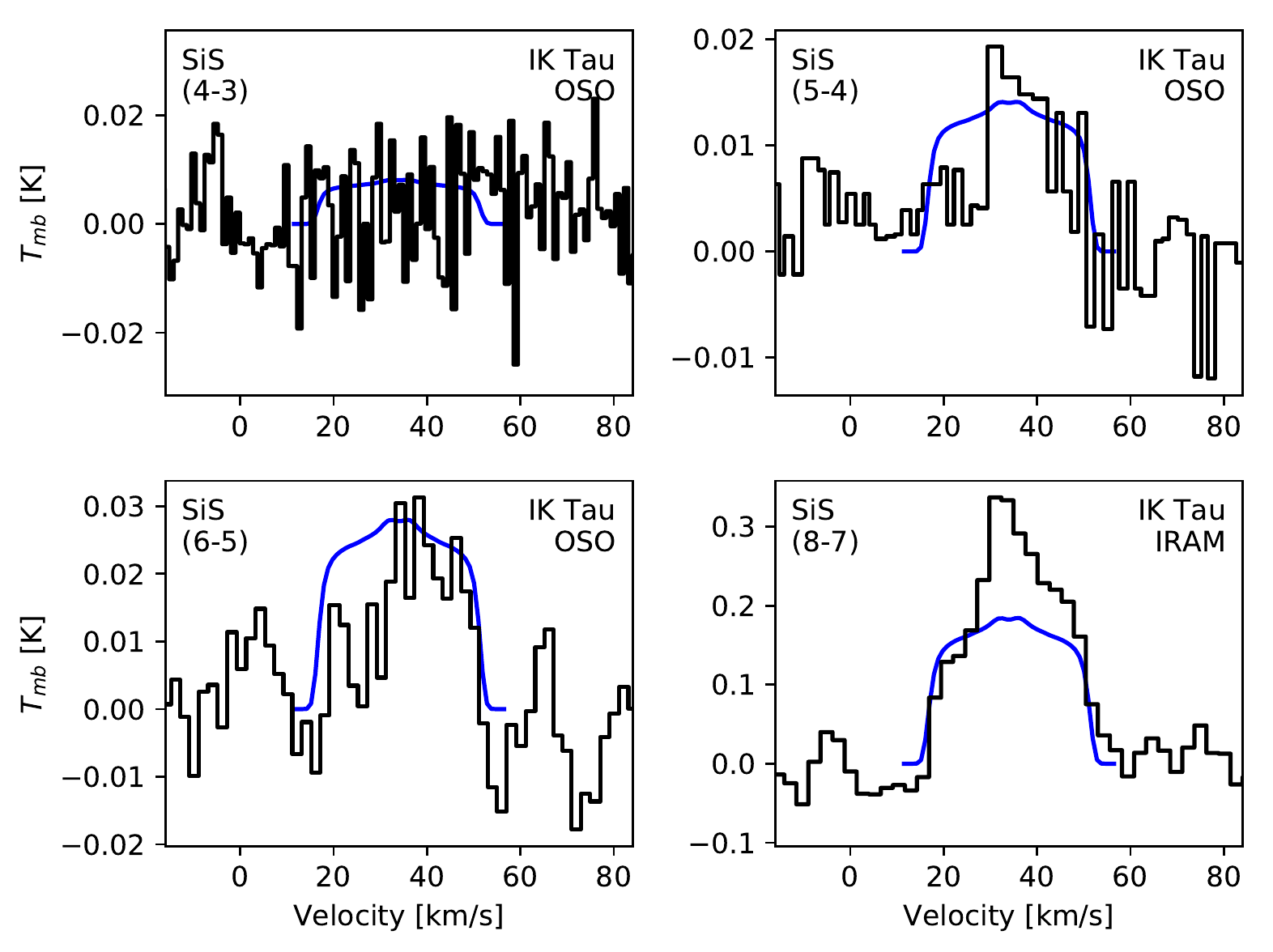}
\includegraphics[width=0.49\textwidth]{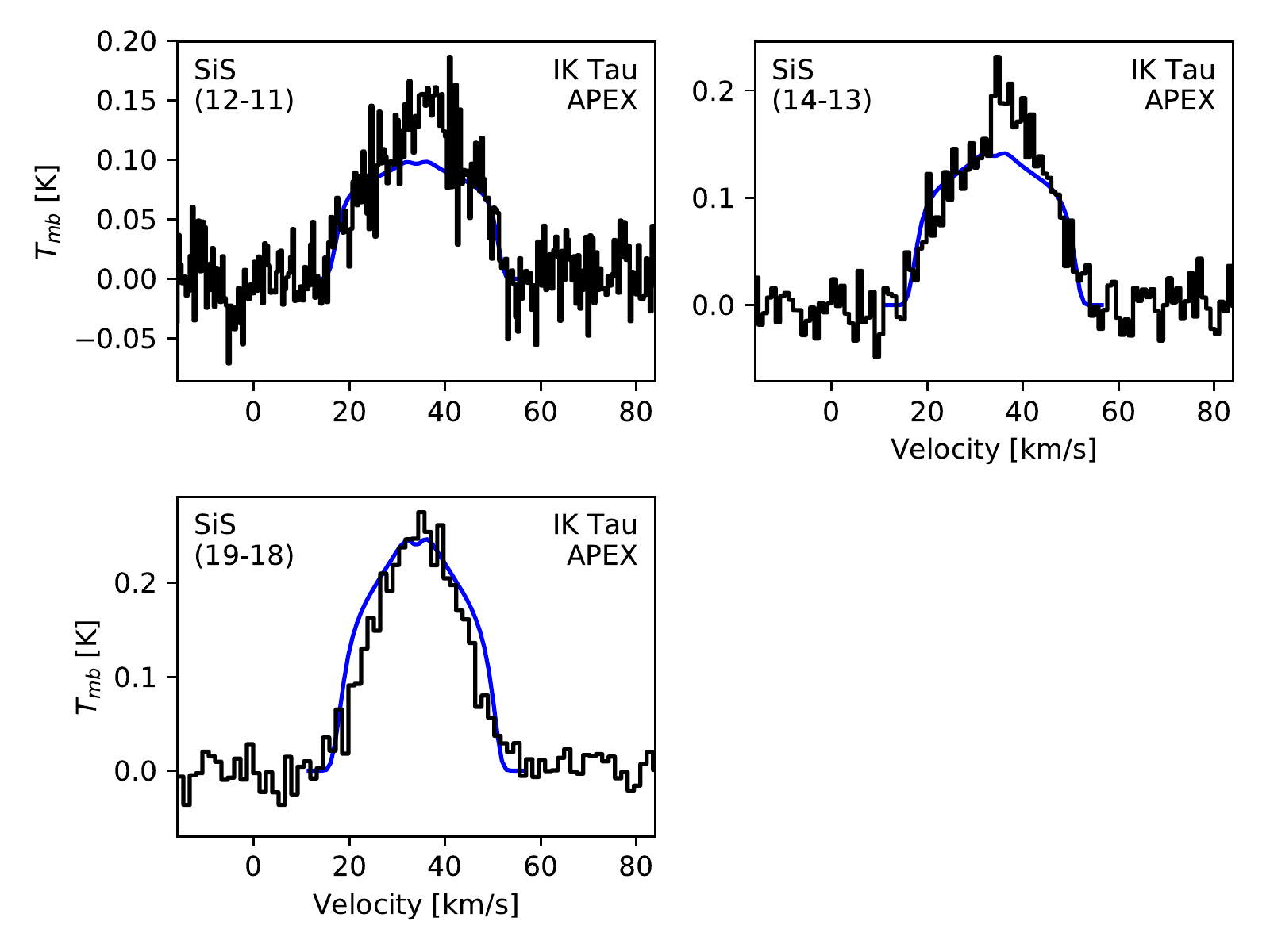}
\caption{Observations (black histograms) and model results (blue lines) for SiS towards the oxygen-rich star IK~Tau, plotted with respect to LSR velocity.}
\label{SiSMplots-1}
\end{figure}

\begin{figure}
\includegraphics[width=0.49\textwidth]{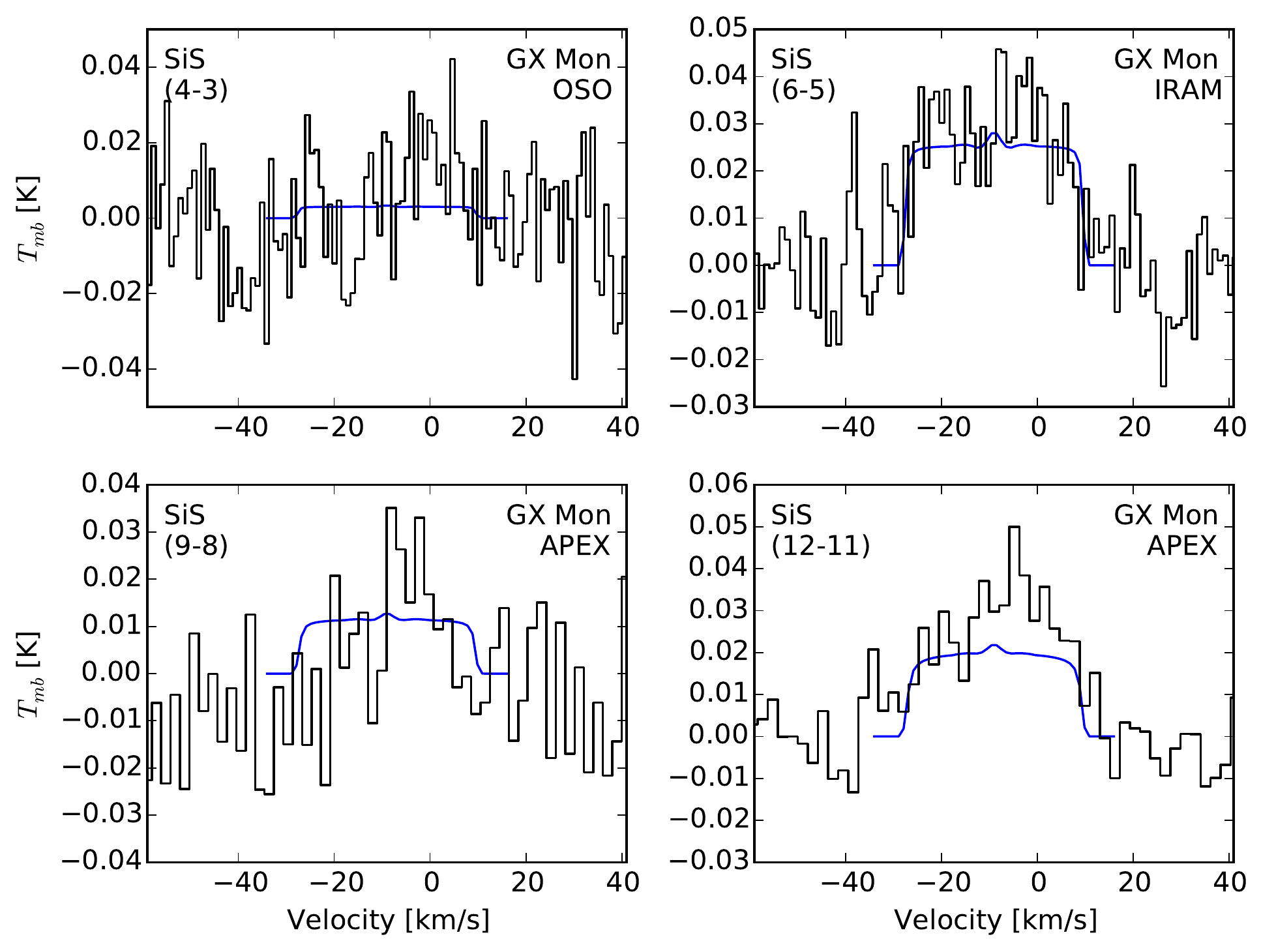}
\includegraphics[width=0.49\textwidth]{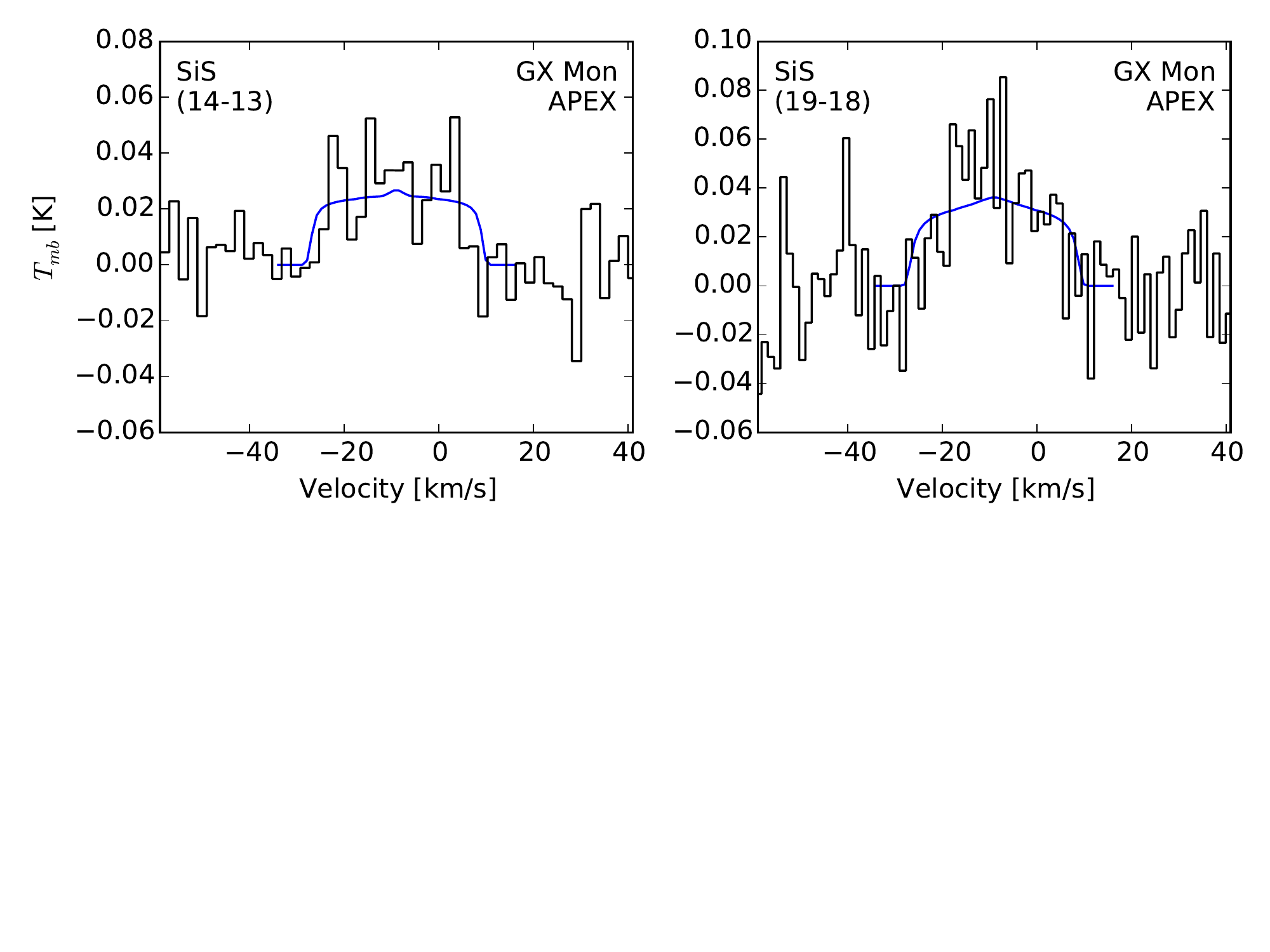}
\caption{Observations (black histograms) and model results (blue lines) for SiS towards the oxygen-rich star, GX~Mon plotted with respect to LSR velocity.}
\label{SiSMplots-2}
\end{figure}

\begin{figure}[t]
\includegraphics[width=0.49\textwidth]{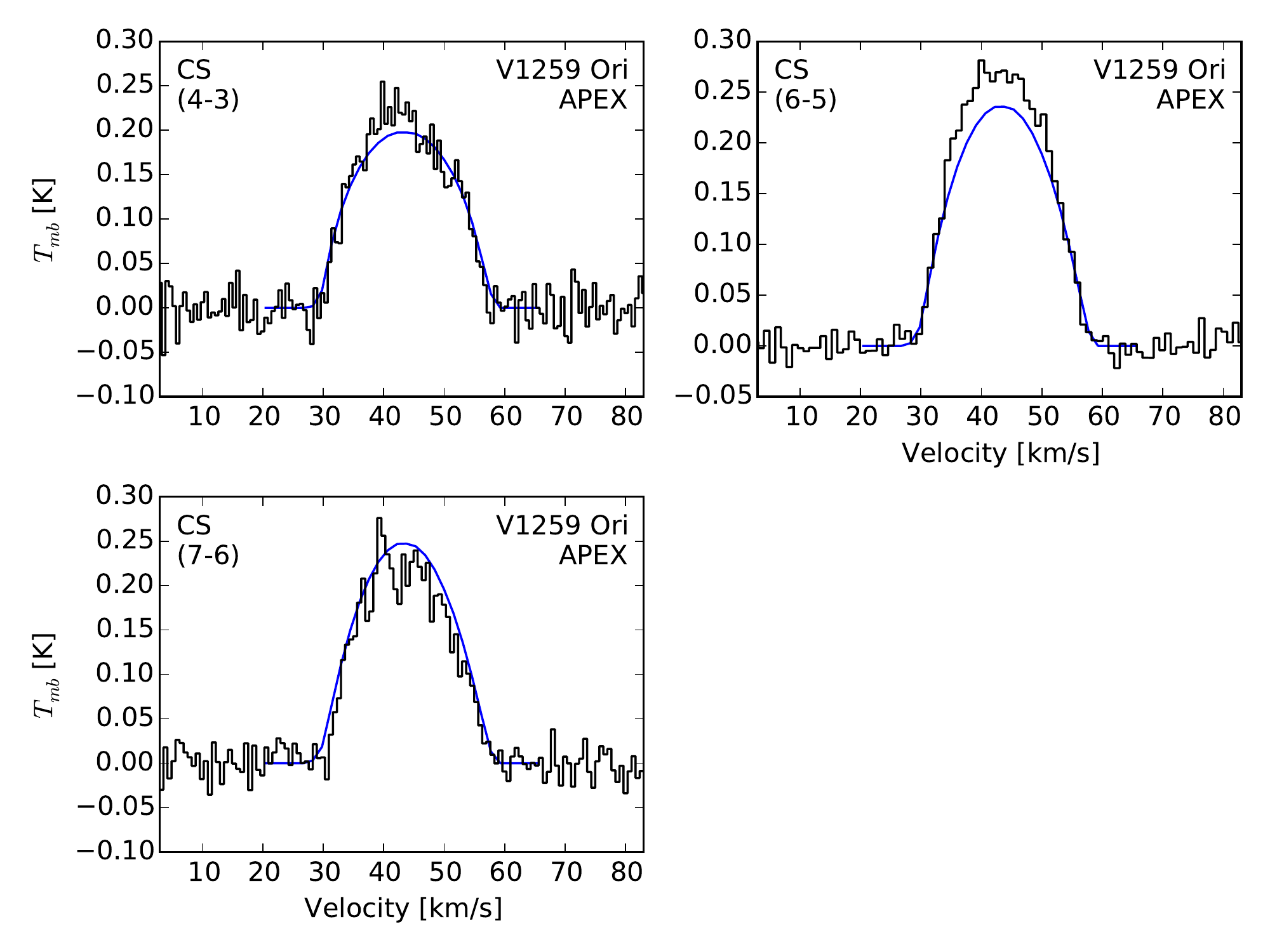}
\caption{Observations (black histograms) and model results (blue lines) for CS towards the carbon star V1259~Ori, plotted with respect to LSR velocity.}
\label{CSCplots-1}
\end{figure}

\begin{figure}[t]
\includegraphics[width=0.49\textwidth]{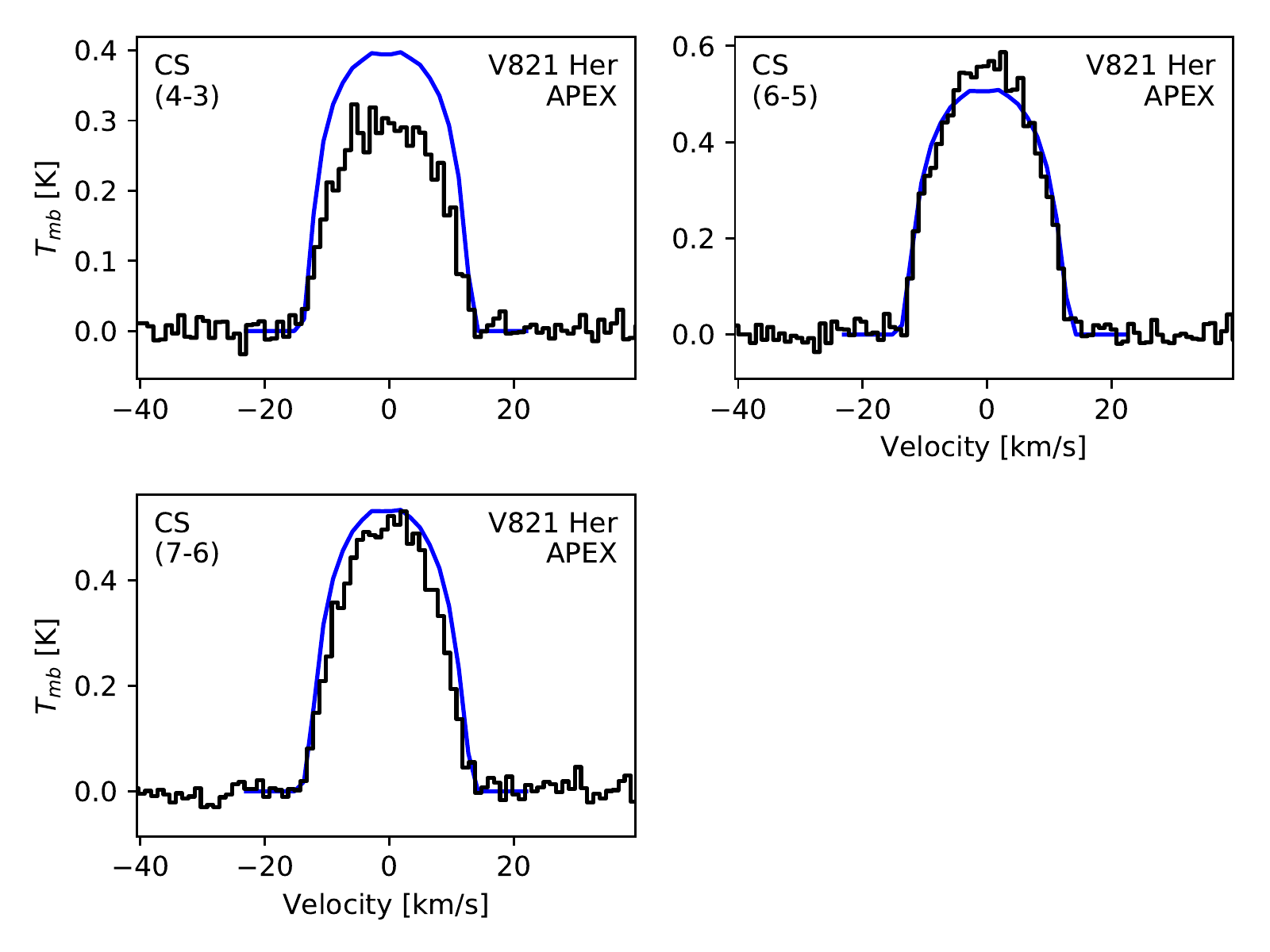}
\caption{Observations (black histograms) and model results (blue lines) for CS towards the carbon star V821~Her, plotted with respect to LSR velocity.}
\label{CSCplots-2}
\end{figure}

\begin{figure}[t]
    \includegraphics[width=0.49\textwidth]{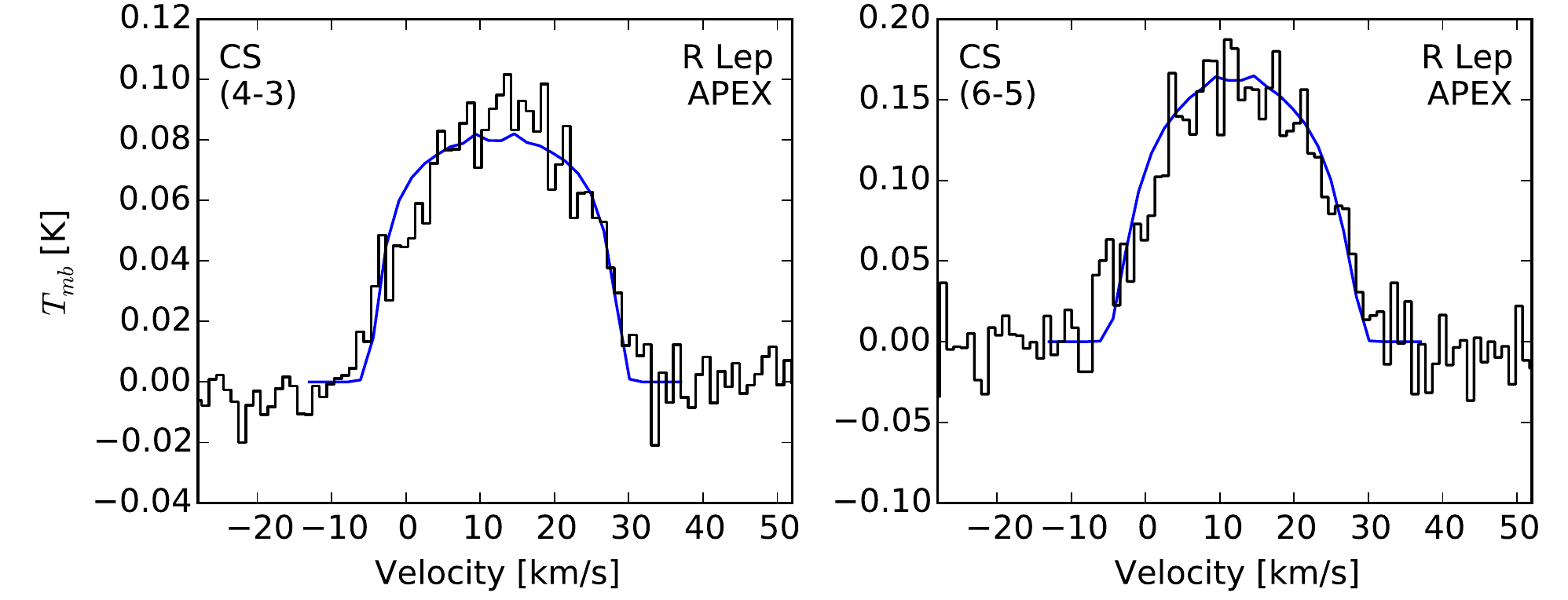}
\caption{Observations (black histograms) and model results (blue lines) for CS towards the carbon star R~Lep, plotted with respect to LSR velocity.}
\label{CSCplots-3}
\end{figure}

\begin{figure}[t]
        \includegraphics[width=0.49\textwidth]{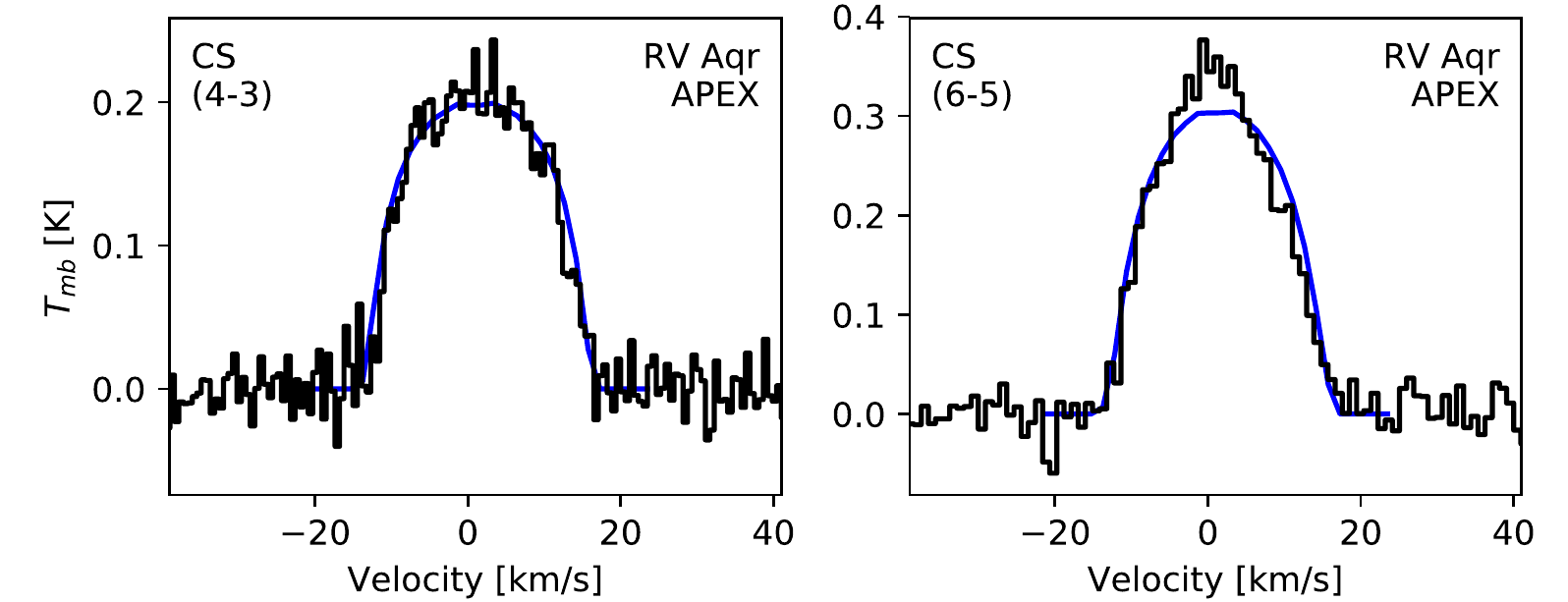}
\caption{Observations (black histograms) and model results (blue lines) for CS towards the carbon star RV~Aqr, plotted with respect to LSR velocity.}
\label{CSCplots-4}
\end{figure}

\begin{figure}[t]
\includegraphics[width=0.49\textwidth]{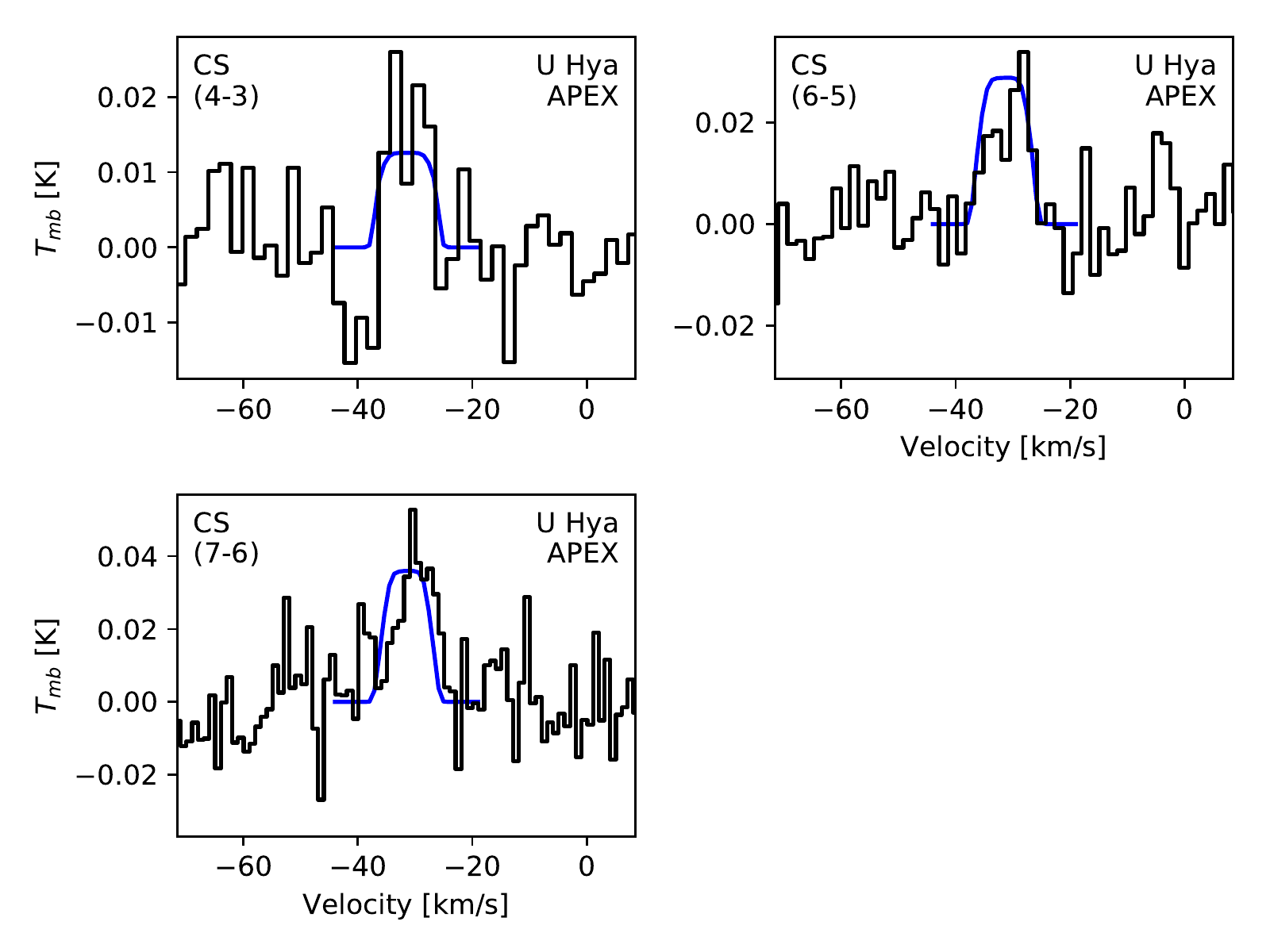}
\caption{Observations (black histograms) and model results (blue lines) for CS towards the carbon star U~Hya, plotted with respect to LSR velocity.}
\label{CSCplots-5}
\end{figure}

\begin{figure}[t]
\includegraphics[width=0.49\textwidth]{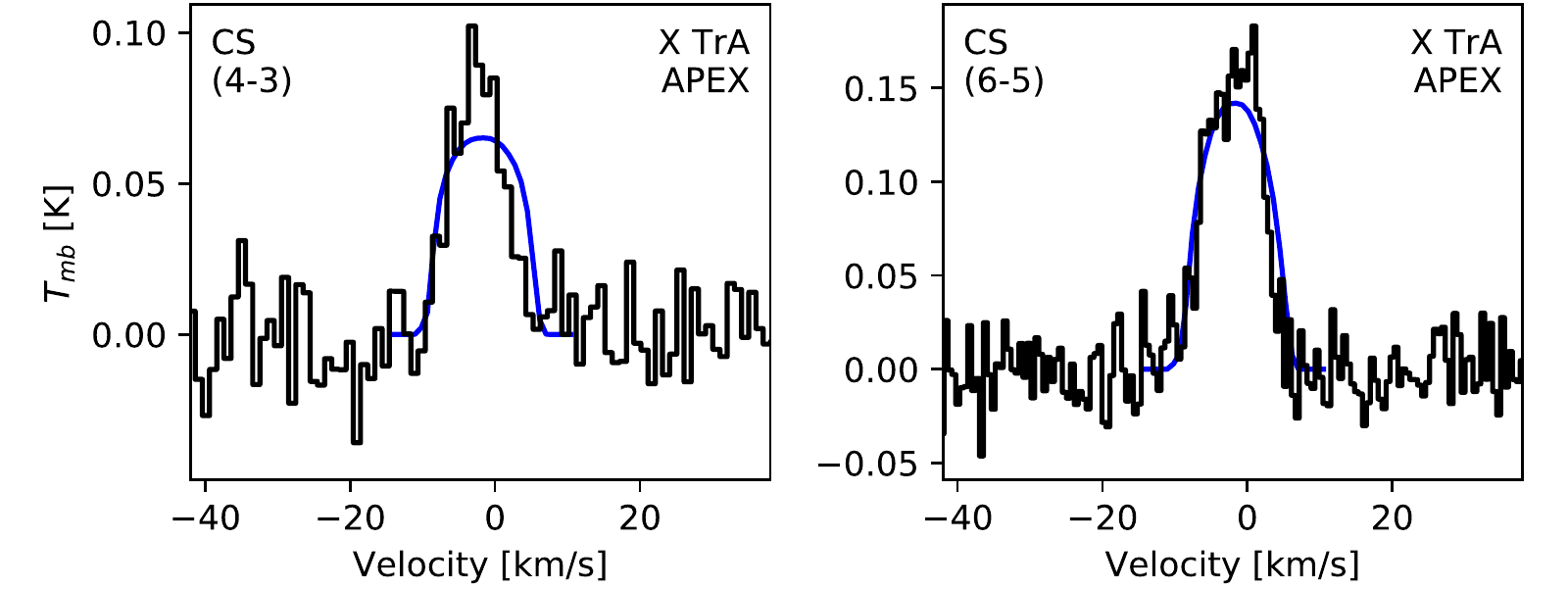}
\caption{Observations (black histograms) and model results (blue lines) for CS towards the carbon star X~TrA, plotted with respect to LSR velocity.}
\label{CSCplots-6}
\end{figure}

\begin{figure}[t]
\includegraphics[width=0.49\textwidth]{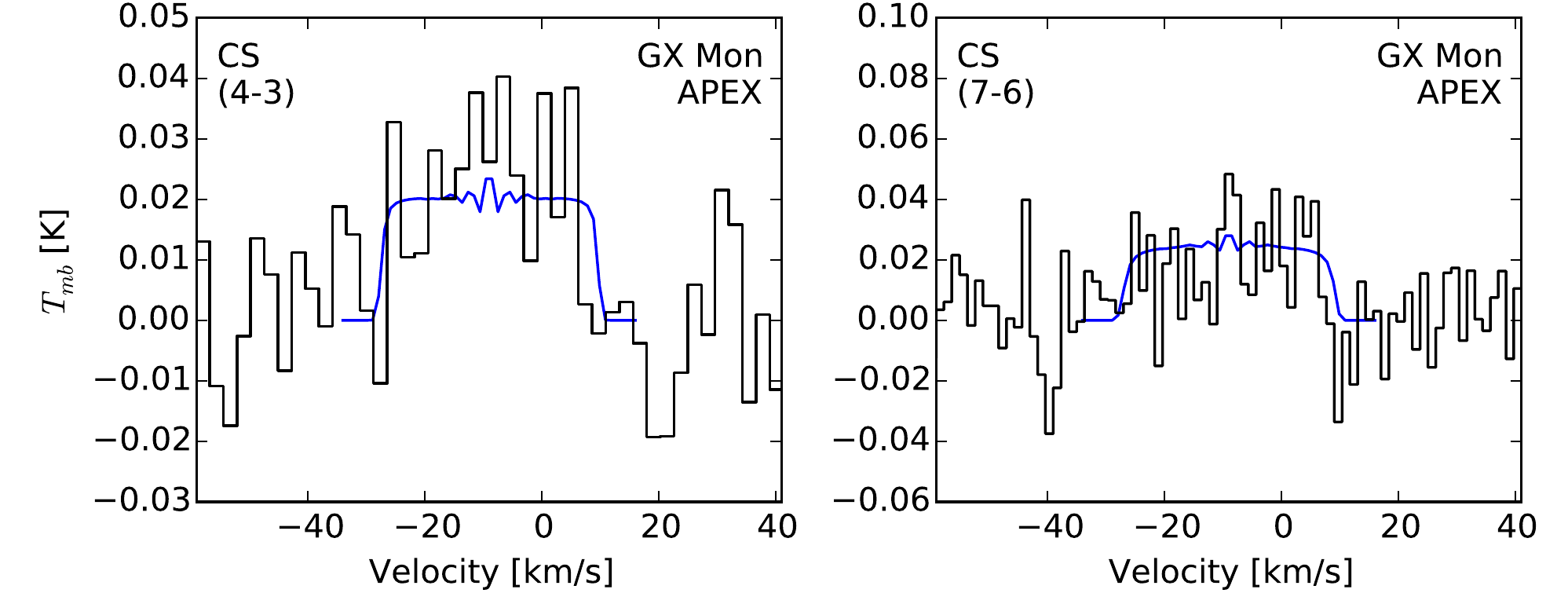}
\caption{Observations (black histograms) and model results (blue lines) for CS towards the oxygen-rich star GX~Mon, plotted with respect to LSR velocity.}
\label{CSMplots-1}
\end{figure}

\begin{figure}
\includegraphics[width=0.49\textwidth]{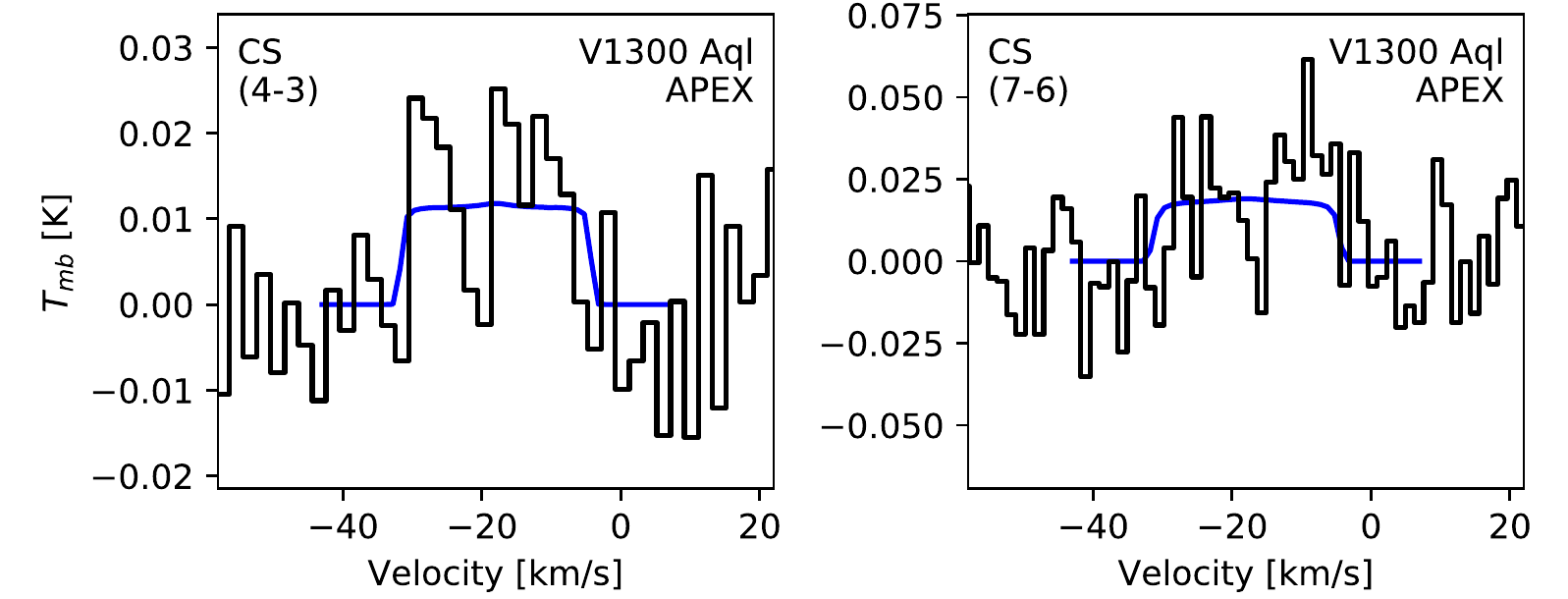}
\caption{Observations (black histograms) and model results (blue lines) for CS towards the oxygen-rich star V1300~Aql, plotted with respect to LSR velocity.}
\label{CSMplots-2}
\end{figure}

\begin{figure}[t]
\begin{center}
\includegraphics[width=0.49\textwidth]{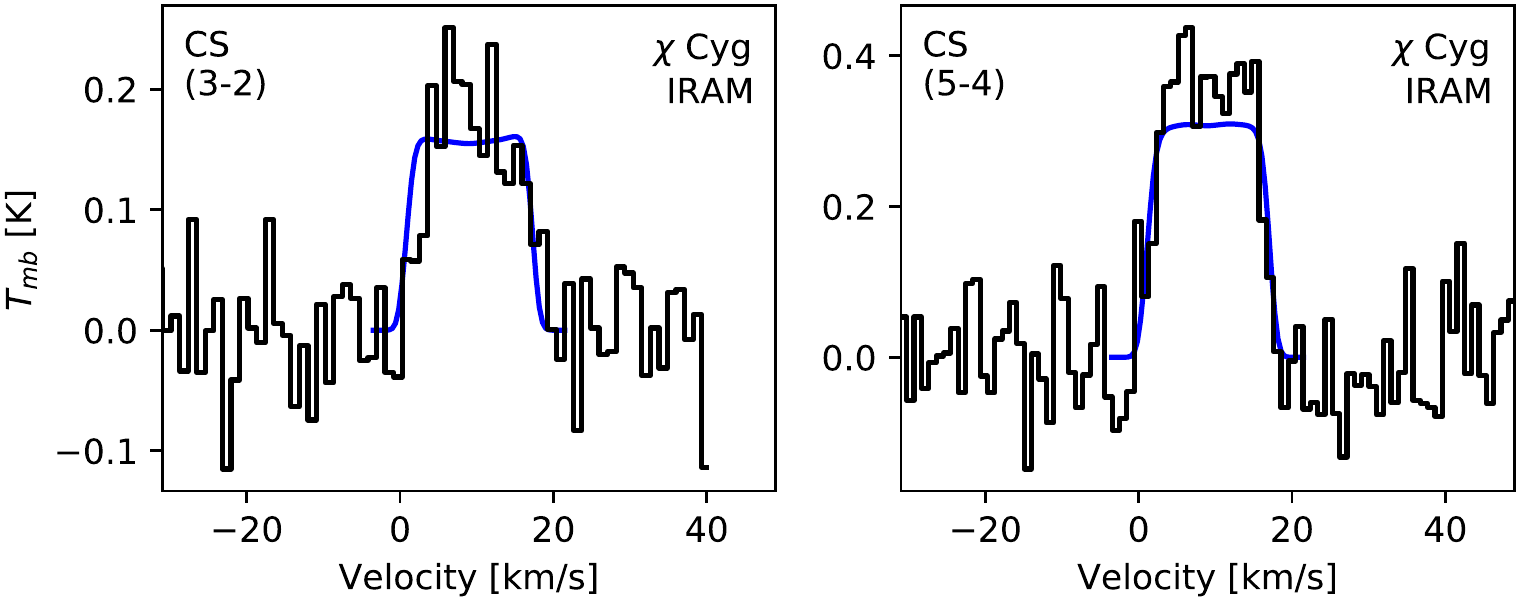}
\caption{Observations (black histograms) and model results (blue lines) for CS towards $\chi$~Cyg, an S-type star, plotted with respect to LSR velocity.}
\label{CSSplots}
\end{center}
\end{figure}

SiS models and observations for the carbon stars are plotted in {Figures \ref{SiSCplots-1}, \ref{SiSCplots-2}, and \ref{SiSCplots-3} with the same plotted for CS in Figures \ref{CSCplots-1}, \ref{CSCplots-2}, \ref{CSCplots-3}, \ref{CSCplots-4}, \ref{CSCplots-5}, and \ref{CSCplots-6}}. The SiS model results and observations for the M-type stars are plotted in {Figures \ref{SiSMplots-1} and \ref{SiSMplots-2}, with the same for CS plotted in \ref{CSMplots-1} and \ref{CSMplots-2}}. The CS model and observations for $\chi$ Cyg, an S-type star, are plotted in Fig. \ref{CSSplots}.


\clearpage


\section{Further discussion of modelling}

We encountered some modelling issues that only pertained to a few stars. For IRC~-10401, we recalculated some key circumstellar parameters, including the mass-loss rate, based on an updated estimation of the period. For three carbon stars (V1259~Ori, AI~Vol, and II~Lup) we encountered various issues when we were modelling the molecular emission.
We discuss these in detail below.

\subsection{IRC -10401}\label{irc-10401}

For IRC -10401, we recalculated the mass-loss rate based on newly available data. When \cite{Ramstedt2009} first calculated the mass-loss rate of this S-type star, the period was not known and they assumed $L_* = 4000~\lsol$ to determine the distance based on dust radiative transfer modelling of the spectral energy distribution. However, now the period has been determined to be 480 days \citep{Kazarovets2016} and, based on the period luminosity relation of \cite{Whitelock2008}, this gives $L_* = 7400~\lsol$, which in turn gives an updated distance of 585 pc (larger than the previous value of 430 pc). Using these updated parameters, we remodelled the same CO observations used by \cite{Ramstedt2009}, using the same procedure, to find an  updated higher mass-loss rate of $2\e{-6}\spy$ (compared with the earlier result of $3.5\e{-7}$). This updated result is what we base our SiS and CS models on in this work.

\subsection{V1259~Ori}\label{v1259ori}

For V1259~Ori we have clear CS detections with high signal-to-noise ratios that are nevertheless equally well fit by the model listed in Table \ref{results} and by models with significantly larger $e$-folding radii, including $R_e > R_{1/2}(\mathrm{CO}) = 2.6\e{17}$~cm, the half abundance radius of CO. We do not expect CS to have a larger extent than CO, which is self-shielding and has a lower photodissocation rate than CS \citep{Heays2017}.
%photodissocation energy of 11.1~eV compared with 7.4~eV for CS \citep{Herzberg1989}, 
This discrepancy in modelling CS is partly caused by the high optical depth of the observed CS lines, which reach optical depths of 20, 40, and 50 in the inner CSE for the ($4\to3$), ($6\to5$), and ($7\to6$) lines, and remain optically thick throughout most of the emitting region. Our model predicts that the ($1\to0$) line at 48.991~GHz would be optically thin throughout most of the envelope (with an optically thick peak in the inner CSE of only three) and is likely to allow us to properly constrain the CS envelope size. Alternatively, interferometric data that resolves the CS extent would also allow us to very precisely constrain the envelope size.

\subsection{AI Vol}\label{aivol}

For AI~Vol we found that the combined abundances of CS and SiS result in a total S abundance higher than expected based on the assumption that the solar S abundance is representative of the local environment within $\sim1$~kpc. Summing CS and SiS abundances, we find a total S abundance of $4.2\e{-5}$, more than one and a half times the solar abundance. However, the lower limit of $2.6\e{-5}$, based on our uncertainties, is equal to the solar abundance given by \cite{Asplund2009}.  Since the CS model is based on only two observed lines and has errors on the fractional abundance in excess of 50\% of the absolute value, it's likely that a larger dataset with higher-$J$ observations would improve our model results. Similarly, although we have seven lines for the SiS model, the highest energy line is the SiS ($19\to18$) with an upper level energy of 166~K, which is not expected to be a good probe of the inner regions of the CSE. Hence higher-$J$ SiS observations would also improve our SiS model.

Another possible cause for our result of an apparent sulphur overabundance is the uncertainty in the mass-loss rate, which is based on several input parameters, each of which have their own uncertainty, as discussed in more detail in \cite{Danilovich2015a}. For example, uncertainty in the distance would affect the calculation of the mass-loss rate (as seen in Sect. \ref{irc-10401} for \mbox{IRC -10401}) and hence the abundances we derive in this work. The distance to AI~Vol was originally taken from \cite{Woods2003} who note a typical uncertainty of up to a factor of two for their distances.

\subsection{II Lup}\label{iilup}

\begin{figure*}[t]
\begin{center}
\includegraphics[width=\textwidth]{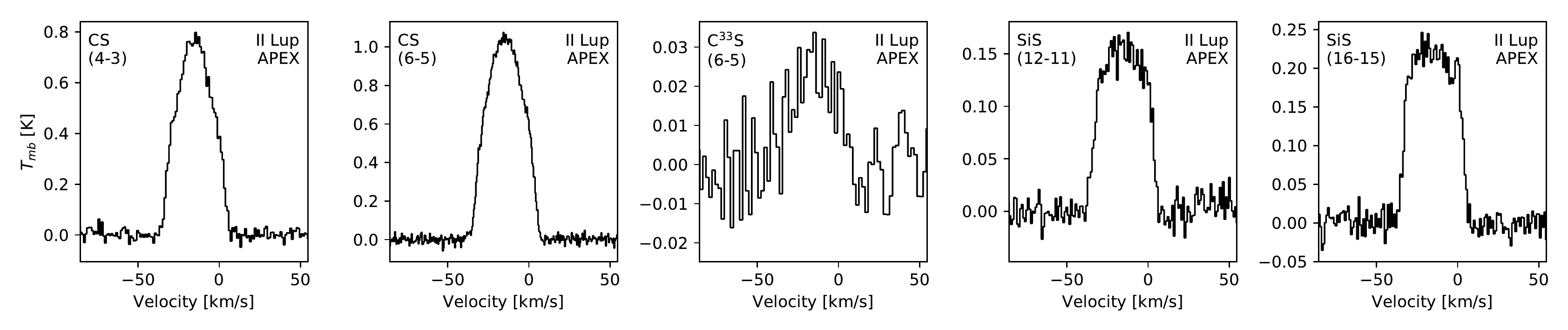}
\caption{Observations of CS and SiS towards II Lup, plotted with respect to LSR velocity.}
\label{iilupobs}
\end{center}
\end{figure*}

We were unable to satisfactorily model the CS and SiS emission towards II~Lup. It was not possible to find a smoothly accelerating model with a constant mass-loss rate which agreed with the available observations for either CS or SiS. Some irregularities can be tentatively seen in Fig. \ref{iilupobs} --- and become clearer when attempting to fit models. These have also been seen in other molecular lines, such as CO, as discussed in more detail in \cite{Smith2014}.

When attempting to model the data acquired in our APEX survey and the two lines --- CS ($7\to6$) and SiS ($19\to18$) --- retrieved from the APEX archive (see Table \ref{supobs}) it was not possible to find a model which agreed with all three available lines for CS nor for SiS. We cannot constrain the molecular envelope size from the data, as we were able to do with most of the other observed stars, and radii calculated from Eqs. \ref{resis} and \ref{recs} do not give a satisfactory fit to the data. In the case of both molecules, a model which fits the two lower-$J$ lines well over-predicts the higher-$J$ line by a factor of approximately two. In the case of SiS, the ($19\to18$) is also narrower than the model prediction by $\sim 6~\kms$, even when taking an accelerating wind into account. These characteristics are suggestive of more complex conditions in the CSE than those taken into account by our models.
To properly untangle the contributions from various components in the wind of II~Lup, we require more observations of low- and high-energy emission from II~Lup, preferably including some spatially resolved observations.

\section{Isotopologues}\label{isotopologues}

As noted in Tables \ref{SiSobs} and \ref{CSobs}, in addition to the main isotopologues, our observations also {covered a line from each of} \up{29}Si\up{32}S, \up{28}Si\up{34}S, and \up{13}C\up{33}S. Of these, we only {tentatively} detected one \up{28}Si\up{34}S line, towards AI~Vol (for which another \up{28}Si\up{34}S line was also available, see Table \ref{supobs}), and one \up{13}C\up{33}S line towards II~Lup. 

\begin{figure}[t]
%\begin{center}
\includegraphics[width=0.49\textwidth]{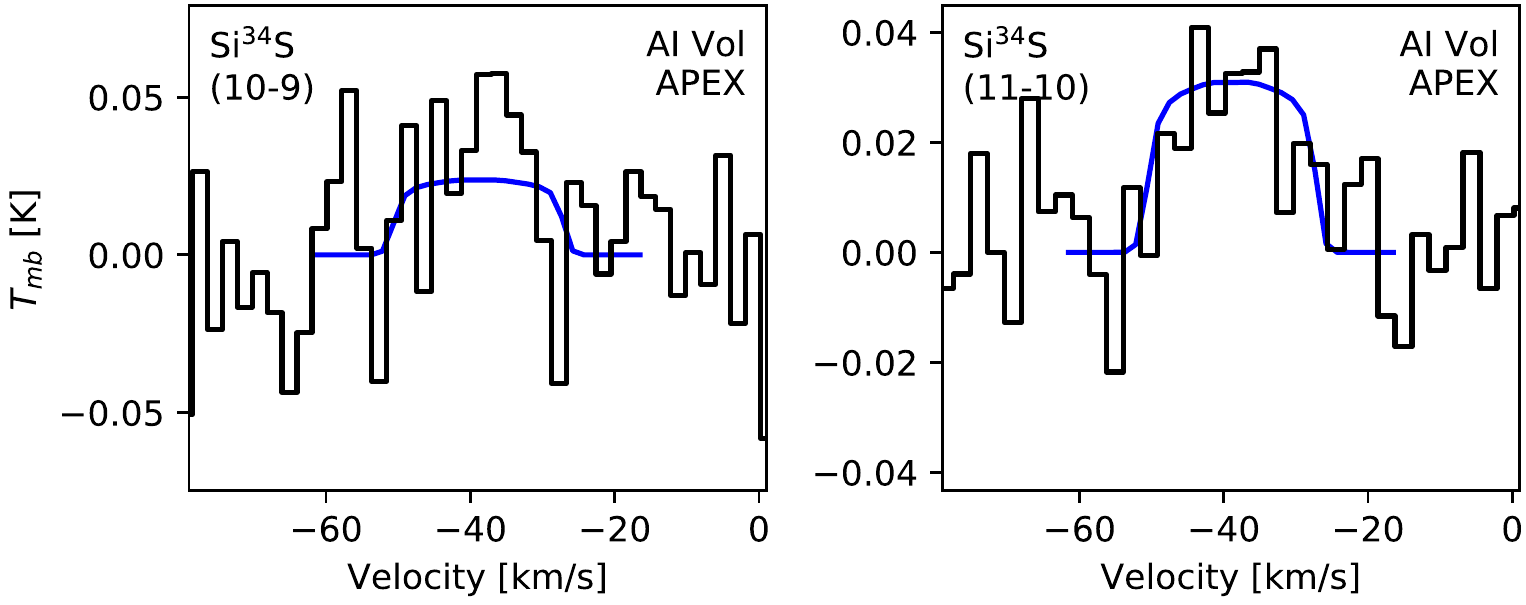}
\caption{Observations (black histograms) and model results (blue lines) for Si\up{34}S towards AI~Vol, plotted with respect to LSR velocity.}
\label{Si34Splots}
%\end{center}
\end{figure}

To model the \up{28}Si\up{34}S emission towards AI~Vol, we used a molecular data file constructed equivalently to the main isotopologue, with the same quantum-numbered energy levels and radiative transitions included. The same collisional rates were used as for the main isotopologue (see Sect. \ref{moldat}). We adopted the $R_e$ found for the main isotopologue as a fixed parameter, and varied the peak abundance, $f_0$, to find a model that fit the isotopologue lines. The data and best-fitting model are plotted in Fig. \ref{Si34Splots}.
We find a peak Si\up{34}S abundance of $(2.1^{+1.9}_{-1.6})\e{-6}$. Our result gives an Si\up{32}S/Si\up{34}S  of $11.4 \pm 10.3$, smaller than the solar system value for \up{32}S/\up{34}S of 22.1 \citep{Asplund2009}.

For II Lup we could not reliably model \up{12}C\up{33}S due to the problems we had modelling \up{12}C\up{32}S (discussed further in Sect. \ref{iilup}). The observed isotopologue line is plotted in Fig. \ref{iilupobs}.

\end{document}